\documentclass[twocolumn,showpacs,amssymb,aps,prd,floatfix,groupedaddress,nofootinbib]{revtex4-1}
\usepackage{amsmath}
\usepackage{graphicx}
\usepackage{color}
\usepackage{float}

\usepackage{tabularx}
\usepackage{pbox}

\usepackage{comment}

\usepackage{pifont}

\newcommand{\beq}{\begin{equation}}
\newcommand{\eeq}{\end{equation}}
\newcommand{\beqnarr}{\begin{eqnarray}}
\newcommand{\eeqnarr}{\end{eqnarray}}
\newcommand{\eexp}{\mathrm{e}}
\newcommand{\I}{\mathrm{i}}

\newcommand{\rmd}{\mathrm{d}}

\begin{document}

\title{Comparison of algorithms for solving the sign problem in
the O(3) model in 1+1 dimensions at finite chemical potential}

\author{
S. D. Katz$^{1,2}$,
F. Niedermayer$^{3}$,
D. N\'ogr\'adi$^{1,2}$,
Cs. T\"or\"ok$^{1,2}$
}

\affiliation{$^1$ \small{\it Institute for Theoretical Physics, E\"otv\"os University, \\
P\'azm\'any P\'eter s\'et\'any 1/A, H-1117 Budapest, Hungary} \\
$^2$ \small{\it MTA-ELTE "Lend\"ulet" Lattice Gauge Theory Research Group, \\
P\'azm\'any P\'eter s\'et\'any 1/A, H-1117 Budapest, Hungary} \\
$^3$ \small{\it Institute for Theoretical Physics, Albert Einstein Center for Fundamental Physics Bern University, \\
Sidlerstrasse 5, CH-3012 Bern, Switzerland}
}

\date{\today}

\begin{abstract}
We study three possible ways to circumvent the sign problem in the 
O(3) nonlinear sigma model in 1+1 dimensions. We compare the results of the worm algorithm to complex Langevin 
and multiparameter reweighting. Using the worm algorithm, the thermodynamics of the model is investigated, 
and continuum results are shown for the pressure at different $\mu / T$ values in the range $0-4$. 
By performing $T=0$ simulations using the worm algorithm, the Silver Blaze phenomenon is reproduced.
Regarding the complex Langevin, we test various implementations of discretizing the complex Langevin equation. 
We found that the exponentialized Euler discretization of the Langevin equation gives wrong results 
for the action and the density at low $T/m$. By performing a continuum extrapolation,
we found that this discrepancy does not disappear and depends slightly on temperature.
The discretization with spherical coordinates performs similarly at low $\mu / T$ but breaks down
also at some higher temperatures at high $\mu / T$. However, a third discretization that uses a constraining force to 
achieve the $\phi^2 = 1$ condition gives correct results for the action but wrong results for the 
density at low $\mu / T$.
\end{abstract}

\maketitle


\section{Introduction\label{sec:intro}}

Monte Carlo simulations of quantum field theories based on the path integral 
formalism play an important role in investigating the physics of various 
models nonperturbatively. However, standard numerical methods fail when the action 
becomes complex, and thus the probability interpretation of the weight $e^{-S}$ 
and importance sampling cannot be applied. 
The problem is present in QCD at finite density or with a theta term and also 
arises in condensed matter physics, e.g. in the simulations of strongly correlated 
electronic systems \cite{Loh:1990zz}. To solve these complex action problems, several 
methods have been devised; for a review of different approaches and further references, 
see Refs. \cite{deForcrand:2010ys, Aarts:2013naa, Aarts:2015kea, Sexty:2014dxa}. 

In the present paper, we compare and test three different methods, namely reweighting, 
the worm algorithm, and the complex Langevin in the case of the 1+1-dimensional O(3) model. 
Our focus is primarily on the applicability of the complex Langevin algorithm, since 
today it may seem that it is a promising approach to simulate even QCD \cite{Sexty:2013ica, Fodor:2015doa}, 
although several problems have not been solved yet. The idea behind complex Langevin 
is stochastic quantization and originates from the work of Parisi, Wu, and Klauder from the 1980s \cite{Parisi:1980ys, 
Parisi:1984cs, Klauder:1983nn}. But soon after its proposal, the first simulations revealed certain problems: 
the instability of the simulations with the absence of convergence (runaway trajectories) \cite{Ambjorn:1985iw}
and that even stable simulations may converge to a wrong limit \cite{Ambjorn:1986fz}. 
These problems hindered reliable calculations, but in the last decade, important improvements 
have been achieved. Runaway trajectories can now be eliminated e.g. using adaptive step size \cite{Aarts:2009dg}, 
and a formal justification of the algorithm as well as necessary and sufficient conditions for 
convergence to correct results have been established \cite{Aarts:2009uq, Aarts:2011ax, Aarts:2013uza}.
Roughly speaking, these suggest that if the probability distributions of the complexified 
variables fall sufficiently fast, then the results are correct. In order to reach this for gauge theories, 
the gauge cooling procedure was developed \cite{Seiler:2012wz}, which works perfectly in some models or at 
a certain parameter range, but may fail in other models and parameter ranges \cite{Makino:2015ooa, Bloch:2015coa, 
Seiler:2012wz, Aarts:2013nja}. In Ref. \cite{Aarts:2013nja}, where heavy dense QCD (HDQCD) was studied, 
it was argued that failure happens below a specific $\beta$ value and by 
increasing the temporal lattice size, one can get correct results at lower temperatures, 
in other words continuum extrapolation may be feasible.
The validity of this statement, however, is not entirely clear and may be model dependent.
On the one hand, the above observation in HDQCD helped in exploring the phase diagram of the model \cite{Aarts:2014kja},
but on the other hand, in the case of full QCD, recent results \cite{Fodor:2015doa} show that,
using $N_t=4,6,8$ lattices, the breakdown of complex Langevin prevents the exploration of the confined region.
We note that e.g. for the 3D XY model the breakdown of the complex Langevin also 
occurred around the phase boundary \cite{Aarts:2010aq}, but in that model the question of continuum limit behavior 
cannot be addressed.

In the present paper, we investigate the 1+1-dimensional O(3) model for this purpose, which is not 
a gauge theory but asymptotically free; thus, the continuum behavior can be analyzed. 
We compare the results of different discretizations of the complex Langevin equation to 
the results of reweighting and the worm algorithm.

From the viewpoint of the sign problem, these two approaches are also interesting and can 
give insight into the properties of the O(3) model.

In this sense, reweighting is a well-defined approach, but with limited efficiency and reliability as the 
sign problem becomes more severe (and also an overlap problem appears). The worm algorithm, 
also called the dual variables approach \cite{Prokof'ev:2001zz}, however, completely eliminates the sign problem 
of the model by introducing new, dual variables. The difficulty in this case is the rewriting of the model 
to these dual variables, but after it has been accomplished, effective simulations using the worm 
algorithm can be performed, and in fact many interesting models have been studied throughout the years 
\cite{Endres:2006zh, Endres:2006xu, Wolff:2008km, Banerjee:2010kc, Mercado:2012yf, Langfeld:2013kno, 
Mercado:2013jma, Herland:2013fv, Gattringer:2015baa, Yang:2015rra}. 
Here, using the worm algorithm we study the thermodynamic properties of the O(3) model.

Although the dual formalism of this model was introduced and studied \cite{Bruckmann:2015sua, Bruckmann:2015hua}
during the finalization of our work on the comparison of the different methods, we also introduce this
formalism in this paper in order to give a consistent introduction to our notations.

In the following sections, after some introductory remarks about the O(3) model in Sec. \ref{sec:form}
and the description of scale setting in Sec. \ref{sec:scale_setting},
we discuss these approaches in more detail: in Sec. \ref{sec:rew} the reweighting, 
in Sec. \ref{sec:worm} the worm algorithm, and in Sec. \ref{sec:CL} the complex Langevin.
In Sec. \ref{sec:comparison}, we compare the results of the simulations.
In the Appendix, we discuss in detail the updating steps of the worm algorithm.

\section{Formulation\label{sec:form}}

The O(3) model in 1+1 dimensions has been widely studied in the past for several reasons, amongst others because it has 
interesting features in common with four-dimensional non-Abelian gauge theories. Over the years, many 
important results have been achieved also numerically and -- since the model is more or less tractable -- analytically.
Nonetheless, we do not give an overview here of the overall history of these results, but mention
only some facts that made this model attractive for us.

First of all, the coupling constant of the theory is dimensionless; thus, the theory is 
perturbatively renormalizable. It is asymptotically free in 1+1 dimensions \cite{Polyakov:1975rr, Brezin:1976qa}, 
which enables us to study the continuum limit of the results obtained at finite lattice spacings.
The O(3) model also has a nonperturbative mass gap generated dynamically.
Moreover, similarly to QCD, the O(3) model also possesses instanton solutions \cite{Novikov:1984ac, Bruckmann:2007zh}.

The Lagrangian of the nonlinear O(3) model is
\beq
  \mathcal{L} = \frac{1}{2g^2} (\partial_{\mu} \phi)^2, 
\eeq
where the fields obey the $\sum_{i=1}^3 \phi_i^2 = 1$ condition in every space-time point.

The discretized action in $1+1$ dimensions with periodic boundary condition is
\beqnarr
  S &=& \frac{1}{g^2} \left( 2 \sum_x \phi_x^2 - \sum_{x,\mu=0,1} \phi_{x+\hat{\mu}} \phi_x \right) \nonumber \\
    &=& 2 \beta V - \beta \sum_{x,\mu=0,1} \phi_{x+\hat{\mu}} \phi_x,
\eeqnarr
where we introduced $\beta = 1/g^2$ and the lattice volume $V=N_x \times N_t$.
After introducing the chemical potential to the rotations in the $(12)$-plane of O(3), 
the action becomes
\beq \label{ON_act,mu}
  S = 2 \beta V - \beta \sum_{x} \left( \phi_{x+\hat{0}} \eexp^{\I \mu a t_{12}} \phi_x + \phi_{x+\hat{1}} \phi_x \right),
\eeq
where $t_{12}$ is the generator of the rotation in the $(12)$-plane of O(3).
\vspace{-0.3cm}

\section{Scale setting} \label{sec:scale_setting}

\begin{figure}[H]
\begin{center}
\includegraphics[scale=0.7]{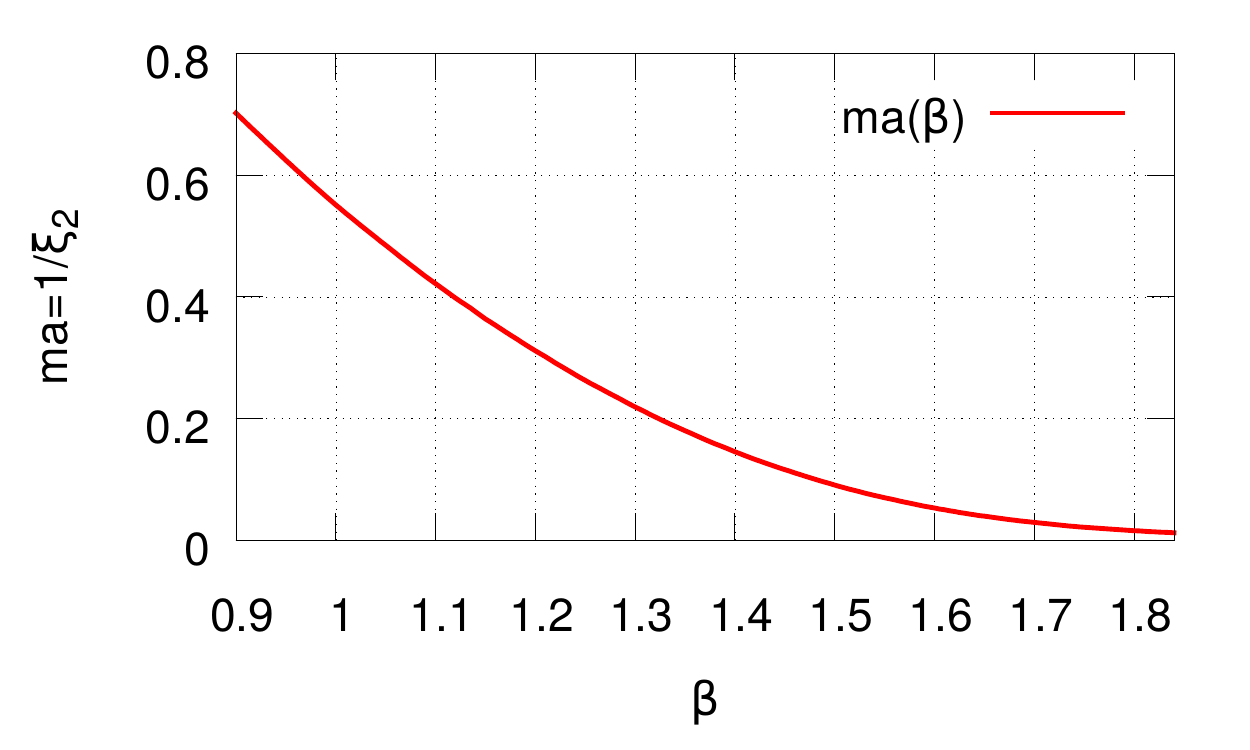}
\vspace{-0.7cm}
\caption{The $ma(\beta)$ scale for the 1+1-dimensional O(3) model. (Errors are smaller than or compatible 
with the width of the line.) The parameters and details of the simulations are summarized in the text.}
\label{O3_scale}
\end{center}
\end{figure}

Since in the later part of the paper we are interested in continuum extrapolations and 
physical quantities computed from the dimensionless quantities measured on the lattice, 
we need to determine the lattice spacing as a function of $\beta$, which we discuss in this section.
In order to achieve this, $\mu=0$ simulations have been performed, for which we used 
the cluster algorithm \cite{Swendsen:1987ce, Wolff:1988uh, Niedermayer:1988kv} and measured 
the second moment correlation length $\xi_2$ at zero temperature. 
$\xi_2$ is defined through
\beq \label{xi_2}
  \frac{1}{\xi_2} = \frac{\sin(\pi / N_t)}{\pi / N_t} \sqrt{ \left( \frac{2 M_0 a^2}{M_2} - \frac{4\pi^2}{N_t^2} \right) },
\eeq
where $M_0$ denotes the zeroth moment and $M_2$ denotes the second moment \cite{Caracciolo:1992nh, Nogradi:2012dj}:
\beq
  M_{2n} = \left( \frac{N_t a}{2\pi}  \right)^{2n} \sum_t \left(2 \sin\left(\frac{\pi t}{N_t a}\right)\right)^{2n} C(t).
\eeq
$C(t)$ is the two-point correlation function, $C(t) = \sum_x \langle \sum_a \phi_a(x,t) \phi_a(0,0) \rangle$.
$\xi_2$ does not equal $\xi=1/ma$ but scales as $\xi$ in the $\beta \rightarrow \infty$ limit
\cite{Caracciolo:1992nh, Caracciolo:1994cc, Caracciolo:1994ud}, and in infinite volume, the ratio $\xi / \xi_2$ is very 
close to 1; it is 1.000826(1) \cite{Balog:1999ww}. The advantage of using $\xi_2$ is that one does not need 
to fit any correlators this way. We can thus estimate the mass gap $ma = 1 / \xi$ 
with the help of $\xi_2$ by running large volume, zero temperature simulations. Actually, we used 
80$\times$80 and 100$\times$100 lattices for $0.9 \le \beta \le 1.57$, 120$\times$120 and 140$\times$140 lattices for 
$1.58 \le \beta \le 1.62$, 250$\times$250 lattices for $1.63 \le \beta \le 1.72$, and 400$\times$400 lattices 
for $1.73 \le \beta \le 1.85$. 
The simulation points were chosen uniformly in the above $\beta$ ranges 
with $\Delta \beta = 0.01$ distance from each other with $10^6$ or $2\times10^6$ cluster updates 
after thermalization, using every tenth for measurement. 
We studied the overlapping $\beta$ regions as well and used larger lattices if deviations larger than 
errors between the smaller and larger volumes had occurred. The results are shown in Fig. \ref{O3_scale}.
\vspace{-0.5cm}
\vspace{-1.0cm}

\section{Reweighting \label{sec:rew}}

Reweighting uses the idea of rewriting the partition function (and the expectation value of observables) 
in such a way, that one needs to do simulations only at zero $\mu$ and determine the configurations that 
can be relevant at finite $\mu$. This is done by measuring the weights of the configurations, 
which enables one to distinguish between them.

In the multiparameter reweighting approach \cite{Fodor:2001au}, one reweights both 
in $\beta$ and $\mu$, as we show it for the partition function of the O(3) model,
\beqnarr
  Z &=& \int \prod_x \rmd \phi_x \delta(\phi_x^2 - 1) \eexp^{-S(\beta, \mu)} \nonumber \\
    &=& \int \prod_x \rmd \phi_x \delta(\phi_x^2 - 1) \eexp^{-S(\beta_0, \mu_0=0)} w(\beta,\mu,\beta_0,\mu_0=0) \nonumber \\
    &=& Z_0 \langle w \rangle_{\beta_0, \mu_0=0},
\eeqnarr
where $w(\beta, \mu, \beta_0, \mu_0=0) = \eexp^{S(\beta_0, \mu_0=0) - S(\beta, \mu)}$ is the 
weight and $Z_0$ is the partition function for $\beta_0$ and $\mu_0=0$.
As one can see, this rewritten partition function can be simulated directly using standard methods 
since the action $S(\beta_0, \mu_0=0)$ is real. 
Using reweighting the expectation value of an $O$ observable is the following:
\beq
    \langle O(\beta, \mu) \rangle_{\beta, \mu} = 
\frac{\langle O(\beta, \mu) w(\beta, \mu, \beta_0, \mu_0=0) \rangle_{\beta_0, \mu_0=0}}
{\langle w(\beta, \mu, \beta_0, \mu_0=0) \rangle_{\beta_0, \mu_0=0}}.
\eeq
Although reweighting can reduce the sign problem, an overlap problem occurs in this case. 
That is, we have different important configurations at our "source" ($\beta_0, \mu=0$) ensemble and at the 
"target" ($\beta, \mu$) ensemble. If the two sets just slightly overlap or do not overlap at all, 
then one rarely reaches the important configurations at $\beta, \mu$ by simulating at $\beta_0, \mu_0=0$.

\vspace{-0.3cm}
\begin{figure}[H]
\begin{center}
\includegraphics[scale=0.6]{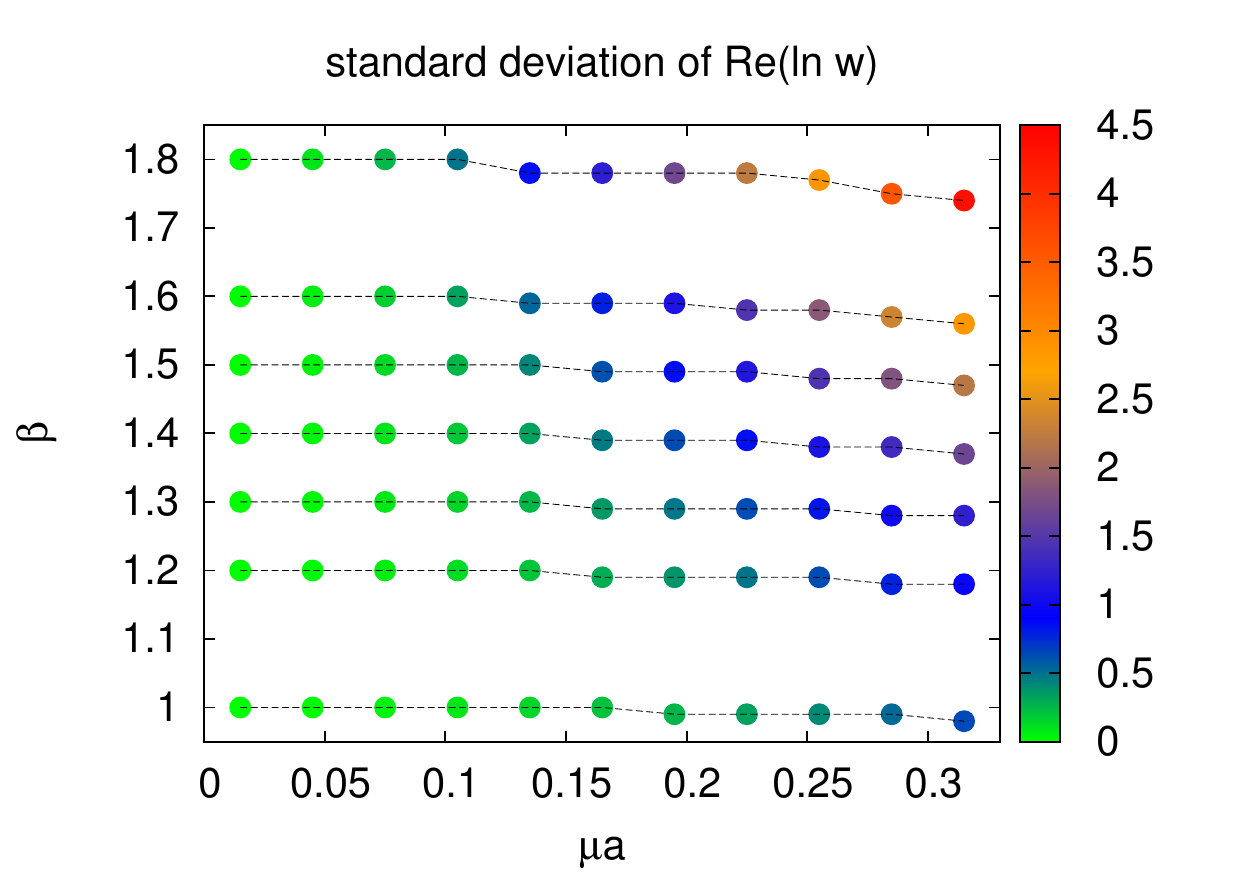}
\vspace{-0.3cm}
\caption{The standard deviation of $\mathrm{Re} (\ln w)$ for reweighting from 
$\beta_0 = 1, 1.2, 1.3, 1.4, 1.5, 1.6, 1.8$ to different $\beta, \mu$ values on 56$\times$14 lattices. 
We define the best reweighting lines (with dashed) as those that have the smallest standard deviations on them.}
\label{fig:rew_01}
\end{center}
\end{figure}
\vspace{-0.8cm}
\begin{figure}[H]
\begin{center}
\includegraphics[scale=0.58]{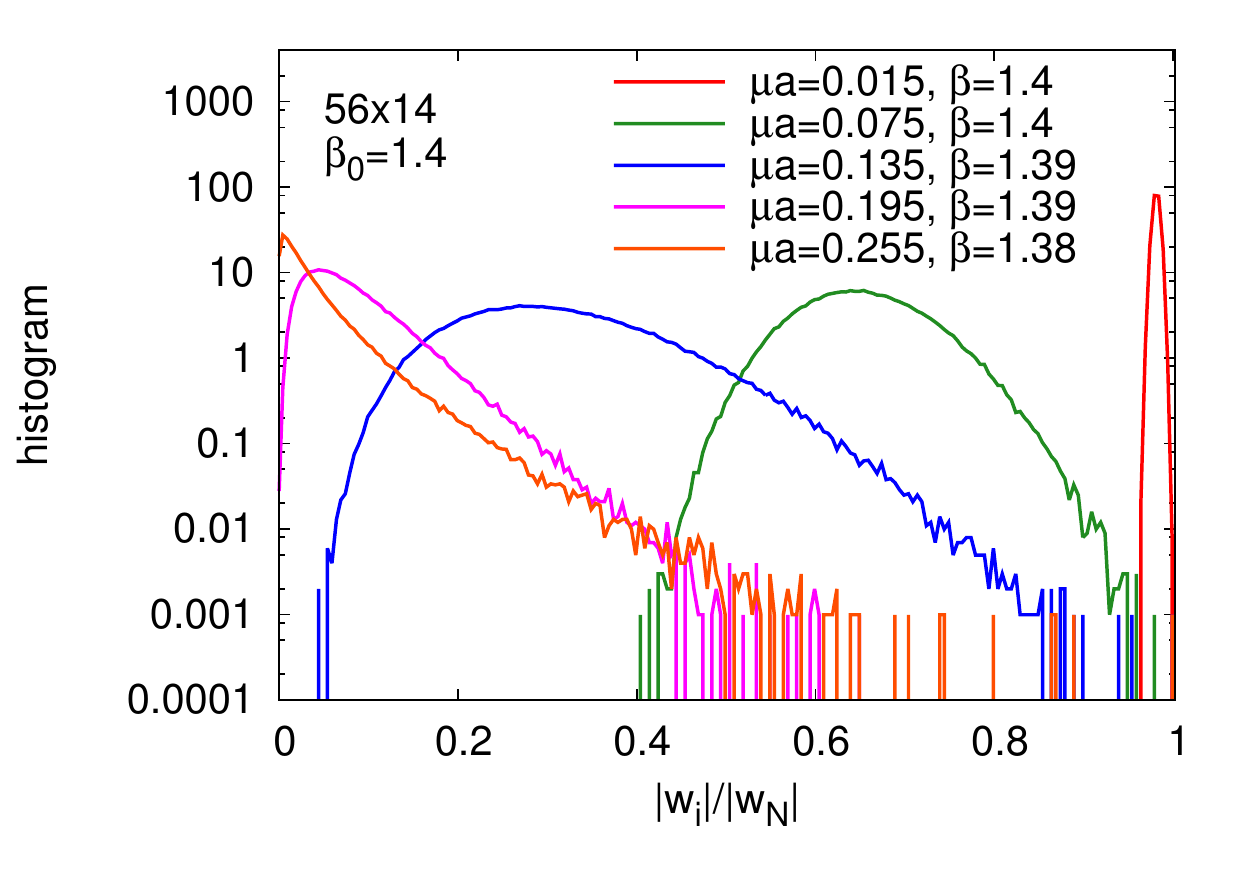}
\includegraphics[scale=0.58]{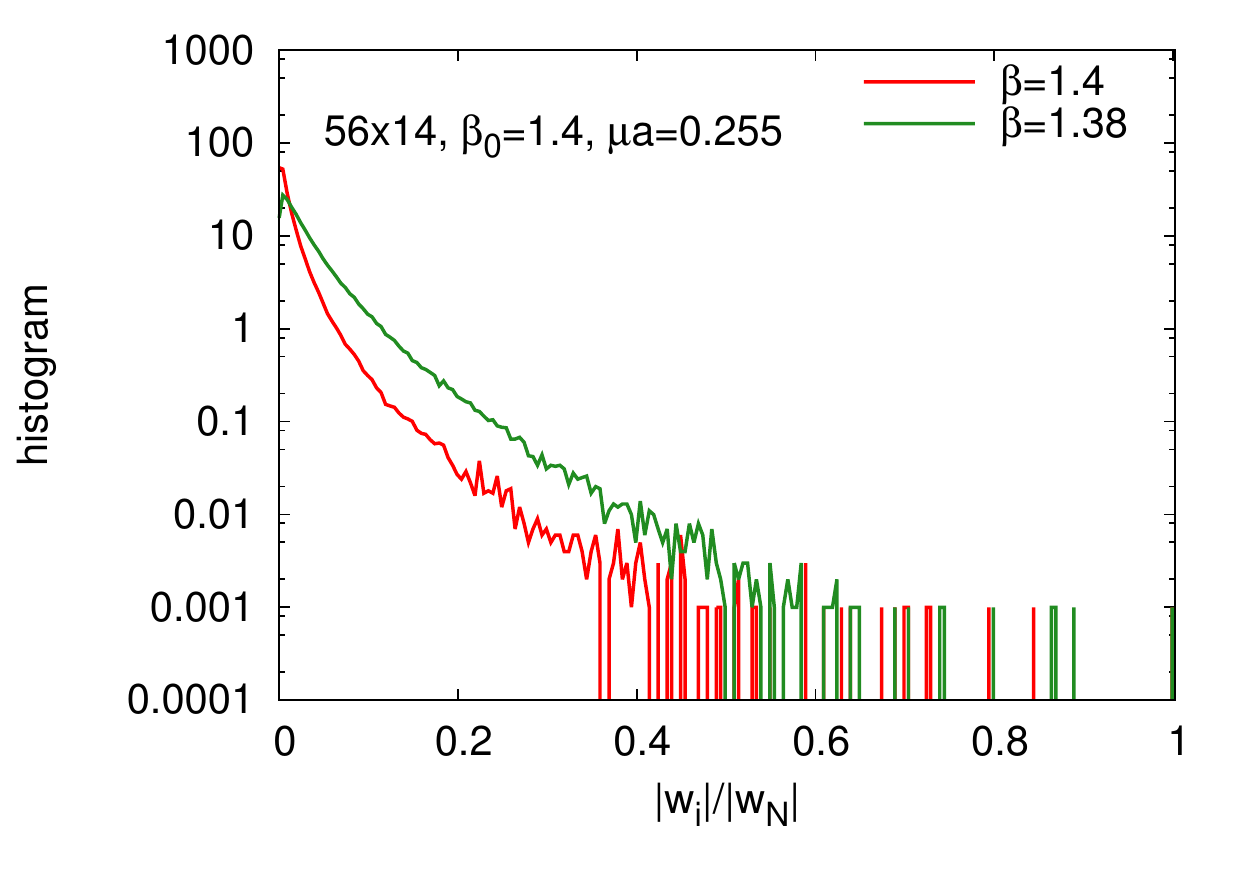}
\vspace{-0.5cm}
\caption{Top: The histogram of normalized weights along the best reweighting line 
starting from $\beta_0=1.4, \mu_0=0$ shows the severeness of the overlap problem. At the horizontal 
axis, one finds $|w_i|/|w_N|$ sorted as $|w_1| < |w_2| \ldots < |w_N|$, 
while at the vertical axis we show $\#/N/\Delta w$, where $\#$ is the number of configurations that have 
normalized weights between $|w_i|/|w_N|$ and $|w_i|/|w_N| + \Delta w$. $N$ is the total number of configurations 
and $\Delta w=0.005$.
Bottom: The figure illustrates how reweighting in $\beta$ can help to enhance overlapping: the green curve 
shows that one has more configurations with greater weights when reweighting from $\beta_0=1.4, \mu_0 a=0$ 
to $\beta=1.38, \mu a=0.255$, than in the case of e.g. standard reweighting.
} \label{fig:hist_rew}
\end{center}
\end{figure}
\begin{figure}[H]
\begin{center}
\includegraphics[scale=0.58]{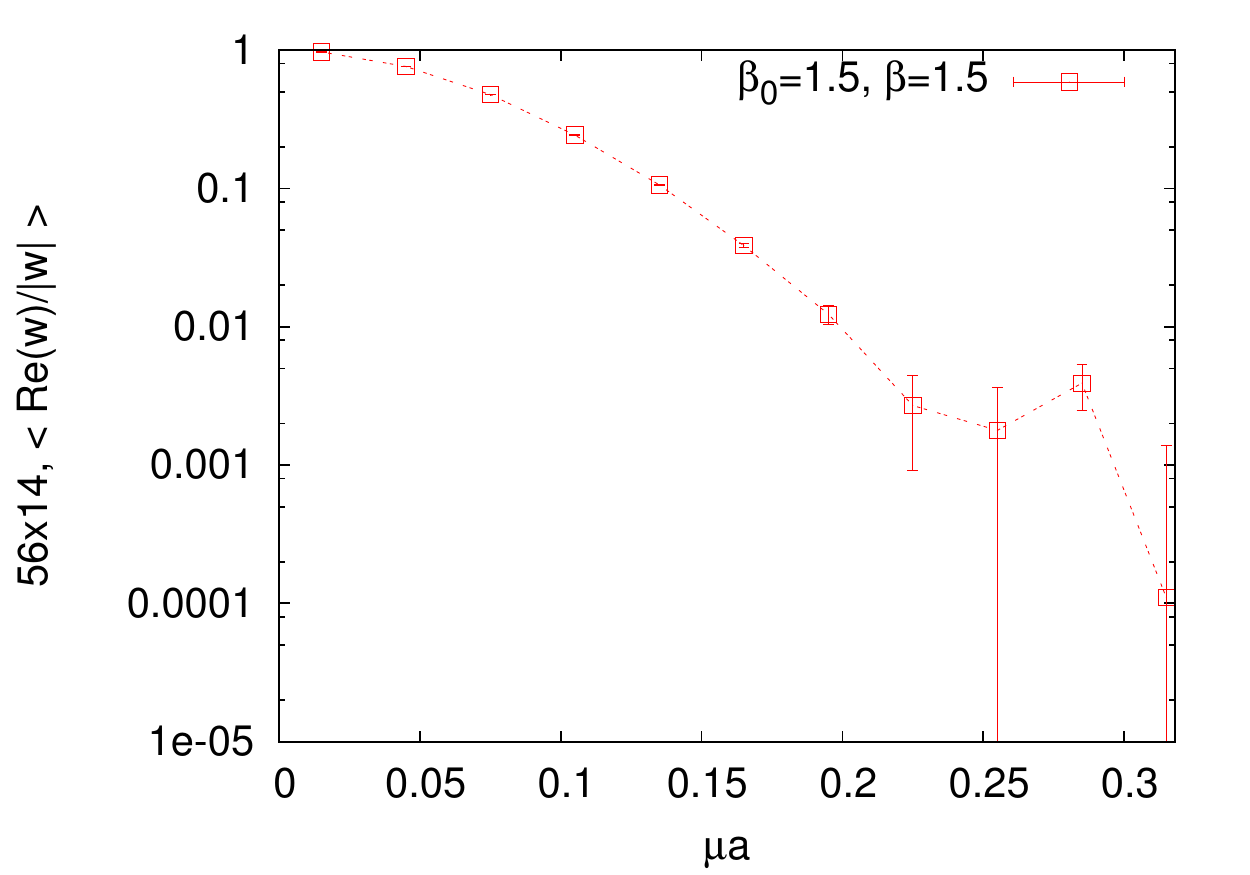}
\includegraphics[scale=0.58]{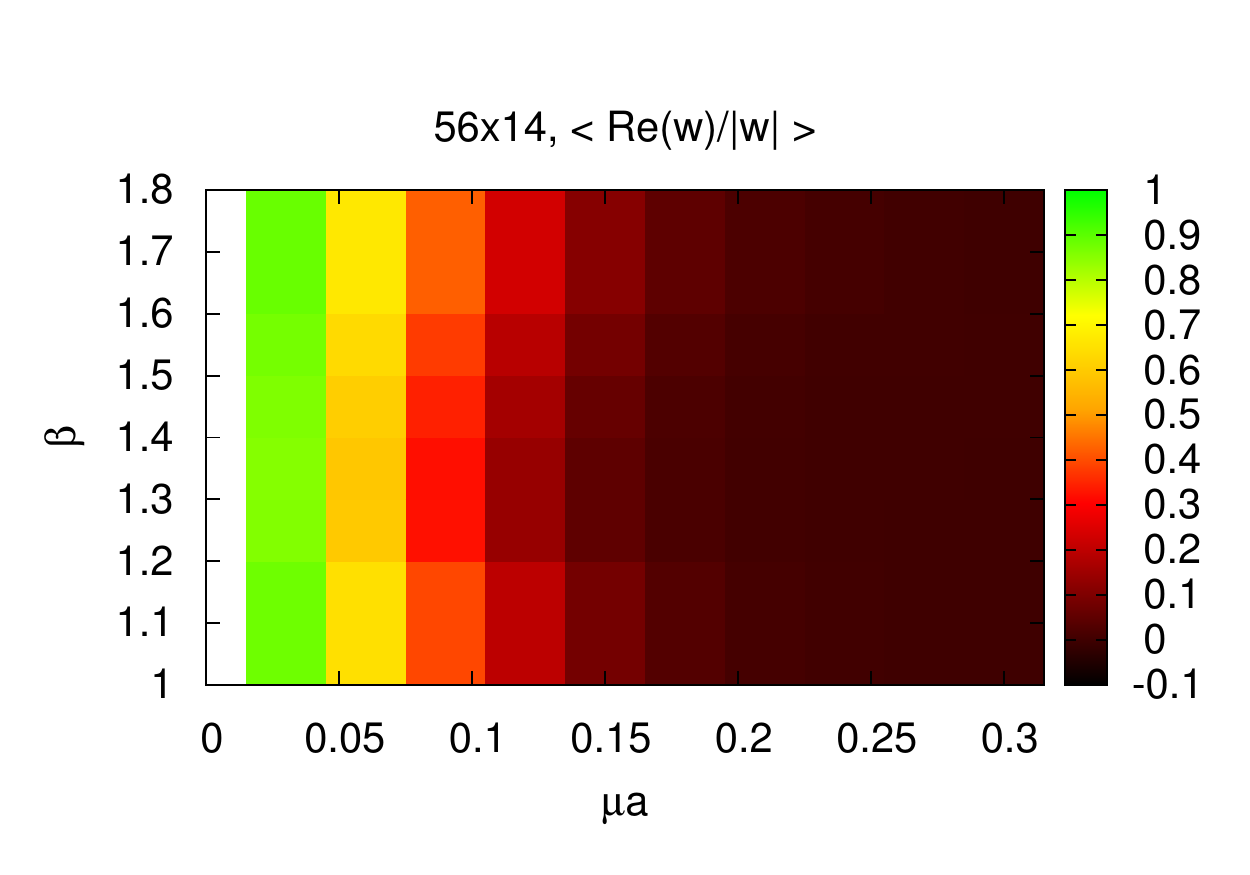}
\vspace{-0.2cm}
\caption{Top: The figure shows the expectation value of the real part of the normalized weight at 56$\times$14 lattice 
when reweighting from $\beta_0=1.5, \mu_0=0$ to $\beta=1.5$ and $\mu a$ values along the horizontal axis.
Bottom: $\langle \mathrm{Re}(w)/|w| \rangle$ shows how hard the sign problem is in different $\beta, \mu$ regions.}
\label{fig:rew_02}
\end{center}
\end{figure}

In these cases, it can happen that one has many relatively small weights at the same order of magnitude, 
and only some with many magnitudes larger, and as a consequence collects only a tiny fraction of useful 
statistics during even long simulations. It was observed that multiparameter reweighting helps to reduce 
the overlap problem in the case of QCD and also can help to reduce the sign problem by doing reweighting 
on the so-called best reweighting lines \cite{Fodor:2002km, Csikor:2004ik}. These are defined as those 
curves that have the smallest standard deviations of Re$(\ln w)$ on them. In the case of the O(3) model,
we illustrate these lines in Fig. \ref{fig:rew_01}, and we show the overlap problem and 
the advantages of multiparameter reweighting in Fig. \ref{fig:hist_rew}'s top and bottom panels, respectively.
We also illustrate the severeness of the sign problem in Fig. \ref{fig:rew_02}, which 
is based on measurements on 56$\times$14 lattices. We used the cluster algorithm to simulate at $\mu_0=0$. 
Further results obtained by the multiparameter reweighting method can be found in Sec. \ref{sec:comparison}, 
where we compare them to the worm and complex Langevin results.
\vspace{-0.2cm}

\section{Worm algorithm}
\label{sec:worm}

Another approach we employed in this study is the worm algorithm.
In order to maintain generality we will review the algorithm and the dual formulation in the O(N) case at $d+1$ dimensions.
In the O(N) case, the fields obey the $\sum_{i=1}^N \phi_i^2 = 1$ condition in every space-time point.
The discretized action in $d+1$ dimensions is
\beq \label{ON_action}
  S = \beta (d+1) V_{d+1} - \beta \sum_{x} ( \phi_{x+\hat{0}} \eexp^{\I \mu a t_{12}} \phi_{x} + \sum_{i=1,..,d} \phi_{x+\hat{i}} \phi_x ),
\eeq
where now $\beta = a^{d-1}/g^2$ and the $d+1$-dimensional lattice volume is $V_{d+1} = N_s^d \times N_t$.

Our goal in lattice simulations is to calculate expectation values. The worm algorithm performs this as counting 
different types of configurations, as we are going to explain later.
The algorithm itself is based on the dual formulation of a model, which means 
in our case the characterization of configurations not with the $\phi(x)$ continuous variables 
at every lattice point, but with a set of discrete variables $\{m_a\}_{a=(l;i=1,\ldots,N)}$, which "live" on links ($l$).
One $\{m_a\}$ configuration matches to a certain, finite partial sum of an infinite sum 
(e.g. the expansion of the partition sum), and the algorithm jumps between these partial sums.

In the following, we first review some conventional notations and integrals over the O(N) sphere, which is helpful 
for the derivation of weights. Then, at first, we restrict ourselves for the dual formulation of the case without chemical 
potential and introduce the worm algorithm; then, we repeat the same procedure for the case with the chemical potential. 
Finally, we present continuum results in Secs. \ref{subs:pressure} and \ref{subs:density}, 
and comparisons to reweighting and the complex Langevin can be found in Sec. \ref{sec:comparison}.

\subsection{Integrals over O(N) sphere}

Consider an $O(N)$ vector $\phi$ of unit length, $\phi^2 = 1$. Averaging over the $O(N)$ sphere, with
the normalization condition $\langle 1 \rangle = 1$, one has $\langle \phi_i \phi_j \rangle = \delta_{ij}/N$,
$ \langle \phi_i \phi_j \phi_k \phi_l \rangle = ( \delta_{ij}\delta_{kl}
+ \delta_{ik}\delta_{jl} + \delta_{il}\delta_{jk} ) / (N(N+2))$.
For the general case of $k$ even number of vectors, one has
\beq \label{SSav}
  \langle \phi_{i_1} \phi_{i_2} \ldots \phi_{i_k} \rangle = C_k^{(N)} (\delta_{i_1 i_2} \delta_{i_3 i_4} \ldots
\delta_{i_{k-1}i_k} + \mathrm{perm}),
\eeq
where 
\beq
C_k^{(N)} = \displaystyle{ \frac{1}{ N(N+2) \ldots (N+k-2) } =
\frac{\Gamma(N/2)}{2^{k/2}\Gamma(N/2+k/2)} }.
\eeq
In the brackets in (\ref{SSav}) there are altogether
\beq
  M_k = (k-1)!! \equiv 1\cdot3\cdot\ldots(k-3)(k-1) = 2^{-\frac{k}{2}+1} \frac{\Gamma(k)}{\Gamma(k/2)}
\eeq
terms for all possible pairings of the indices. This can be obtained e.g. from the recursion relation
$M_{k+2} = M_k + k(k-1)M_{k-2}$, with $M_2=1, M_4=3$. To check this, one can contract in the last two indices
$i_{k-1}, i_k$. The corresponding recursion relation reads
\beq
  C_{k-2}^{(N)} = C_k^{(N)}\left( N + \frac{M_k}{M_{k-2}} - 1 \right) = C_k^{(N)} (N+k-2).
\eeq
It is interesting to note that for a free system, i.e. when the constraint $\delta(\phi^2 - 1)$ is
replaced by the Gaussian $\mathrm{exp}(-\phi^2/2)$ one gets $C_k^{(N)} = 1$ for the weights in (\ref{SSav}).
This is in agreement with the relation
\beq
  \lim_{N \rightarrow \infty} N^{k/2} C_k^{(N)} = \lim_{N\rightarrow \infty} \frac{N^{k/2}}{N(N+2)\ldots(N+k-2)} = 1.
\eeq
Collecting the powers of different components one has
\begin{align} \label{weights}
  w(k&_1,\ldots,k_N) \equiv \langle \phi_1^{k_1} \phi_2^{k_2} \ldots \phi_N^{k_N} \rangle \nonumber \\
  &= \frac{\Gamma(N/2)}{\Gamma((k+N)/2)} \prod_{i=1}^N \frac{\Gamma((k_i+1)/2)}{\Gamma(1/2)} \nonumber \\
  &= \frac{1}{N(N+2)\ldots(N+k-2)} \prod_{i=1}^N (1\cdot3\cdot \ldots \cdot (k_i-1)),
\end{align}
where all $k_i$ are even and $k=k_1+\ldots+k_N$. In the last expression the product is taken only for $i$'s for 
which $k_i > 0$. Also, obviously, $w(0,0, \ldots, 0)=1$.
Some useful relations obtaining the coefficients are (assuming that all powers $k$ are even):
\beq
  \int_{-\infty}^{\infty} \rmd x \,\, \eexp^{-x^2} x^k = \Gamma\left(\frac{k+1}{2}\right),
\eeq
and
\beqnarr
  \int \rmd^N x \,\, &\eexp^{-x^2}& \, x_1^{k_1} \ldots x_N^{k_N} = \prod_{i=1}^N \Gamma\left( \frac{k_i+1}{2} \right) \nonumber \\
     &=& \int_0^\infty \rmd r \, \eexp^{-r^2} r^{N+k-1} S_N \langle \phi_1^{k_1} \ldots \phi_N^{k_N} \rangle \nonumber \\
     &=& \frac{1}{2} \Gamma\left( \frac{N+k}{2} \right) S_N \langle \phi_1^{k_1} \ldots \phi_N^{k_N} \rangle,
\eeqnarr
where $S_N$ is the surface of the $N$-dimensional sphere: $S_N = 2\pi^{N/2}/\Gamma\left(\frac{N}{2}\right)$.

\subsection{Strong coupling expansion without chemical potential} \label{strong_coupl_no_mu}

In the case without chemical potential, for one link between neighbor lattice points $x$ and $y$, one can write
\begin{multline} \label{str_c_no_chem_pot}
  \eexp^{\beta{\phi(x) \phi(y)}} = \sum_{m_1, \ldots, m_N} \frac{\beta^m}{m_1! \ldots m_N!} (\phi_1(x) \phi_1(y))^{m_1} \ldots \\
  \times (\phi_N(x) \phi_N(y))^{m_N},
\end{multline}
where $m = m_1 + \ldots + m_N$. Then one can consider the sum
\begin{align} \label{sum_no_mu}
  \sum_{i=1}^N \sum_{u,v \in \Lambda} &\int_\phi \phi_i(u) \phi_i(v) \eexp^{-S} = \sum_{\mathrm{conf}} W ({u,v,i;m}) \nonumber \\
     &= \eexp^{-\beta (d+1) V_{d+1}}\sum_{\mathrm{conf}} \left( \prod_l \frac{\beta^{m^{(l)}}}{m_1^{(l)}! \ldots m_N^{(l)}!} \right) \nonumber \\
     &\times \left( \prod_x w(k_1(x), \ldots, k_N(x)) \right),
\end{align}
where conf = $\{u,v,i;\{m_j^{(l)}\}^{\textrm{all } l \textrm{ links}}_{j=1,\ldots,N}\}$ is the configuration: 
two distinguished points ($u$ and $v$), a component $i$, 
and the set $\{m_j^{(l)}\}^{\textrm{all } l \textrm{ links}}_{j=1,\ldots,N}$.
During the simulations, the configuration can change in different ways, which we discuss in the Appendix. 
One way is to change $u$ to a neighboring site, meanwhile increasing/decreasing $m_i^{(l)}$ along the $l$ link 
that connects these two.
We start from $u=v$, and the continuous path connecting $u$ and $v$ that appears this way is called the \textsl{worm}.
The weights $w(\ldots)$ of Eq. (\ref{sum_no_mu}) are given by Eq. (\ref{weights}).
The value of $k_j(x)$ depends on the position of $x$: $k_j(x) = \hat{k}_j(x) + (\delta_{xu} + \delta_{xv})\delta_{ij}$,
where $\hat{k}_j(x) = \sum\limits_{x \in \partial l} m_j^{(l)}$. Then, all $k_j(x)$ values must be even.
The ratios of the weights when one of the $k_i$'s is changed by $\pm 2$ are
\beq
  \frac{w(k_1 + 2, k_2, \ldots, k_N)}{w(k_1, k_2, \ldots, k_N)} = \frac{k_1 + 1}{k + N},
\eeq
\beq
  \frac{w(k_1 - 2, k_2, \ldots, k_N)}{w(k_1, k_2, \ldots, k_N)} = \frac{k + N - 2}{k_1 - 1}.
\eeq
With the help of the sum (\ref{sum_no_mu}) defined above, one can see that the partition sum is related to
those configurations where the two ends of the worm coincide ($u=v$); this gives explicitly $V_{d+1}Z$. 
The configurations with one distance between the two ends ($c_1$) divided by the number of configurations 
where the two ends of the worm coincide ($c_2$),
give $\displaystyle{ \frac{c_1}{c_2}= -\frac{2}{\beta}\left( \frac{\langle S \rangle}{V_{d+1}} - \beta (d+1) \right) }$, 
from which one can determine $\langle S \rangle$, the expectation value of the action.

\subsection{Strong coupling expansion with chemical potential}

The action with the chemical potential coupled to $t_{12}$ is obtained from the standard one by replacing 
the interaction terms which couple the fields in the time direction according to 
$\phi(x+\hat{0})\phi(x) \rightarrow \phi(x+\hat{0}) \eexp^{\I \mu a t_{12}} \phi(x)$.
Therefore, the corresponding action is complex.
For the O(2) nonlinear sigma model this problem was avoided in Ref. \cite{Banerjee:2010kc}
using the worm algorithm: the terms in the strong-coupling
expansion are real even in the presence of the chemical potential.
We describe here the extension to the O($N$) case for general $N$.
Let us introduce $\phi_{\pm}=\frac{1}{\sqrt{2}}(\phi_1 \pm \I \phi_2)$.
Expressed through these variables, the scalar product is 
\beqnarr \label{eq:scalar_prod}
  \phi(x&+&\hat{0}) \phi(x) = \phi_-(x+\hat{0}) \phi_+(x) + \phi_+(x+\hat{0}) \phi_-(x) \nonumber \\
  &+& \phi_3(x+\hat{0}) \phi_3(x) + \ldots + \phi_N(x+\hat{0}) \phi_N(x)
\eeqnarr
and the action is
\beqnarr \label{action_mu}
  S &=& \beta (d+1) V_{d+1} - \beta \sum_x \sum_{\nu=0}^{d} \Big( \eexp^{-\mu_{\nu}a}\phi_-(x+\hat{\nu})\phi_+(x) \nonumber \\
  &+& \eexp^{\mu_{\nu}a} \phi_+(x+\hat{\nu})\phi_-(x) + \sum_{j=3}^N \phi_j(x+\hat{\nu}) \phi_j(x) \Big), 
\eeqnarr
where $\mu_{\nu}=\mu \delta_{\nu,0}$, $\nu=0,1,\ldots,d$.
When one integrates over $\phi$ at a given site, the nonvanishing contributions are all real and positive,
\begin{align}
  w(&k_1,k_2,\ldots,k_N) = \langle (\phi_+ \phi_-)^{k_{12}} \phi_3^{k_3} \ldots \phi_N^{k_N} \rangle \nonumber \\
  &= \frac{1}{2^{k_{12}}} \sum_{m=0}^{k_{12}} {k_{12} \choose m} \langle \phi_1^{2m} \phi_2^{2k_{12}-2m} \phi_3^{k_3} \ldots \phi_N^{k_N} \rangle \nonumber \\
  &= \frac{\Gamma(N/2)}{\Gamma((k+N)/2)} 2^{-k_{12}} \Gamma(k_{12}+1) \prod_{i=3}^N \frac{\Gamma((k_i+1)/2)}{\Gamma(1/2)},
\end{align}
where $k=2k_{12}+k_3+\ldots+k_N$ and $k_3, \ldots, k_N$ are even.

The strong-coupling expansion for spatial neighbor sites is the same as in (\ref{str_c_no_chem_pot}), and for temporal 
neighbor sites, it is
\begin{align}
  &\eexp^{\beta {\bf S}^T \eexp^{\I \mu a t_{12}} {\bf S}^\prime} = \sum_{m_+, m_-, m_3, \ldots, m_N} \frac{\beta^m}{m_+!m_-!m_3! \ldots m_N!} \nonumber \\
  &\times \left( \eexp^{\mu a} S_- S_+^\prime \right)^{m_+} \left( \eexp^{-\mu a} S_+ S_-^\prime \right)^{m_-} \left( S_3 S_3^\prime \right)^{m_3} \ldots \left( S_N S_N^\prime \right)^{m_N},
\end{align}
where $m = m_+ + m_- + m_3 + \ldots + m_N$ and ${\bf S} \equiv \phi(x), {\bf S^\prime} \equiv \phi(x+\hat{0})$.
Consider then
\begin{align} \label{sum_mu}
  &\sum_{i} \sum_{u,v \in \Lambda} \int_\phi \phi_i(u) \phi_i(v) \eexp^{-S} = 
\sum_{\mathrm{conf}} W(\{u,v,i;m\};\mu) \nonumber \\
  &= \eexp^{-\beta(d+1)V_{d+1}} \sum_{\mathrm{conf}} \left( \prod_l \frac{ \beta^{m^{(l)}} \eexp^{(\mu a)^{(l)}m_+^{(l)}} \eexp^{-(\mu a)^{(l)} m_-^{(l)}} }{m_+^{(l)}! m_-^{(l)}! \ldots m_N^{(l)}!} \right) \nonumber \\
  &\times \left( \prod_x w(k_{12}(x), k_3(x), \ldots, k_N(x)) \right),
\end{align}
where $i=+,-,3, \ldots, N$, and $(a\mu)^{(l)} = a\mu$ for timelike links and 0 for spatial links.
In these expressions,
\beq
  k_j(x) = \hat{k}_j(x) + \delta_{xu}\delta_{ji_u} + \delta_{xv}\delta_{ji_v} \quad j=+,-,3, \ldots, N,
\eeq
where $(i_u, i_v) = (-,+), (3,3), \ldots, (N,N)$, and different $\hat{k}$'s are defined as 
\beq
  \hat{k}_+ (x) = \sum_{\nu=0}^{d} \left( m_+^{(x-\hat{\nu},x)} + m_-^{(x,x+\hat{\nu})} \right)
\eeq
\beq
  \hat{k}_- (x) = \sum_{\nu=0}^{d} \left( m_-^{(x-\hat{\nu},x)} + m_+^{(x,x+\hat{\nu})} \right)
\eeq
\begin{multline}
  \hat{k}_j (x) = \sum_{x \in \partial l} m_j^{(l)} = \sum_{\nu=0}^{d} \left( m_j^{(x-\hat{\nu},x)} + m_j^{(x,x+\hat{\nu})} \right) \\ j=3, \ldots, N
\end{multline}
The nonzero terms in (\ref{sum_mu}) are those in which the same number of $\phi_+(x)$ and $\phi_-(x)$ factors 
are present; i.e. $k_+(x) = k_-(x) = k_{12}$, and the number of $\phi_j(x)$ factors, $k_j(x)$ (for $j=3,\ldots,N$), 
is even at all sites $x$.
The detailed steps of the worm algorithm based on these prescriptions are discussed in the Appendix.
In the following subsections, we leave the general formalism and consider the O(3) model in $1+1$ dimensions.

\subsection{Numerical results obtained with the worm algorithm} \label{subsec:num_res_worm}

\subsubsection{Check of the algorithm \label{subs:alg_check}}

In order to check the reliability of our algorithm, we have studied the spectrum of the O(3) model.
The energy levels are characterized by the isospin quantum numbers $I, I_3$ and the momentum $p^\prime$.
Let us consider the $p^\prime=0$ case. 
Then the smallest energy at zero $\mu$ in a given sector is denoted by $E(I)$.
The chemical potential splits the $2I+1$-fold degeneracy and the energy levels become
$E(I,I_3;\mu)=E(I)-\mu I_3$, which has a minimum at $I=I_3=q(\mu)$.
By increasing $\mu$, larger $q$ values are expected.
Using the worm algorithm and counting the number of $+$ and $-$ link variables connecting two time slices, 
one can determine $q$ for that interval.

The two ends of the worm divide the periodic time direction into two parts: 
an interval with length $\tau$ (where $0 \le \tau < N_t$) 
and charge $I_3=q$, and an interval with length $N_t-\tau$ and charge $I_3=q-1$.
Then, these give the leading contribution to the correlator 
$C(\tau;q,q-1) \approx A_{q,q-1}\exp\{-E(q;\mu)\tau-E(q-1;\mu)(N_t-\tau)\} \propto \exp\{-(E(q)-E(q-1)-\mu)\tau\}$,
and thus by fitting the correlator, the energy differences can be determined.
Choosing $\mu\approx E(q)-E(q-1)$ one obtains a long plateau in the effective mass plot.
This way, one can follow the signal over a large interval in $\tau$ to measure energies of higher excitations.
These energy differences provide a strong consistency check, since we measure the same difference with different $\mu$ values.
In particular, we have measured the energy differences on $16\times200$ lattices at $\beta=1.779$ using several values of $\mu$ 
and obtained $E(1)-E(0)=0.0662(1), E(2)-E(1)=0.1284(3), E(3)-E(2)=0.1867(3)$.
Note that these agree roughly with the (approximate) rotator picture which is expected to hold for small spatial volumes
\cite{Leutwyler:1987ak, Hasenfratz:2009mp, Niedermayer:2010mx}.
The mass gap $E(1)-E(0)$ agrees within the statistical error with the value cited in Ref. \cite{Balog:2009np}.

\subsubsection{Pressure \label{subs:pressure}}

Similar to what was mentioned at the end of Sec. \ref{strong_coupl_no_mu}, the Eq. (\ref{sum_mu}) sum
is related to the partition function if $u=v$, when it gives $V_{2}Z$.
For the action, one needs to calculate the ratio of two terms. In the denominator there is $Z$, while in the numerator, there is 
$-\frac{\beta}{2}(\eexp^{\mu a} \times \#_1 + \eexp^{-\mu a} \times \#_2 + \#_3 + \#_4)$, where $\#_1$ is the number 
of configurations with $v=u+\hat{0}$ and $i_u=-, i_v=+$; $\#_2$ is the number of configurations with 
$u-\hat{0} = v$ and $i_u=-, i_v=+$; $\#_3$ is the number of configurations with $u\pm\hat{1} = v$ 
and $i_u=-, i_v=+$; and $\#_4$ is the number of configurations with $u\pm\hat{\nu} = v$ with $\nu=0,1$
and $i_u, i_v=3$. The value of the numerator is constructed in the way that is suitable 
for the action (\ref{action_mu}). [The first term of (\ref{action_mu}), which is independent of the dual variables 
was added to the averages at the end.]

For calculating thermodynamic quantities, we used lattices with $N_x > N_t$ and measured 
the action after each worm movement. Since it is divergent as $a \rightarrow 0$, we renormalized it 
by subtracting $\langle S(\beta, T=0, \mu=0) \rangle$. For the latter, we used large symmetric lattices, in particular 
those that were used for determining the scale (see Sec. \ref{sec:scale_setting}). 
In order to eliminate the finite-size effects, 
we chose box sizes of $ma(\beta_{\mathrm{pc}}) N_x \geq 5$, where $\beta_{\mathrm{pc}}$ is the inflection point of the pressure.
Since no phase transition is expected in this model, this $\beta_{\mathrm{pc}}$ is only a pseudocritical quantity.
The chosen box sizes correspond to the aspect ratio $N_x / N_t = 4$, so we have used 
$32\times8, 40\times10, 56\times14, 64\times16, 72\times18, 80\times20, 120\times30$ lattices for finite-temperature simulations. 
\vspace{-0.0cm}
\begin{figure}[H]
\begin{center}
\includegraphics[scale=0.60]{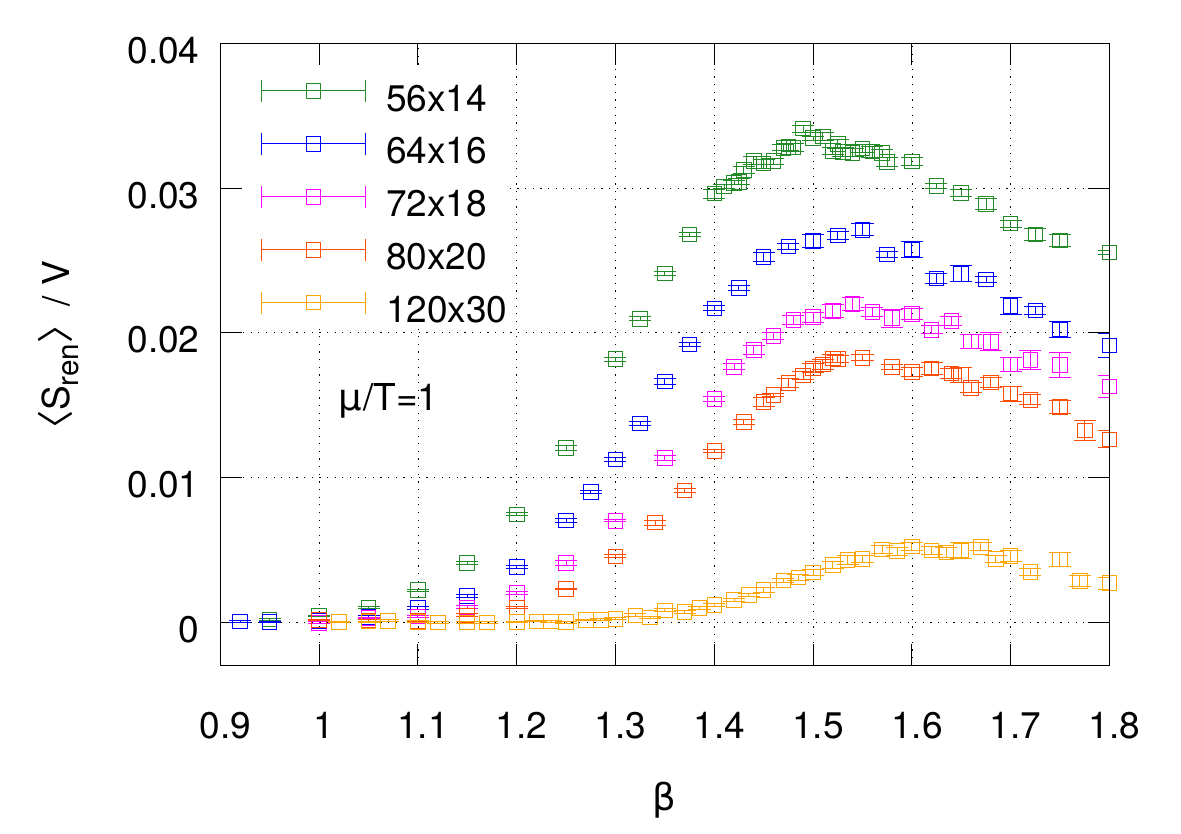}
\vspace{-0.2cm}
\caption{The renormalized action divided by the lattice volume ($V=N_x \times N_t$), measured by the worm algorithm.}
\label{renormalized action}
\end{center}
\end{figure}
We used around $4\times10^{10} \ldots 1.2\times10^{11}$ local worm updates on these lattices, after $(3\ldots5)\times10^5$ 
thermalization steps. Note that these numbers refer to the local change of the configuration.
In order to compare the amount of updates to those of the Langevin simulations, 
one should divide them with the two-dimensional lattice volume.  
After calculating the action, we used the \textsl{integral method} \cite{Boyd:1996bx} to obtain the pressure $p(T)$:
\begin{multline} \label{pressure}
  \frac{p}{T^2} = \frac{N_t}{N_x} \log Z = \frac{N_t}{N_x} \int_{\beta_0}^{\beta} \rmd \beta^{\prime}  \frac{\partial \log Z}{\partial \beta^\prime} \\
  = \frac{N_t}{N_x} \int_{\beta_0}^{\beta} \rmd \beta^{\prime} \left\langle - \frac{\partial S}{\partial \beta^\prime} \right\rangle.
\end{multline}
Since we defined $S$ with $\beta$ included, $\partial S/\partial \beta$ is simply $S/\beta$. The pressure is 
also divergent, so we need to renormalize it using the expectation value of the 
renormalized action $\langle S_{\mathrm{ren}}(\beta, T, \mu) \rangle = 
\langle S(\beta, T, \mu) \rangle - \langle S(\beta, T=0, \mu=0) \rangle$ in the integrand of formula (\ref{pressure}).
In the following, we denote the renormalized pressure with $p$.
Figure \ref{renormalized action} shows the renormalized action density at $\mu/T=1$, while Figs. \ref{fig:pressure, mu=0} 
and \ref{fig:pressure, mu nonzero} show the results for the renormalized pressure.
\vspace{-0.2cm}
\begin{figure}[h]
\begin{center}
\includegraphics[scale=0.62]{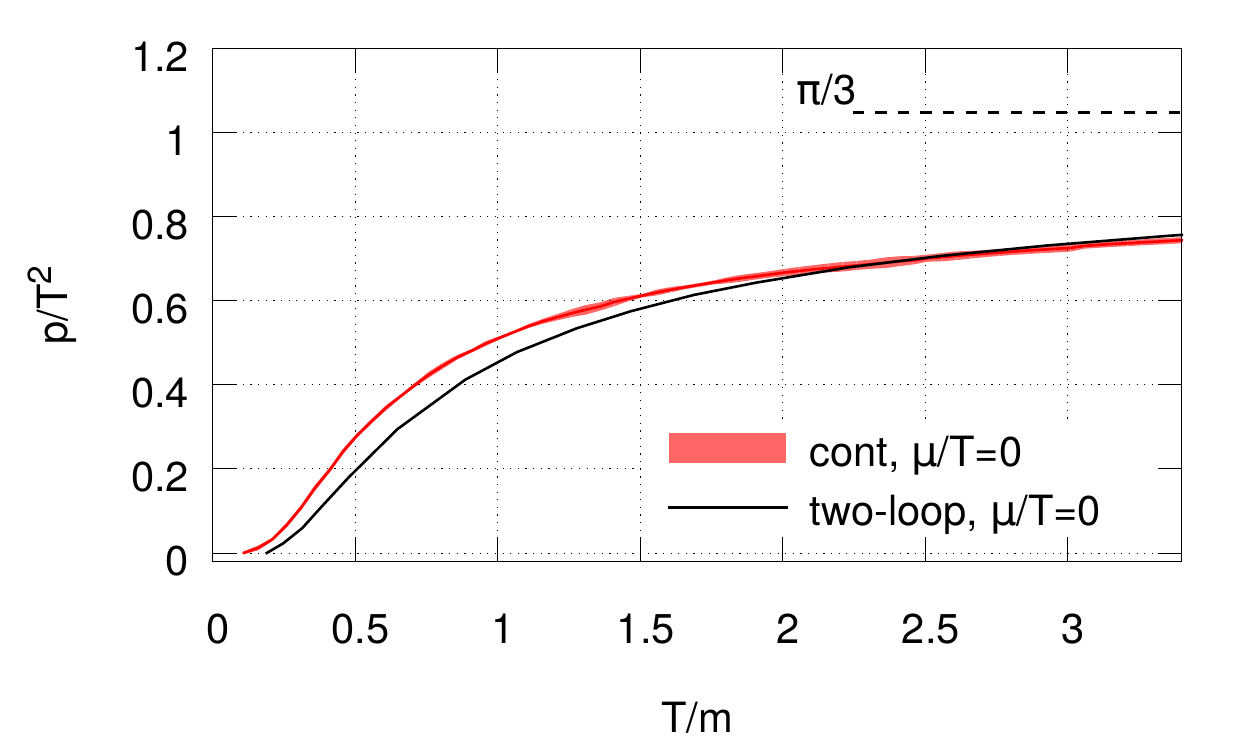}
\vspace{-0.3cm}
\caption{Comparison of our lattice results and the two-loop calculation of Ref. \cite{Seel:2012vj} for $p/T^2$ 
at $\mu=0$ in the continuum. The dashed line at $\pi/3$ shows the asymptotic limit at high temperature.}
\label{fig:pressure, mu=0}
\end{center}
\end{figure}
\vspace{-1.0cm}
\begin{figure}[h]
\begin{center}
\includegraphics[scale=0.62]{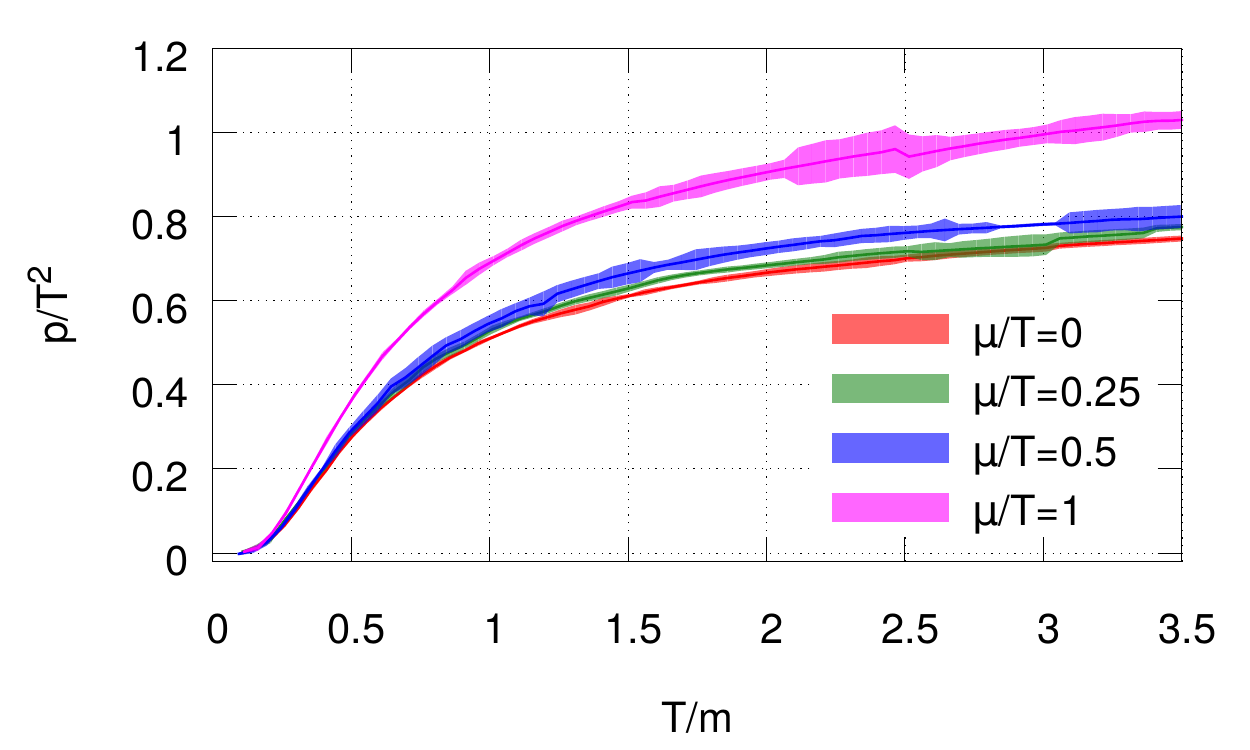}
\includegraphics[scale=0.62]{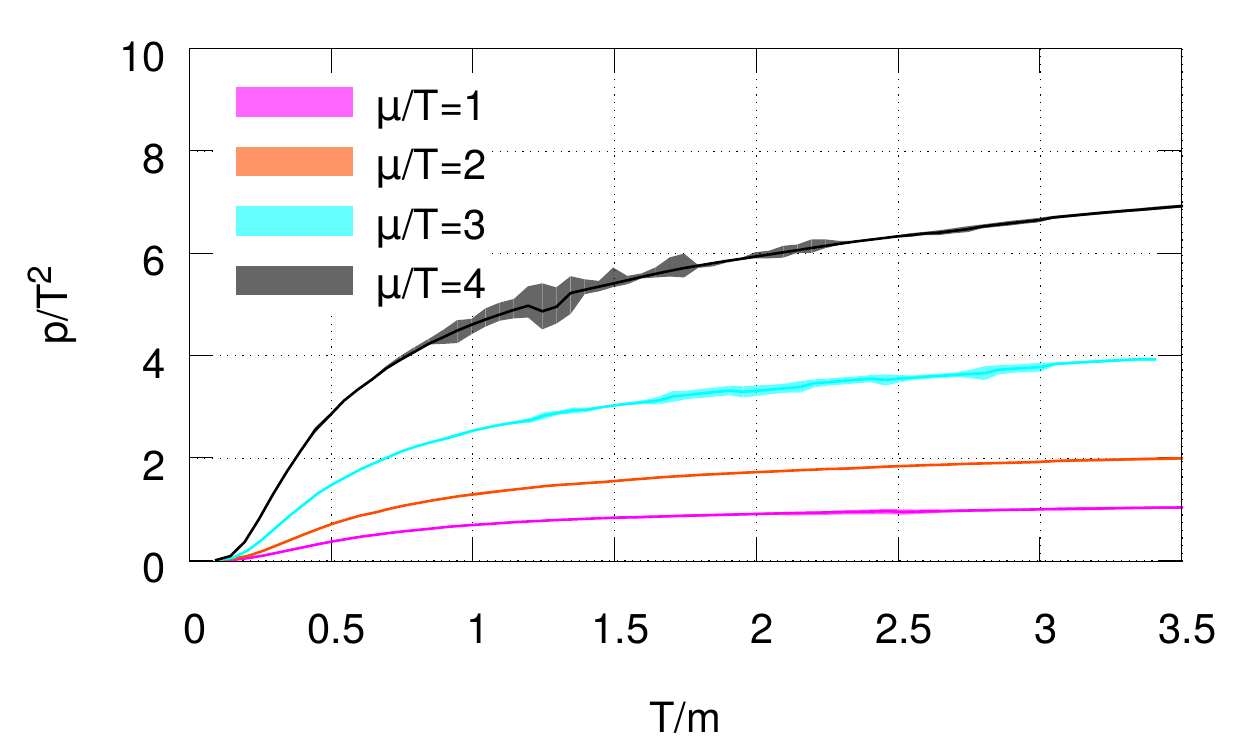}
\vspace{-0.3cm}
\caption{Continuum results for $p/T^2$ as a function of $T/m$ for different $\mu/T$ values.}
\label{fig:pressure, mu nonzero}
\end{center}
\end{figure}

\subsubsection{Trace anomaly} \label{subs:trace_anom}

Another quantity of interest is the trace anomaly (also called interaction measure):
\beq \label{tr_anom}
  \frac{\theta}{T^2} = \frac{\epsilon-p}{T^2} = - \frac{N_t}{N_x} a \frac{\partial \log Z}{\partial a} = \frac{N_t}{N_x}\frac{(am)}{\frac{\partial (am)}{\partial \beta}} \left\langle \frac{S}{\beta} \right\rangle.
\eeq
This quantity is also divergent; thus, renormalization is needed to obtain a finite value in the continuum, 
which is achieved simply by using the renormalized action $S_{\mathrm{ren}}$ instead of $S$ in the above formula.
Below, we show the continuum results for different $\mu / T$ values as a function of temperature (Fig. \ref{worm, trace anom} ).
\begin{figure}[h!]
\begin{center}
\includegraphics[scale=0.62]{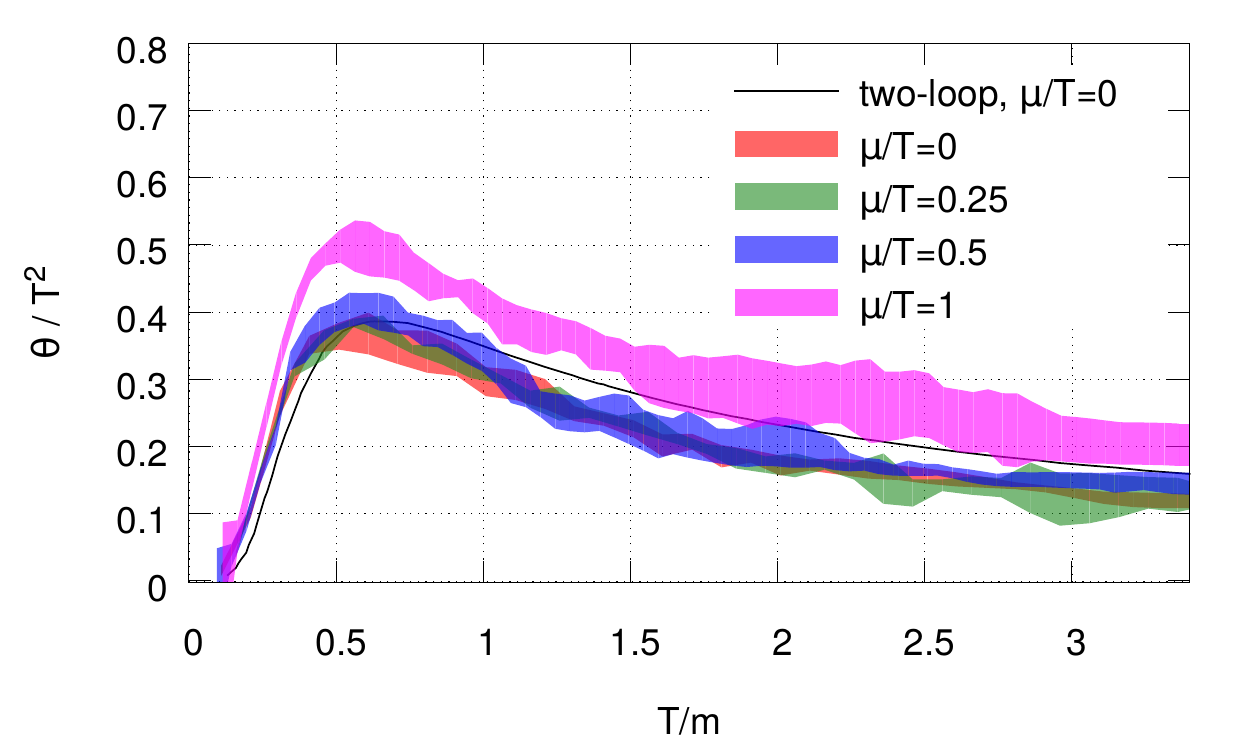}
\includegraphics[scale=0.62]{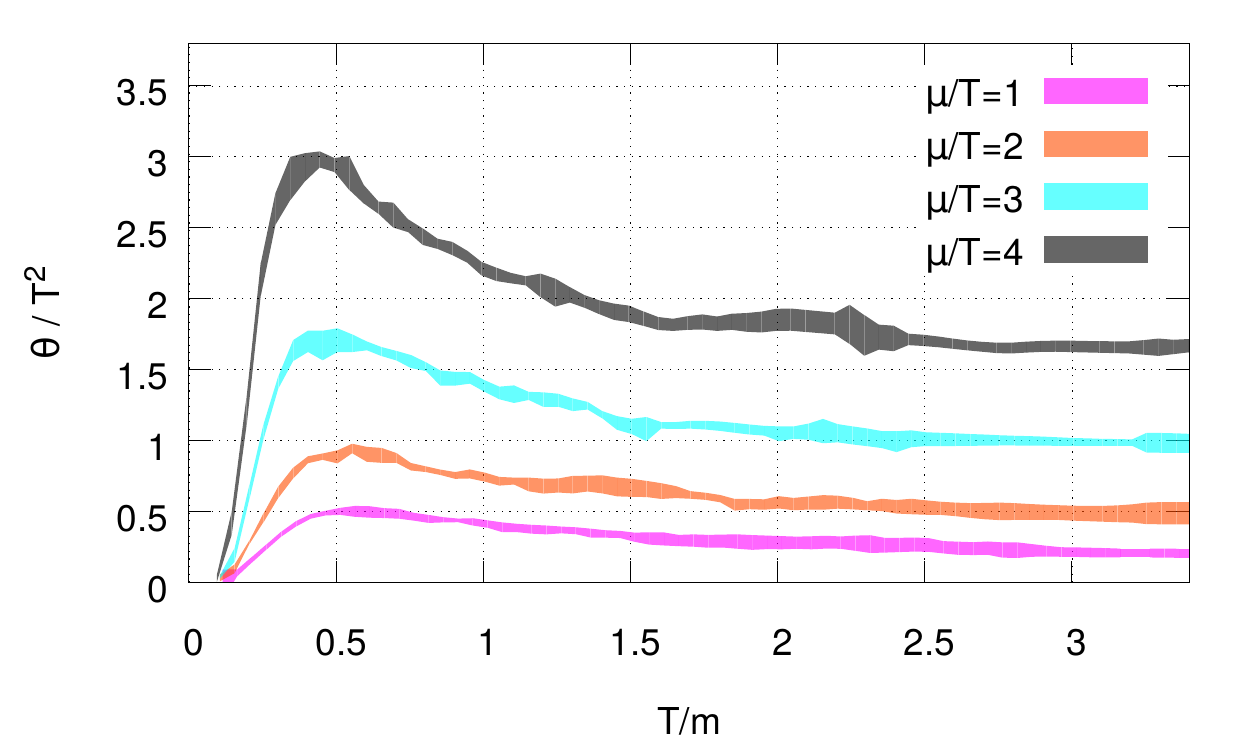}
\caption{Continuum results for $\theta/T^2$ as a function of $T/m$ for different $\mu/T$ values.
The two-loop result for $\mu=0$ is from Ref. \cite{Seel:2012vj}.}
\label{worm, trace anom}
\end{center}
\end{figure}

We note that, similarly to the inflection point of the pressure, the peak position of the trace anomaly 
can also serve as a definition of a pseudocritical temperature ($T_{\mathrm{pc}}$) characterizing the transition in 
the O(3) model.
\footnote{Together with the $T_{\mathrm{pc}}$ calculated from the inflection point of the pressure, we
show how the pseudocritical temperatures depend on $\mu / T$ in Fig. \ref{T_threshold with dens}.}

\subsubsection{Density} \label{subs:density}

Another quantity we measured during the simulations was the isospin charge density, which is defined through
\begin{multline} \label{dens_def}
  n = \frac{T}{V_{sp}} \frac{\partial\log Z}{\partial \mu} = \frac{T}{V_{sp}} \frac{1}{Z} \frac{\partial Z}{\partial \mu} 
    = \frac{T}{V_{sp}} \left\langle - \frac{\partial S}{\partial \mu} \right\rangle \\
    = m \frac{1}{N_t N_x} \frac{1}{am} \left\langle - \frac{\partial S}{\partial (\mu a)} \right\rangle,
\end{multline}
where $V_{sp}$ is the spatial volume which is simply $N_x a$ in our case.
The density does not need to be renormalized, because the divergent part of the action is independent 
of $\mu$. In the figures, we show the dimensionless ratio $n/m$.
With the worm algorithm, the density can be calculated again as a ratio, which has $Z$ in its 
denominator, and in the numerator, there is $\frac{\beta}{2} (\eexp^{\mu a} \times \#_1 - \eexp^{-\mu a} \times \#_2)$, 
where $\#_1$ is the number of configurations with $v=u+\hat{0}$ and $i_u=-, i_v=+$ and
$\#_2$ is the number of configurations with $u-\hat{0} = v$ and $i_u=-, i_v=+$.
The value of the numerator is constructed in the way that is suitable for $\partial S/\partial (\mu a)$
[see the derivative of (\ref{action_mu}) w.r.t. $\mu a$].

\begin{figure}[H]
\begin{center}
\includegraphics[scale=0.60]{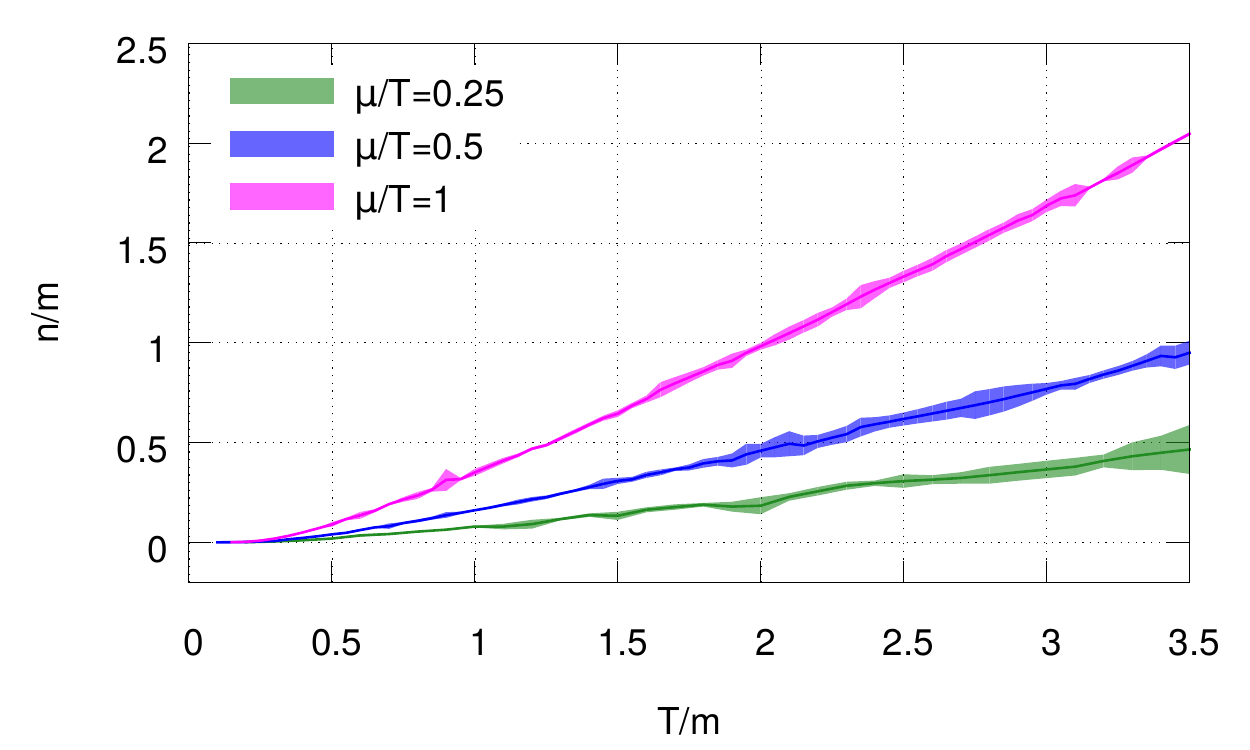}
\includegraphics[scale=0.60]{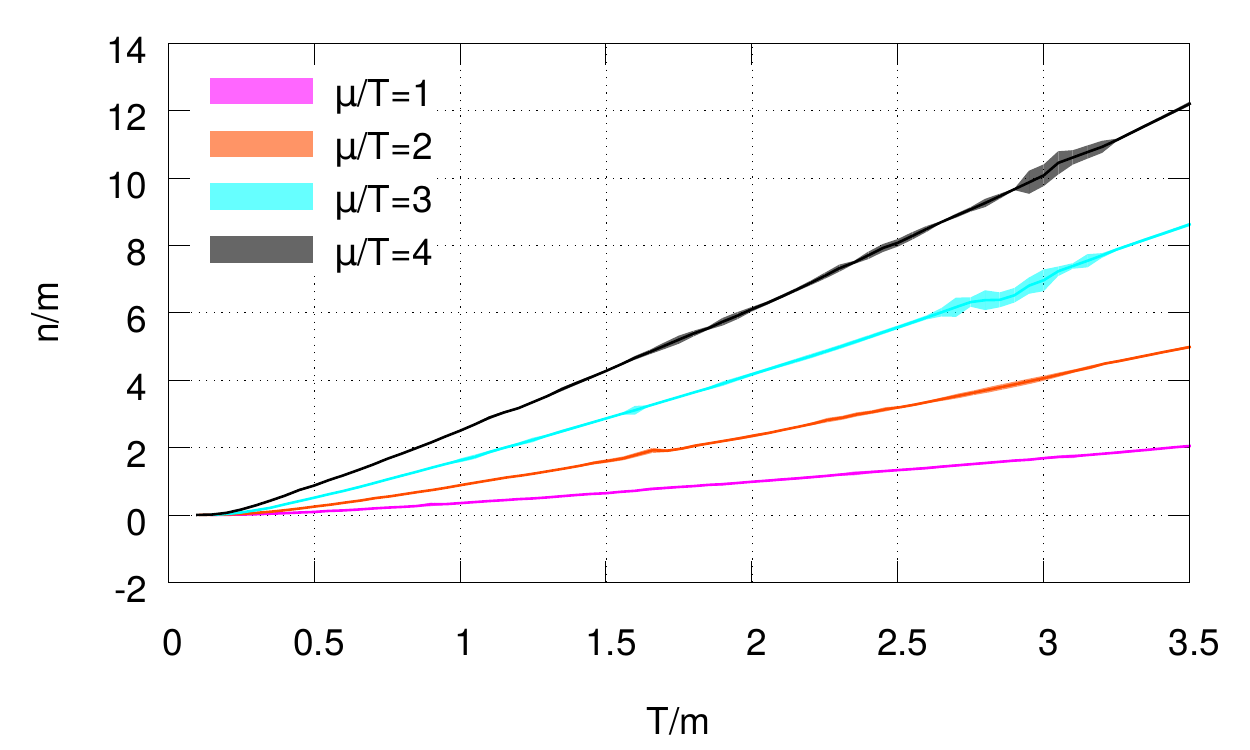}
\vspace{-0.3cm}
\caption{The isospin charge density divided by $m$ at finite temperature in the continuum limit.}
\label{fin_temp_dens}
\end{center}
\end{figure}

As one can observe in Fig. \ref{fin_temp_dens}, $n/m$ depends almost linearly on $T/m$.
Although we did not perform continuum extrapolation above $T/m \approx 3.5$, the numerical data 
from $56\times14$ lattices show that this linear behavior also holds at higher temperature, 
at least up to $T/m \approx 4.6$.
The configurations used for the finite density calculation were the same as for the pressure.

\begin{figure}[H]
\begin{center}
\includegraphics[scale=0.60]{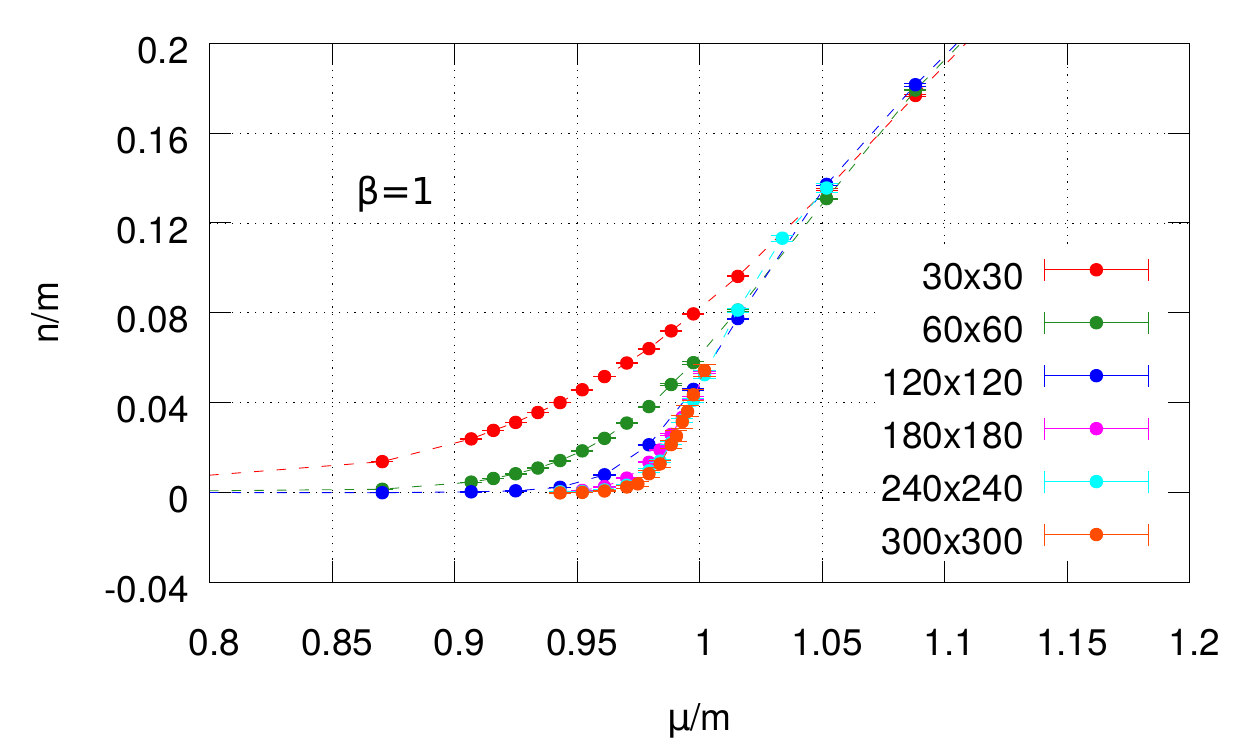}
\includegraphics[scale=0.60]{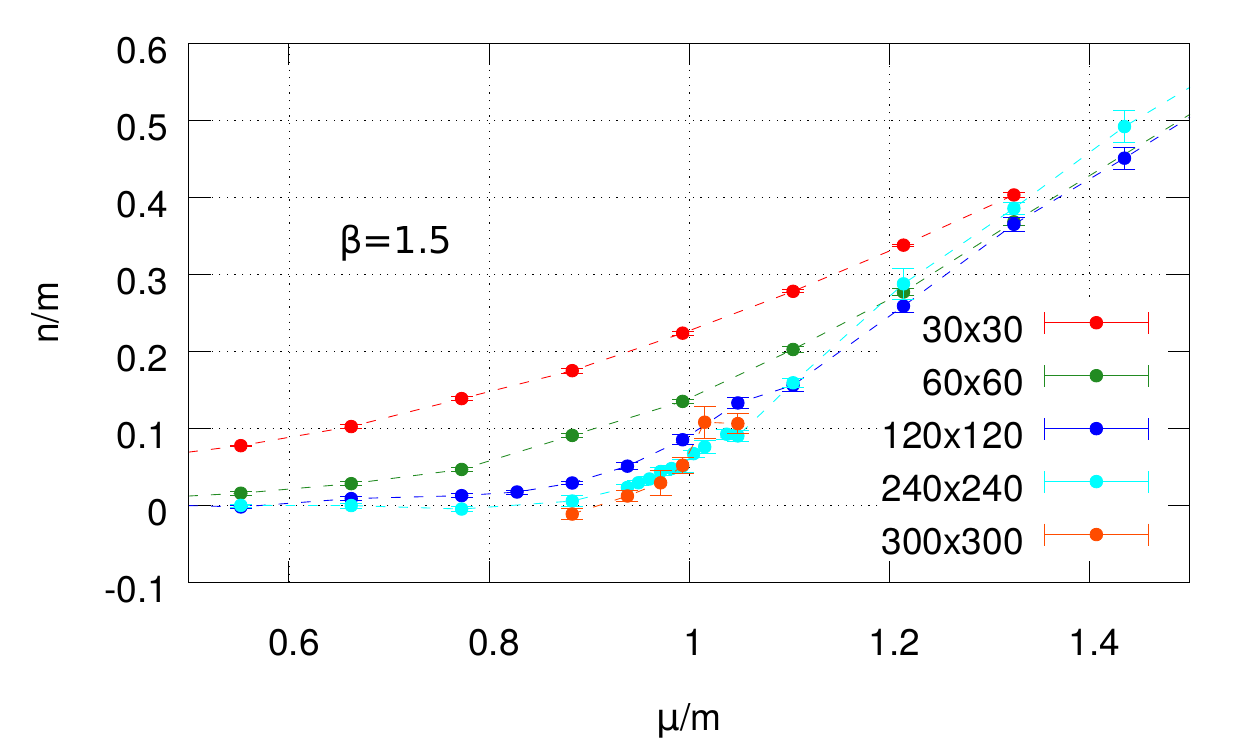}
\vspace{-0.3cm}
\caption{The isospin charge density over $m$ at low temperatures at $\beta=1$ and at $\beta=1.5$ for different lattice sizes.
We used the $T \rightarrow 0$ limit extrapolation based on these results to obtain the continuum limit.
Although the analytical behavior of $n/m$ is known at $T=0$ and infinite volume near $\mu \sim m$ \cite{Bruckmann:2016txt},
our lattice results do not show this directly, because we are either far from the continuum (upper panel) or the temperature 
is not so small to reproduce this precisely (lower panel).}
\label{silver blaze}
\end{center}
\end{figure}

\begin{figure}[H]
\begin{center}
\includegraphics[scale=0.61]{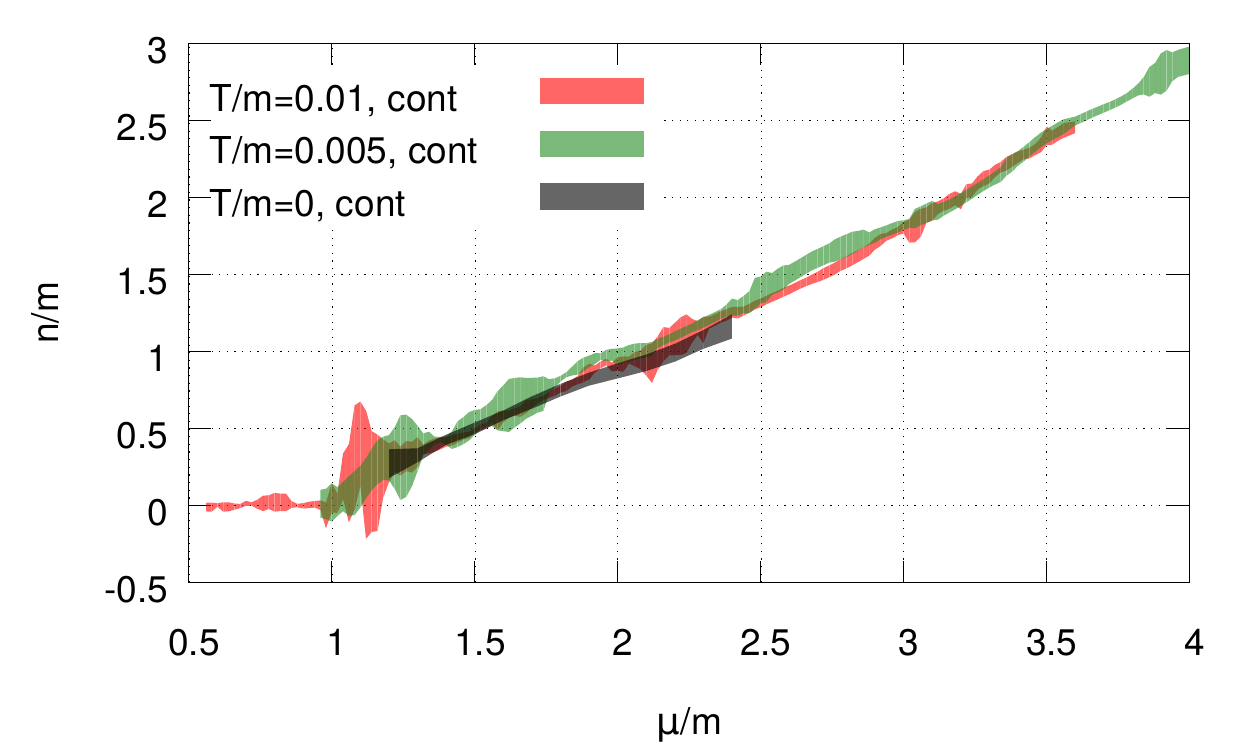}
\vspace{-0.3cm}
\caption{Comparison of $n/m$ continuum results at $T/m=0.01, 0.005$ and $T/m=0$.}
\label{lowT_comp}
\end{center}
\end{figure}

We have also analyzed the low temperature behavior of the density as a function of $\mu/m$, where we 
observed the well-known Silver Blaze phenomenon (Fig. \ref{silver blaze}). We approached the $T=0$ continuum physics by running 
simulations at fixed $\beta$ values increasing the volume of the symmetric lattice; then, 
we extrapolated these $T=0$ results to the continuum (Fig. \ref{T=0, dens contlim 1} and \ref{T=0, dens contlim 2}). 
Another way of obtaining the continuum results would be to run simulations 
at fixed low temperatures and take the continuum limit first, then extrapolate these 
finite temperature continuum results to $T=0$. We did not analyze in full detail this latter case but only
performed simulations to obtain the density at two low temperatures: $T/m=0.01$ and $T/m=0.005$. We compared 
these to the $T=0$ results in Fig. \ref{lowT_comp}.

\vspace{-0.3cm}
\vspace{-0.8cm}

The parameters for these low temperature simulations can be found in Table \ref{low_temp_worm_runs}.
We note that the thermalization took significantly more steps at low temperature as one increased 
the lattice size and $\beta$.
For example in the case of $N_t=500, \beta=1.3234$ for $N_xma=10$ thermalization took $3\times10^9$ steps, 
but for $N_xma=40$, it was $\sim4.4$ times longer, and for $N_xma=100$ (symmetric lattice),
it was $\sim19$ times longer than for $N_xma=10$.
Thermalization was analyzed using the values of density during the simulations.

\vspace{0.2cm}
\begin{figure}[H]
\begin{center}
\includegraphics[scale=0.61]{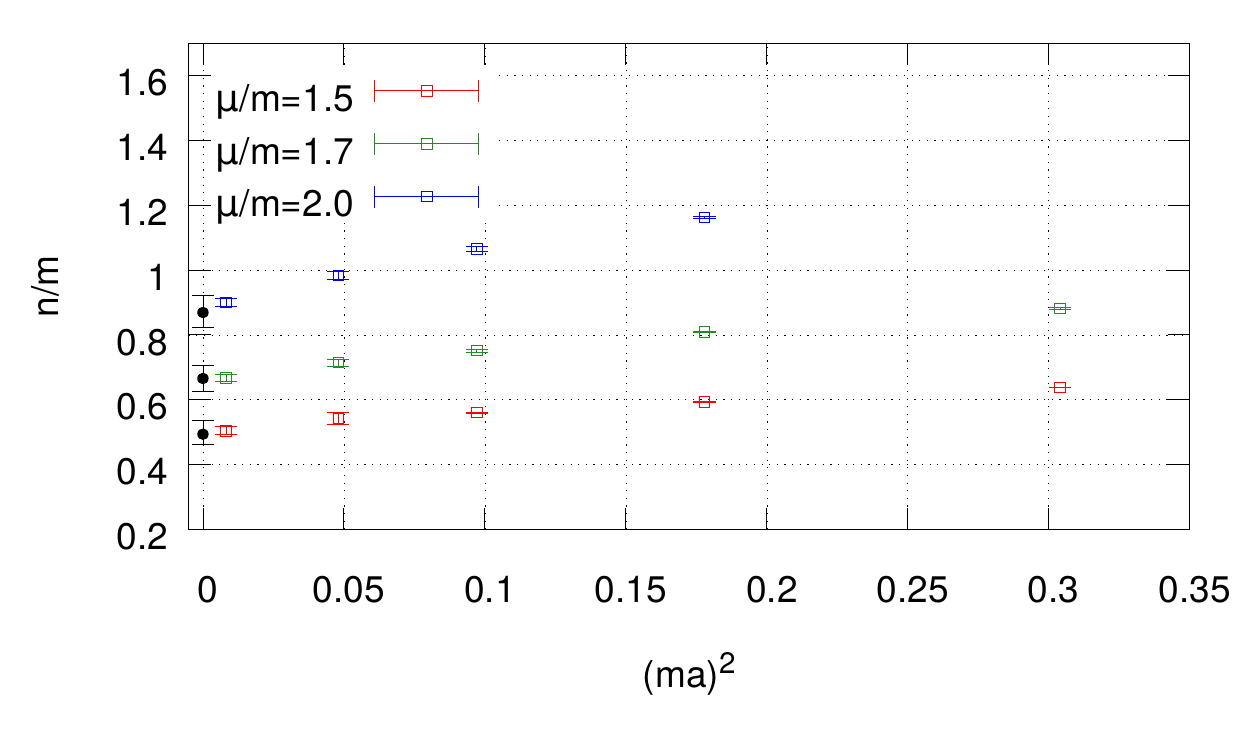}
\vspace{-0.3cm}
\caption{Continuum extrapolation at some $\mu / m$ at $T/m=0$.}
\label{T=0, dens contlim 1}
\end{center}
\end{figure}

\vspace{-0.2cm}
\begin{figure}[H]
\begin{center}
\includegraphics[scale=0.61]{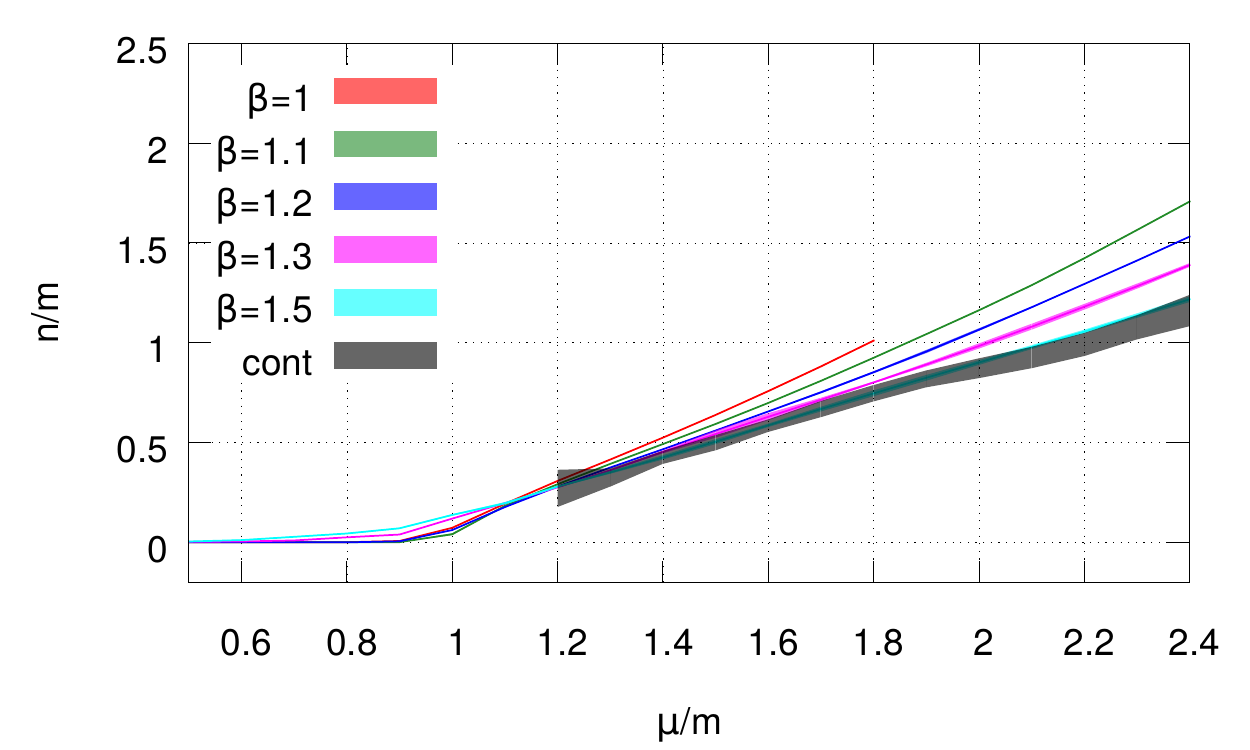}
\vspace{-0.3cm}
\caption{Continuum limit for $n/m$ at $T/m=0$ using lattices with fixed $\beta$ values.}
\label{T=0, dens contlim 2}
\end{center}
\end{figure}
\vspace{-0.2cm}

\begin{table}[h]
\begin{center}
\begin{tabular}{ |c|c|c|c| }
\hline
$T/m$ & $\beta$ & $ma$ & $N_t = N_x$ \\
\hline
0 & 1 & 0.551 & 30, 60, 120, 180, 240, 300, 360 \\
  & 1.1 & 0.422 & 30, 60, 120, 180, 240, 300, 360 \\
  & 1.2 & 0.312 & 30, 60, 120, 180, 240, 300, 360, 500 \\
  & 1.3 & 0.219 & 30, 60, 120, 180, 240, 300, 360, 500 \\
  & 1.5 & 0.091 & 30, 60, 120, 180, 240, 300, 360 \\
\hline
\end{tabular}
\begin{tabular*}{0.416\textwidth}{@{\extracolsep{\fill}} |c|c|c|c|c| }
\hline
$T/m$ & $\beta$ & $ma$ & $N_t$ & $N_x ma$ \\
\hline
0.01 & 1.1789 & 0.333 & 300 & approx. 10, 20, 40, 60, \\
     & 1.2644 & 0.25 & 400 & 100 (symmetric lattices) \\
     & 1.3234 & 0.2 & 500 & \\
     & 1.3682 & 0.167 & 600 & \\
\hline
0.005 & 1.2644 & 0.25 & 800 & approx. 10, 20, 30, 40, \\
      & 1.2963 & 0.222 & 900 & 50, 60, 70 \\
      & 1.3234 & 0.2 & 1000 & \\
      & 1.3473 & 0.182 & 1100 & \\
\hline
\end{tabular*}
\caption{The set of parameters for low and zero temperature runs with the worm algorithm.
We have run simulations at several $\mu a$ using the above parameters.
The number of used worm configurations for these runs was around $(3\ldots 9)\times 10^{10}$
after thermalization.}
\label{low_temp_worm_runs}
\end{center}
\end{table}

\section{Complex Langevin algorithm}
\label{sec:CL}

We proceed with the realization of the complex Langevin algorithm for the O(3) model.
The continuum complex Langevin equation for each component $i=1,2,3$ of a three-component scalar field variable is
\beq \label{cont Langevin}
  \frac{\partial \phi_{x,i}(\tau)}{\partial \tau} = - \frac{\delta S[\phi; \tau)}{\delta \phi_{x,i}(\tau)} 
  + \eta_{x,i}(\tau) ,
\eeq
where $\tau$ is the simulation time and $\eta_{x,i}(\tau)$ is a Gaussian noise obeying the following relations:
\beq \label{noise1}
  \langle \eta_{x,i}(\tau) \eta_{x^\prime,j} (\tau^\prime) \rangle = 2 \delta_{ij}\delta_{xx^\prime}\delta(\tau-\tau^\prime),
  \quad \langle \eta_{x,i}(\tau) \rangle = 0.
\eeq
The simplest discretization for Eq. (\ref{cont Langevin}) is the so-called Euler (Euler--Maruyama) discretization:
\beq \label{E discr Langevin}
  \phi_{x,i}^{(n+1)} = \phi_{x,i}^{(n)} 
                       - \varepsilon \frac{\delta S}{\delta \phi_{x,i}}^{(n)} 
                       + \sqrt{\varepsilon} \eta_{x,i}^{(n)},
\eeq
where we denote the simulation steps with $n$, and $\varepsilon$ is a finite step size.
However, Eq. (\ref{E discr Langevin}) does not preserve the length of 
the $\phi_x$ vector, so in order to simulate the O(3) model, we must somehow include the
constraint $\sum_i \phi_{x,i}^2$ = 1, because the partition function for the O(3) model is
\begin{multline} \label{CL_Z}
  Z = \int \prod_x \rmd \phi_x \delta(\phi_x^2 - 1) \eexp^{-S[\phi]} \\
    = \int \prod_x \rmd \phi_x \eexp^{- (S[\phi] - \sum_x \ln \delta( \phi_x^2 - 1 ))}.
\end{multline}
Usually the integration measure is not considered explicitly during the integration: one does not 
use the force arising from the constraint, but uses other (general) coordinates or specific integration. 
For example, suitable general coordinates in our case are spherical coordinates, and an example for a 
specific integration scheme in Cartesian coordinates is the so-called Euler discretization in group space, 
which is used for example in complex Langevin (CL) simulations of SU(N) gauge groups \cite{Seiler:2012wz}.
In the following we will study these approaches to integrate CL equations.

\subsection{Use of spherical coordinates} \label{subs:spherical}

Using spherical coordinates $\phi_x = (\sin\vartheta_x \cos\varphi_x, \\ 
\sin\vartheta_x \sin\varphi_x, \cos\vartheta_x)$, $Z$ becomes
\begin{align}
  Z &= \int \prod_{x_1} \rmd\varphi_{x_1} \prod_{x_2} \rmd\vartheta_{x_2} \eexp^{-\left( S[\varphi,\vartheta]-\sum_x \ln \sin\vartheta_x \right)} \nonumber \\
    &= \int \prod_{x_1} \rmd\varphi_{x_1} \prod_{x_2} \rmd\vartheta_{x_2} \eexp^{-S_{\mathrm{eff}}[\varphi, \vartheta]},
\end{align}
where 
\begin{align}
  S_{\mathrm{eff}}&[\varphi, \vartheta] = 2 \beta V - \beta \sum_{x,\nu} \Big( \sin\vartheta_{x+\hat{\nu}} \sin\vartheta_x \cos(\varphi_{x+\hat{\nu}} - \varphi_{x} \nonumber \\
  &- \I \mu a \delta_{\nu,0} ) + \cos\vartheta_{x+\hat{\nu}} \cos\vartheta_x \Big) - \sum_x \ln \sin\vartheta_x .
\end{align}
From this expression one can deduce the drifts: 
\begin{align}
  - \frac{\delta S_{\mathrm{eff}}}{\delta \varphi_x} &= \beta \sum_{\nu} \Big( \sin\vartheta_x \Big[ \sin\vartheta_{x+\hat{\nu}} \sin(\varphi_{x+\hat{\nu}} - \varphi_x \nonumber \\
  &- \I \mu a \delta_{\nu,0} ) - \sin\vartheta_{x-\hat{\nu}} \sin( \varphi_x - \varphi_{x-\hat{\nu}} - \I \mu a \delta_{\nu,0}) \Big] \Big),
\end{align}
and
\begin{align}
  - \frac{\delta S_{\mathrm{eff}}}{\delta \vartheta_x} &= \beta \sum_{\nu} \Big( \cos \vartheta_x \Big[ \sin\vartheta_{x+\hat{\nu}} \cos( \varphi_{x+\hat{\nu}} - \varphi_x \nonumber \\
  &- \I \mu a \delta_{\nu,0} ) + \sin\vartheta_{x-\hat{\nu}} \cos(\varphi_x - \varphi_{x-\hat{\nu}} - \I \mu a \delta_{\nu,0} ) \Big] \nonumber \\
  &- \sin\vartheta_x (\cos\vartheta_{x+\hat{\nu}} + \cos\vartheta_{x-\hat{\nu}}) \Big) - \frac{1}{\tan\vartheta_x}.
\end{align}
Then the discretized complex Langevin steps are
\beq
  \varphi_x(n+1) = \varphi_x(n) + \varepsilon_n K^{(\varphi)}_x(n) + \sqrt{\varepsilon_n} \eta^{(\varphi)}_x(n),
\eeq
\beq
  \vartheta_x(n+1) = \vartheta_x(n) + \varepsilon_n K^{(\vartheta)}_x(n) + \sqrt{\varepsilon_n} \eta^{(\vartheta)}_x(n),
\eeq
where $K^{(\varphi)}_x = -\delta S_{\mathrm{eff}} / \delta \varphi_x$ and $K^{(\vartheta)}_x = -{\delta S_{\mathrm{eff}}}/{\delta \vartheta_x}$.
In these equations $\eta^{(\varphi)}_x$ and $\eta^{(\vartheta)}_x$ are real, Gaussian noises, and the finite 
step size $\varepsilon_n$ is determined adaptively, so it also depends on $n$.
As one can observe, the force -$\delta S_{\mathrm{eff}}[\varphi, \vartheta]/\delta \vartheta_x$ 
is singular because of the $1/\tan \vartheta_x$ term. In order to avoid overflow during the simulations 
due to the singular forces, one can do a constrained simulation and truncate the configuration 
space to avoid the values of $\vartheta_x$ near zero and $\pi$. 
If the step size is not small enough, this is needed in order to get stable simulation runs.

\noindent We achieved this by reflecting the trajectories in the following way:
\beq
\vartheta_x(n+1):=\begin{cases}
2\vartheta_{\text{LIM}} - \vartheta_x(n+1), \\ 
	\qquad \qquad \text{ if } \vartheta_x(n+1) < \vartheta_{\text{LIM}}, \\
\vartheta_x(n+1), \\
	\qquad \qquad \text{ if } \vartheta_{\text{LIM}} < \vartheta_x(n+1) < \pi-\vartheta_{\text{LIM}}, \\
2(\pi - \vartheta_{\text{LIM}}) - \vartheta_x(n+1), \\
	\qquad \qquad \text{ if } \pi-\vartheta_{\text{LIM}} < \vartheta_x(n+1).
\end{cases}
\eeq
The threshold value for $\vartheta$ was defined with a parameter $\vartheta_{\textrm{LIM}}$. Results shown later 
support the expectation that if $\varepsilon$ and $\vartheta_{\textrm{LIM}}$ are small enough, 
then the results are independent of their values.

\subsection{Integration in group space} \label{subs:cl,group}

Using ideas from the CL equation on SU(N) gauge groups \cite{Seiler:2012wz}, one can write a specific integration 
scheme that uses Cartesian coordinates, but takes the constraint into account: the Euler discretization 
(which we call below the exponentialized Euler-Maruyama discretization).
For the $O(3)$ group elements $O_x$, this is the following
\beq \label{O3_group_CL}
O_x(n + 1) = R_x(\varepsilon) O_x(n).
\eeq
Since all $\phi_x$ can be written with some $\phi_0$ constant unit vector
and with an $O_x$ rotation matrix as $\phi_x = O_x \phi_0$, the above time evolution can turn into the 
time evolution of the original $\phi_x$ variables. \\
The $R_x(\varepsilon)$ in Eq. (\ref{O3_group_CL}) can be written in different ways. It can be e.g. 
\beq
R^{(1)}_x(\varepsilon) = \exp \left( {\sum_a t_a ( \varepsilon K_{ax} + \sqrt{\varepsilon} \eta_{ax} )} \right),
\eeq
or 
\beq
R^{(2)}_x (\varepsilon) = \prod_{a \in (1,2,3)} \exp\left(t_a( \varepsilon K_{ax} + \sqrt{\varepsilon} \eta_{ax} )\right), 
\eeq
or 
\beq
R^{(3)}_x (\varepsilon) = \prod_{a \in (1,2,3)} \exp\left(t_a \varepsilon K_{ax}\right) \exp\left(t_a \sqrt{\varepsilon} \eta_{ax}\right).
\eeq
Since $\eexp^{A+B} \neq \eexp^A \eexp^B$, when $[A,B] \neq 0$, these are not equivalent to 
each other at finite $\varepsilon$, but the difference in the simulation results is not detectable at 
the numerical precision and parameter set we used. Due to this, during 
our simulations we used $R^{(2)}_x$, which is not the computationally cheapest version 
[it is $R^{(1)}_x(\varepsilon)$], but the cheapest version that evolves the system in each direction in the tangent 
space individually one after another.
In the expressions above, the $t_a$s are the three generators of O(3) in the three-dimensional representation.
The drift $K_{ax}$ is 
\begin{align}
  K_{ax} &= - D_{ax} S[O] = -\partial_\alpha S[\eexp^{\alpha t_a} O_x]|_{\alpha=0} \nonumber \\
         &= \beta (  \phi_0^T O_{x+\hat{0}}^T \eexp^{\I\mu a t_3} t_a O_x \phi_0 
                    - \phi_0^T O_x^T t_a \eexp^{\I\mu a t_3} O_{x-\hat{0}} \phi_0 \nonumber \\
                    &\quad + \phi_0^T O_{x+\hat{1}}^T t_a O_x \phi_0 
                    - \phi_0^T O_x^T t_a O_{x-\hat{1}} \phi_0) \nonumber \\
         &= \beta (  \phi_{x+\hat{0}}^T \eexp^{\I\mu a t_3} t_a \phi_x 
                    - \phi_x^T t_a \eexp^{\I\mu a t_3} \phi_{x-\hat{0}} \nonumber \\
                    &\quad + \phi_{x+\hat{1}}^T t_a \phi_x 
                    - \phi_x^T t_a \phi_{x-\hat{1}} ) .
\end{align}
Here, $\phi^T$ and $O^T$ denote the transpose of $\phi$ and $O$, and $\eta_{ax}$ is the usual Gaussian noise.
The time evolution determined by Eq. (\ref{O3_group_CL}) then can be written with $\phi_x$ as
\beq
  \phi_x^{(n+1)} = \prod_{a\in (1,2,3)} \exp \left[ ( \varepsilon_n K_{ax} + \sqrt{\varepsilon_n} \eta_{ax} ) t_a \right] \phi_x^{(n)}.
\eeq
In particular, we performed the updates by varying the order of the three matrix multiplication randomly.
Higher order integrations, like Runge-Kutta \cite{Batrouni:1985jn}, are also possible,
\begin{align}
  \phi_x^\prime &= \exp \left[ ( \varepsilon_n K_{ax}[\phi^{(n)}] + \sqrt{\varepsilon_n} \eta_{ax} ) t_a \right] \phi_x^{(n)} \nonumber \\
  \phi_x^{(n+1)} &= \exp \Big[ \frac{\varepsilon_n}{2} \left(1+\frac{C_A \varepsilon_n}{6}\right) \left( K_{ax}[\phi^{(n)}] + K_{ax}[\phi^\prime] \right) t_a \nonumber \\
  &\quad + \sqrt{\varepsilon_n} \eta_{ax} t_a \Big] \phi_x^{(n)},
\end{align}
where $C_A=2$ is the Casimir invariant for the three-dimensional representation of O(3).

\subsection{Direct method to include the constraint in Cartesian coordinates} \label{subs:cl,std}

Using Cartesian coordinates, the constraining force can be considered by using a 
term arising from $-\sum_x \ln \delta(\phi^2_x-1)$, see Eq. (\ref{CL_Z}). This is also singular, so one can 
attempt to approximate the Dirac $\delta$ with a sharp Gaussian, $(\sqrt{2\pi}b)^{-1} \eexp^{-x^2/(2b^2)}$
$\rightarrow$ $\delta(x)$, as $b \rightarrow 0$.
The force is then
\begin{align}
  K_x &= -\frac{\delta}{\delta \phi_x}\left(S[\phi]-\sum\limits_y \ln \delta (\phi_y^2-1)\right) \nonumber \\
      &= \beta \left( \phi_{x+\hat{0}} \eexp^{\I\mu a t_3} + \eexp^{\I\mu a t_3}\phi_{x-\hat{0}}
+ \phi_{x+\hat{1}} + \phi_{x-\hat{1}} \right) \nonumber \\
      &\quad - \frac{2}{b^2} (\phi_x^2-1)\phi_x,
\end{align}
where the last term helps to keep the length of $\phi_x$ \mbox{near 1.} Then the fields evolve according to
\beq \label{direct discr Langevin}
  \phi_{x,i}^{(n+1)} = \phi_{x,i}^{(n)} + \varepsilon_n K_{x,i}^{(n)} + \sqrt{\varepsilon_n} \eta_{x,i}^{(n)},
\eeq
where again $\varepsilon_n$ is the finite step size determined adaptively, and the noise is Gaussian 
with $\sqrt{2}$ width. Note that using this time evolution, the $\sum_i \phi_{x,i}=1$ condition is no 
longer true during the simulations, but the force can push the field into this direction. We refer 
to this time evolution as 'standard Euler-Maruyama discretization with Dirac $\delta$' in the following.

In the next section, we discuss the results obtained using the various algorithms.

\section{Comparison of results}
\label{sec:comparison}

As was discussed in the Introduction, the complex Langevin algorithm may provide 
a feasible way to study sign problems in different models, but may converge to wrong results, which 
would lower the reliability of the simulation when there are no alternative results in the problematic parameter range. 
The conventional reasoning of explaining the wrong results is that 
the justification of complex Langevin \cite{Aarts:2009uq, Aarts:2011ax, Aarts:2013uza} 
is not correct in that parameter range, because some observables develop long tailed distributions. 
(For details, we refer to Ref. \cite{Aarts:2011ax}.) However, in some models 
(e.g. in a random-matrix model \cite{Mollgaard:2013qra, Mollgaard:2014mga}) 
it was observed that using different variables in describing the model can help to eliminate this problem. 
This can imply that the failure of the algorithm is not because physics has changed, but because of 
some unknown algorithmic details. The source of these can be quite broad. One can think of systematic errors 
originating from e.g. step size to zero limit, low randomness in the random number generator, 
floating point round-off errors, or not taking the continuum limit, or sampling errors because low 
autocorrelation for example due to some distinguished region in configuration space, etc.
In the following we analyze some of these aspects in the case of the 1+1-dimensional O(3) model.
The results are compared to the worm results, which are referred to as the correct ones in the text.

Although we used adaptive step size in all our simulations, this cannot replace the completion 
of the $\varepsilon \rightarrow 0$ extrapolation of the results.
(Of course, its effect depends on the used numerical precision and the algorithm under study 
as well as other subtle circumstances. For example, as we will see, the simulations with spherical 
coordinates do not depend in a detectable way on $\varepsilon$ at the simulations with the used 
set of parameters for e.g. the action variable.)
In the following, first, we discuss the results for the action $S$ and the trace anomaly $\theta$ and then for the density $n/m$.

\subsection{Action and trace anomaly}

Using {\bf spherical coordinates} to parametrize the model, we ran simulations at $56\times14$ lattices,
at several chemical potentials (see Table \ref{Table: CL spherical}).
We analyzed the step size and $\vartheta_{\textrm{LIM}}$ dependence of the results.
Although we found that there was no detectable step size dependence of the results, 
we extrapolated to zero step size at $\mu a = 0.071$ and $0.143$ using four step sizes. 
We also found that the results do not depend on $\vartheta_{\textrm{LIM}}$ if it was chosen quite small. 
To test this at $\mu a = 0.071$, we used $\vartheta_{\textrm{LIM}} = 10^{-3}, 10^{-5}, 10^{-7}, 10^{-8}, 10^{-11}$ 
and $\varepsilon = 0.0005$.
\begin{figure}[H]
\begin{center}
\includegraphics[scale=0.62]{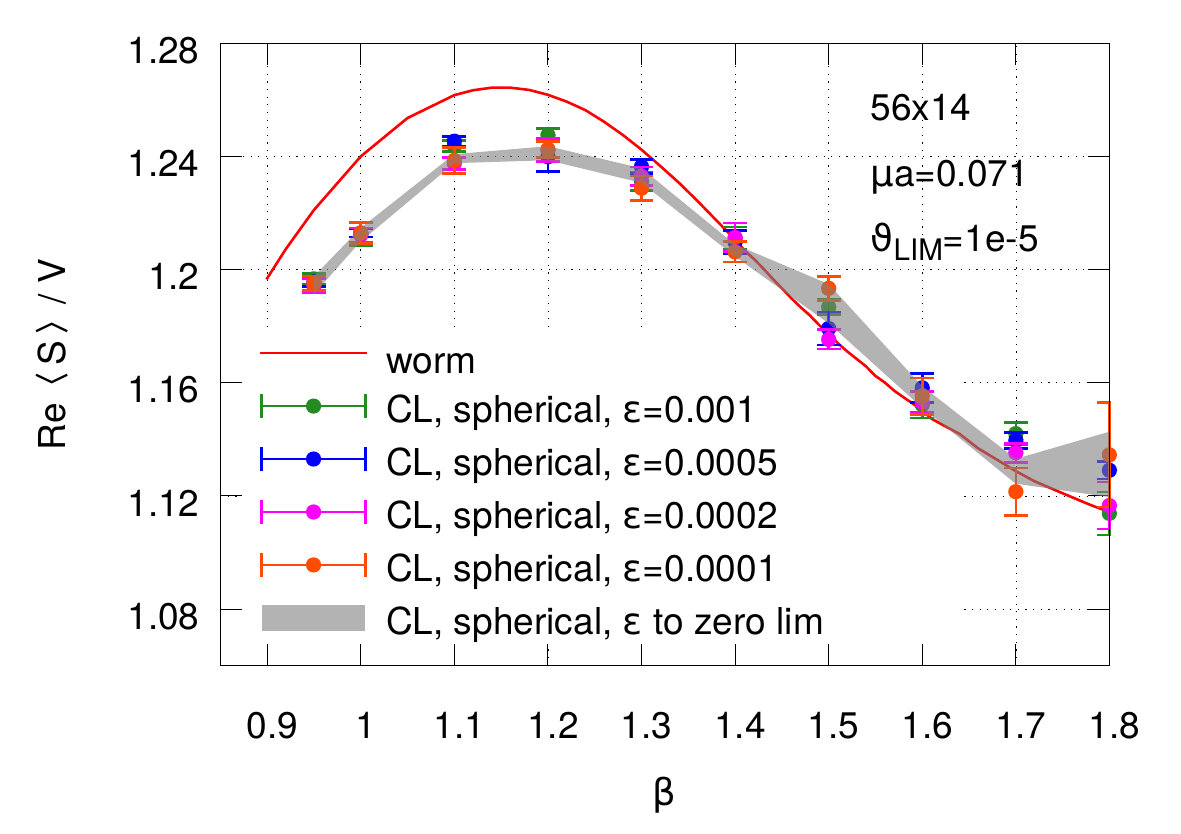}
\includegraphics[scale=0.62]{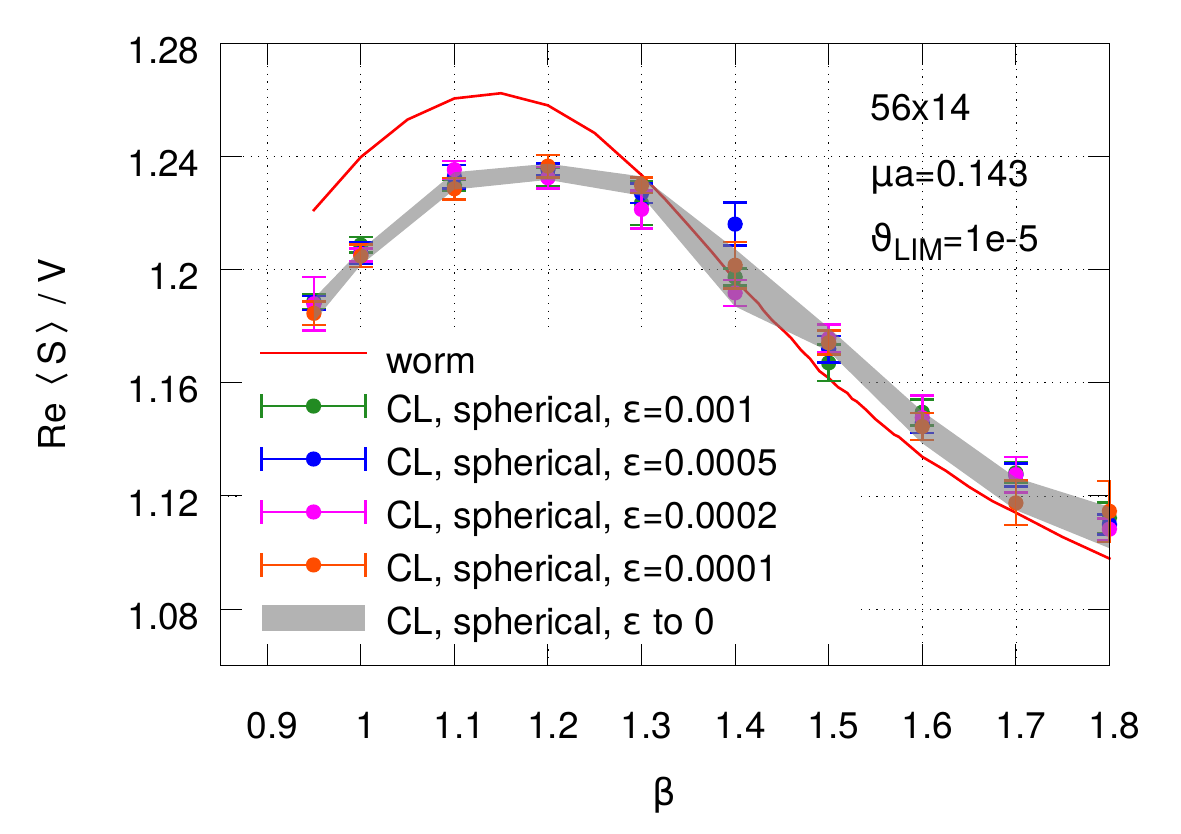}
\vspace{-0.3cm}
\caption{CL results for ${\textrm{Re}}\langle S \rangle/V$ with spherical coordinates. 
Top: $\mu a$ = 0.071 ($\mu$/T = 1), bottom: $\mu a$ = 0.143 ($\mu$/T = 2).}
\label{spherical}
\end{center}
\end{figure}
We found that at $\mu a = 0$, spherical CL results agree completely with the correct results. 
At $56 \times 14, \mu a = 0.018$ they deviate below $\beta \sim 1.3$. 
This clearly shows that the wrong convergence property is not the consequence of the severeness 
of the sign problem, because it was very mild at $\mu a = 0.018$ (see Fig. \ref{fig:rew_01}, 
right panel). We note that CL results obtained with spherical coordinates seem to slightly deviate from the correct ones 
in the high $\beta$ region as well, but these differences are not significant statistically 
(they are within 2 sigma). At high $\mu / T$ however, the deviations are significant, so at e.g. $\mu a = 0.286$, 
$56 \times 14$, results for the action are wrong at all $\beta$ values. Since this discretization was 
a bit problematic due to the singularity in the force, and deviations at high $\mu / T$ seemed discouraging, 
we did not test so carefully its continuum behavior or possible improvements. However, we mention that using 
$72 \times 18$ lattices did not show any improvement at $\mu / T = 1$.
We show some results in Fig. \ref{spherical}.

We have investigated the {\bf group integration approach} (Subsection \ref{subs:cl,group}) more carefully. 
First, we compared simulation results of using $R^{(1)}$, $R^{(2)}$, or $R^{(3)}$ 
at $56\times14$ lattices at $\mu a = 0.071, \varepsilon=0.0005$ 
using $2000$ Langevin trajectories at several $\beta$ in the range $0.9 \ldots 1.8$. We found complete 
agreement using these parameters. Then we used $R^{(2)}$ during our further simulations with the exponentialized 
Euler-Maruyama discretizaton (abbreviated in the following and in the figures as {\textsl{exp. E-M}}). 
We carried out simulations on several lattice sizes and chemical potentials in the $\beta$ range
0.9$\ldots$1.8. The parameters for these simulations can be found in Table \ref{Table: CL group int}.
Several initial step sizes were used during these simulations and we extrapolated these to zero step size.
(However, we did not find significant step size dependence of the results obtained with step sizes $10^{-4}$ and smaller.) 
\vspace{-0.1cm}
\begin{figure}[h]
\begin{center}
\includegraphics[scale=0.62]{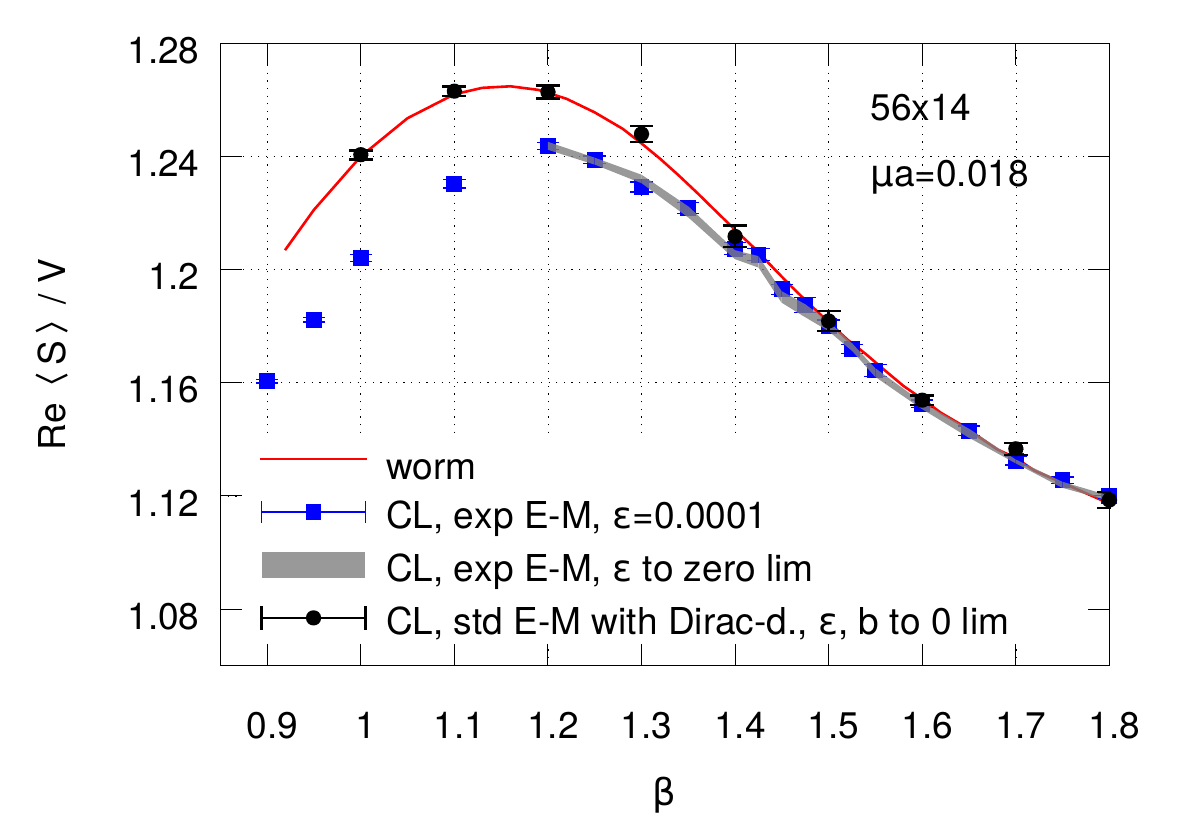}
\includegraphics[scale=0.62]{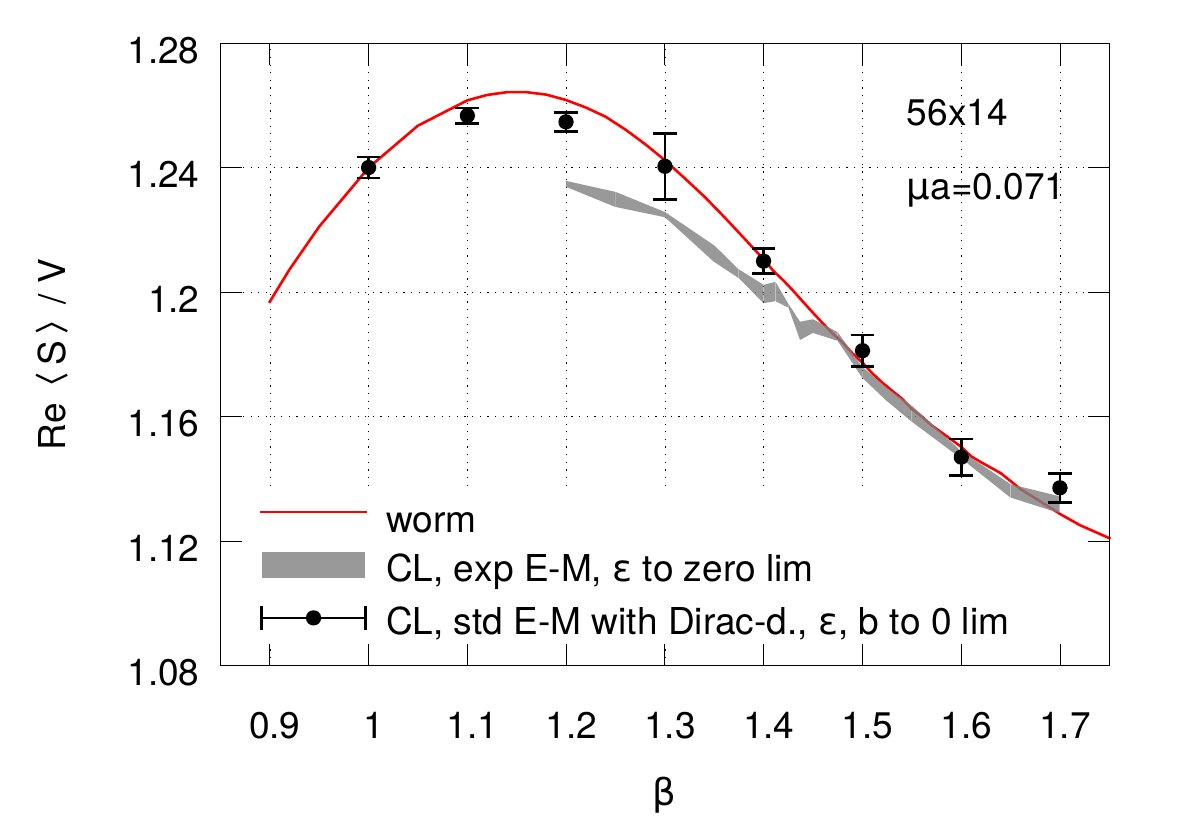}
\vspace{-0.3cm}
\caption{The comparison of ${\textrm{Re}}\langle S \rangle/V$ results for different algorithms: 
worm, CL with exponentialized E-M, CL with standard E-M discretization with Dirac $\delta$.
Top: $\mu a = 0.018$ ($\mu/T=0.25$), bottom: $\mu a = 0.071$ ($\mu/T=1$). 
Note that the exp. E-M method is wrong at low $\beta$ even when the sign problem is still very mild.}
\label{action, comp}
\end{center}
\end{figure}
\vspace{-0.4cm}

\begin{figure}[h!]
\begin{center}
\includegraphics[scale=0.6]{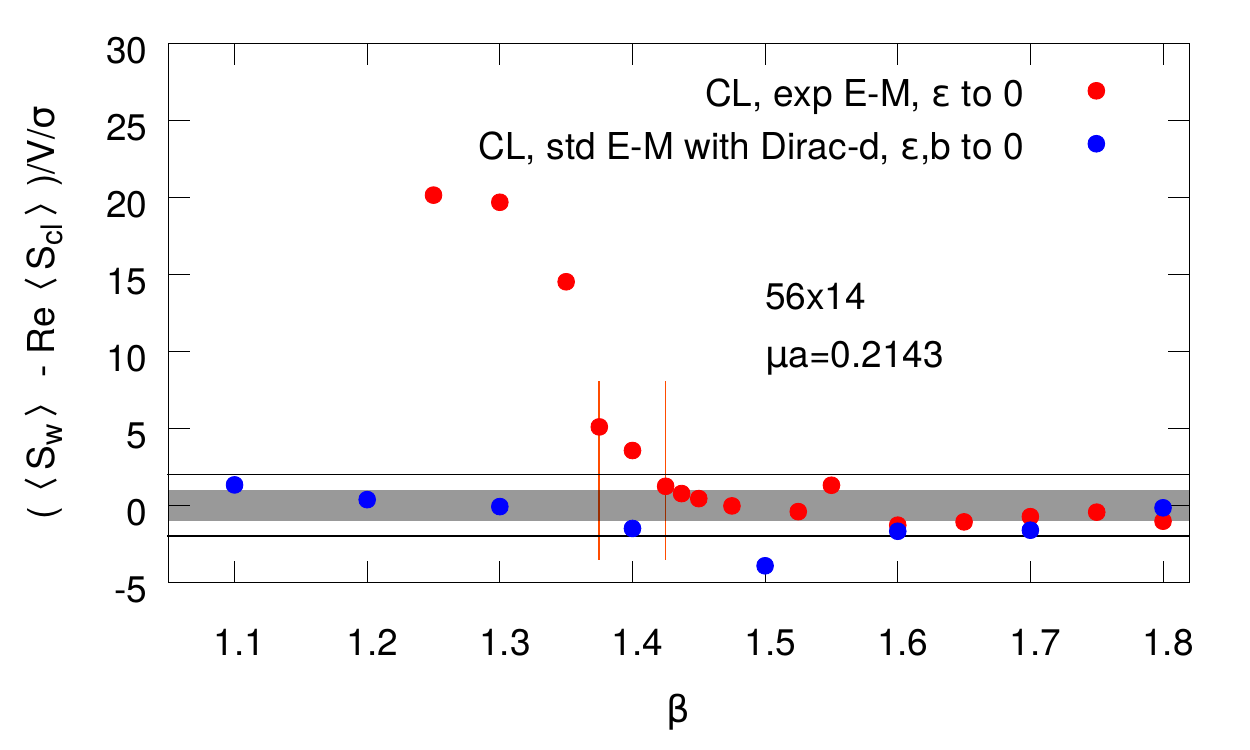}
\vspace{-0.4cm}
\caption{$(\langle S_{\mathrm{w}} \rangle - \textrm{Re}\langle S_{\mathrm{cl}} \rangle) / V / \sigma$ at $\mu a = 0.143$ ($\mu/T=2$) (top) and at $\mu a = 0.2143$ ($\mu/T=3$) (bottom). The gray band shows the 1 sigma interval, and the black lines show the 
2 sigma interval. The orange lines show the beta threshold range, below which exp. E-M complex Langevin fails.}
\label{2sigma}
\end{center}
\end{figure}

We came to similar conclusions as with spherical coordinates: at $\mu / T = 0$ complete agreement was found, then 
at $\mu / T$ nonzero -- even at $\mu / T = 0.25$ --, a discrepancy in the low $\beta$ region. We note however, 
that the exp. E-M. results do not deviate from the correct results at higher $\beta$ values.
In order to quantify the deviations and investigate their continuum behavior, we first calculated 
$(\langle S_{\mathrm{w}} \rangle - \textrm{Re}\langle S_{\mathrm{cl}} \rangle) / V / \sigma$, where 
$\langle S_{\mathrm{w}} \rangle$ is the worm result, $\langle S_{\mathrm{cl}} \rangle$ is the exp. E-M result 
in the $\varepsilon \rightarrow 0$ limit, $V=N_x \times N_t$ is 
the lattice volume and $\sigma = \sqrt{\Delta_{\mathrm{w}}^2 + \Delta_{\mathrm{cl}}^2}$, 
where $\Delta_{\mathrm{w}}$, $\Delta_{\mathrm{cl}}$ are the full errors 
of the worm and CL simulations, respectively. ($\Delta_{\mathrm{w}}$ is just the statistical error, 
but $\Delta_{\mathrm{cl}}$ contains systematic errors because of the step size extrapolation.)
After that, we determined a $\beta$ region at each lattice size and $\mu a$ parameter, when 
$(\langle S_{\mathrm{w}} \rangle - \textrm{Re}\langle S_{\mathrm{cl}} \rangle) / V / \sigma$ started to be above 2.
This definition of the $\beta$ range (the latest point under 2 sigma, and two successive points above 2 sigma) 
is a bit ambiguous in the sense that it depends on the statistics, but we did not find significant 
differences between the results of some smaller statistics runs and the long runs.
Figure \ref{2sigma} shows an example of the determination of this $\beta$ range. 

\vspace{-0.1cm}
\vspace{-0.8cm}
\begin{figure}[H]
\begin{center}
\includegraphics[scale=0.6]{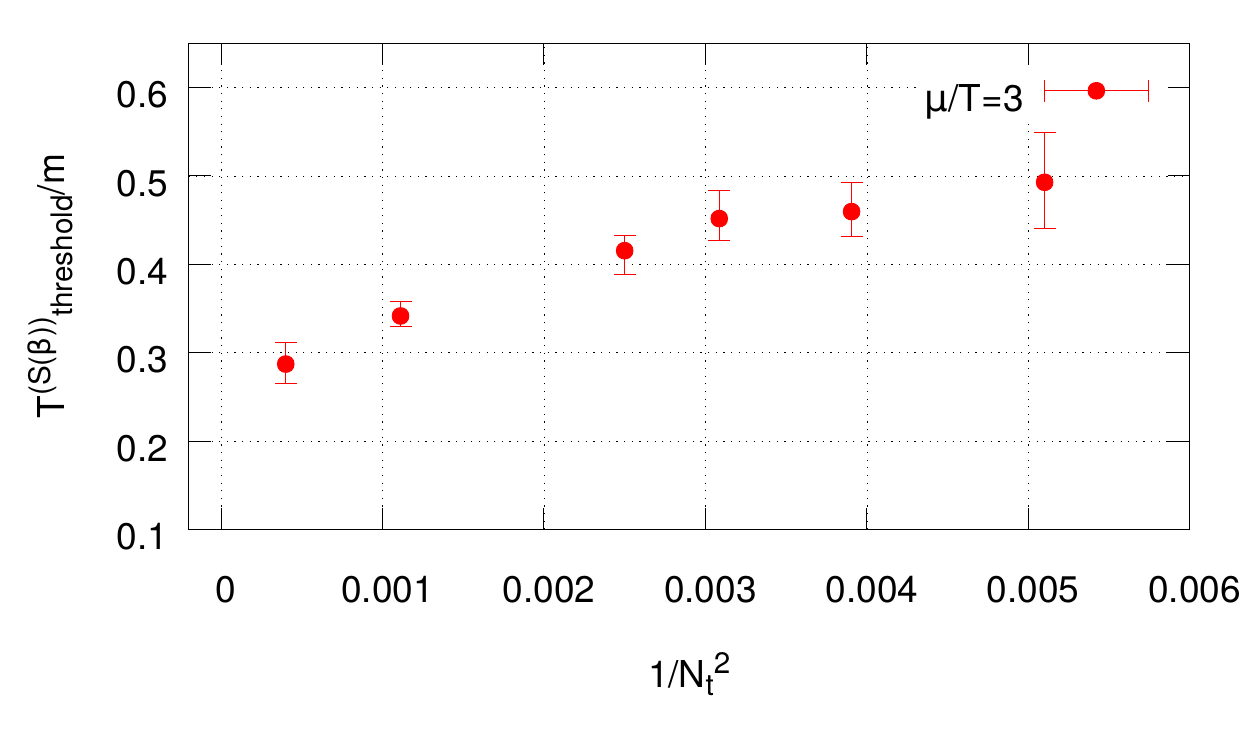}
\vspace{-0.4cm}
\caption{The temperature $T^{(S(\beta))}_{\mathrm{threshold}}/m$ as a function of $1/N_t^2$ at $\mu/T=3$.
Below $T^{(S(\beta))}_{\mathrm{threshold}}/m$ the action densities obtained by the 
exp. E-M discretization start to become wrong.
Using larger lattices this temperature becomes lower, but the continuum extrapolation of this quantity 
cannot be made without assuming some functional dependence of it as a function of $1/N_t$, 
which would hinder drawing the conclusion.}
\label{T_thresh from S(beta)}
\end{center}
\vspace{-0.4cm}
\end{figure}

Then, the middle of this range with errors to cover the whole range was used to calculate 
$T^{(S(\beta))}_{\mathrm{threshold}}/m$, the temperature below which CL converges to wrong results at the lattice under study.
We show how these temperatures depend on the temporal lattice size at $\mu/T=3$ in Fig. \ref{T_thresh from S(beta)}.
In that figure, one can see that these threshold temperatures become lower as $N_t$ increases, but we do not know 
the scaling of this quantity as a function of the lattice spacing or $N_t$; thus, we cannot extrapolate to the continuum 
without assumptions.
In order to avoid these assumptions, we have calculated the continuum limit of 
$(\langle S_{\mathrm{w}} (T,\mu/T) \rangle - \textrm{Re}\langle S_{\mathrm{cl}} (T,\mu/T)  \rangle) / V$ and then determined the 
\textsl{continuum} threshold temperature $T^{(S)}_{\mathrm{threshold}}/m$ below which the continuum results deviate from zero.
The results for these temperatures (with further relevant temperature ranges discussed in the text) are shown 
in Fig. \ref{T_threshold with dens}.

\vspace{-0.2cm}
\begin{figure}[H]
\begin{center}
\includegraphics[scale=0.6]{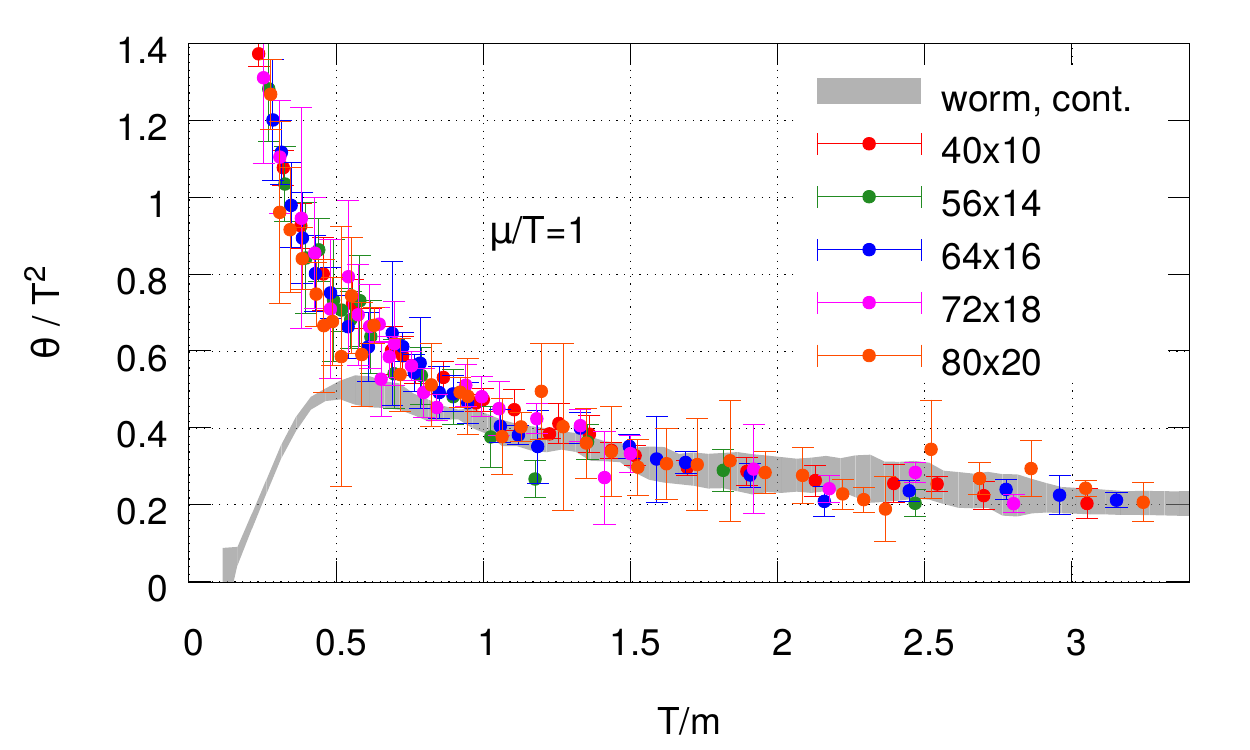}
\vspace{-0.3cm}
\caption{The trace anomaly determined with the complex Langevin algorithm using the exp. E-M discretization
 and the worm algorithm. Although we do not plot the continuum limit of the complex Langevin results here, 
the results suggest that there is no improvement toward the continuum.}
\label{trace_anom, contlim}
\end{center}
\end{figure}
\vspace{-0.8cm}

\begin{figure}[H]
\begin{center}
\includegraphics[scale=0.6]{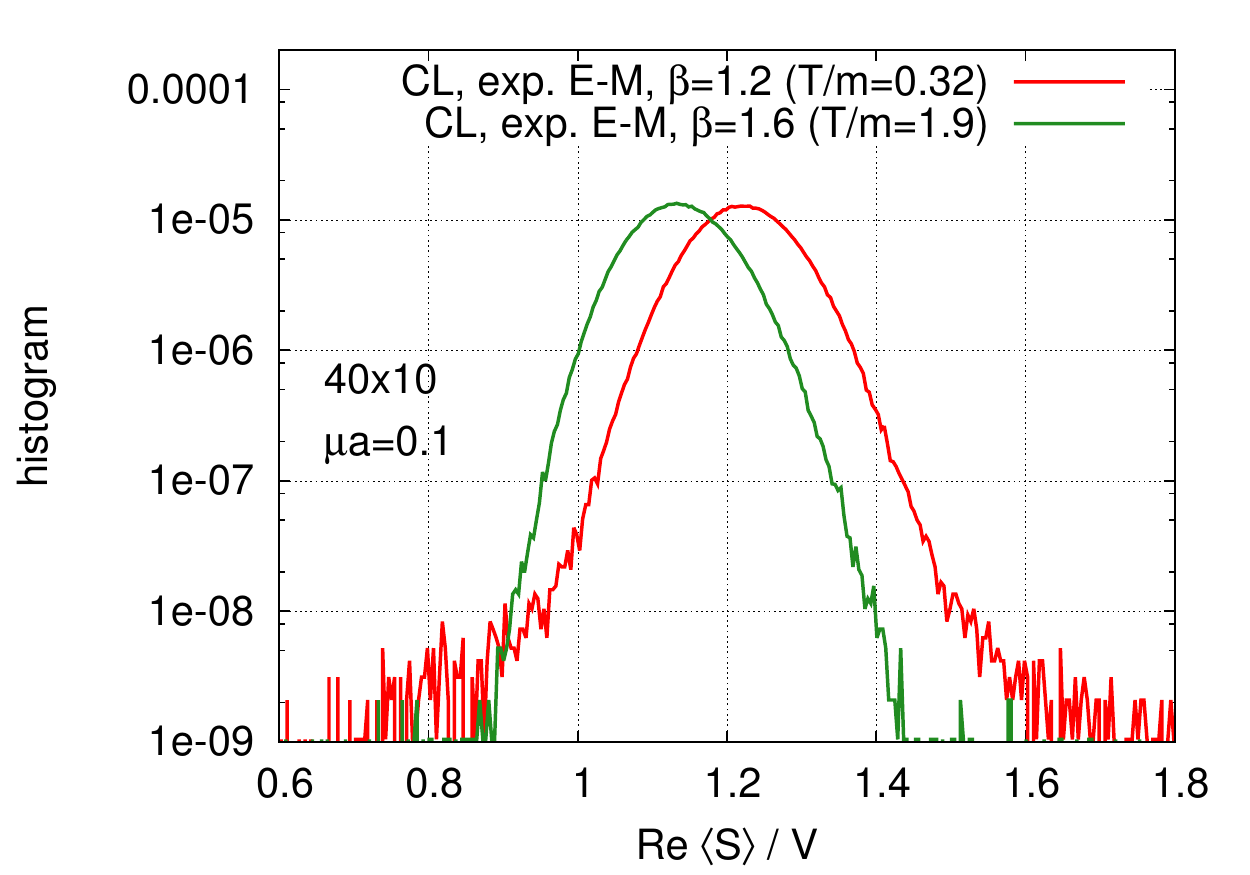}
\includegraphics[scale=0.6]{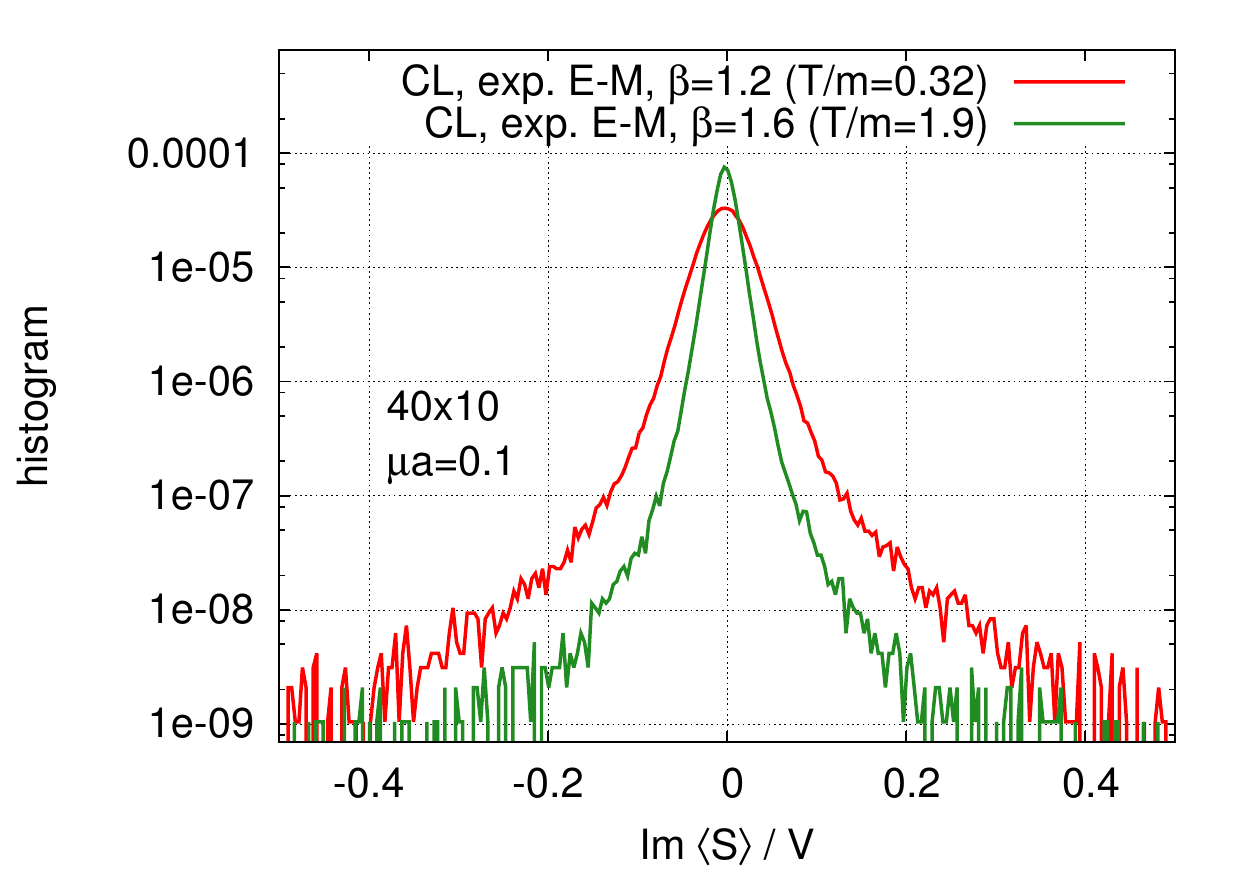}
\vspace{-0.3cm}
\caption{The histogram of $\textrm{Re}\langle S \rangle/V$ (top) and $\textrm{Im}\langle S \rangle/V$ 
(bottom) at $\beta=1.2$ and $\beta=1.6$.
These $\beta$ values correspond to $T/m=0.32$ and $T/m=1.9$; the former is such a temperature where the complex Langevin 
exp. E-M discretization develops wrong results for the action, and the latter is a temperature where it is correct.
One can see that the histograms indeed show the usual concomitant sign of wrong results, that is, the longer tail of 
the distribution of the observables.}
\label{hist, cl, S}
\end{center}
\vspace{-0.8cm}
\end{figure}

Of course, one can discuss the deviations of the correct and the CL action density in terms of a more standard 
physical quantity which has a continuum limit: the trace anomaly.
So we have also used the trace anomaly and investigated what happens with the 
continuum limit of the "wrong" complex Langevin results (Fig. \ref{trace_anom, contlim}), and obtained similar 
threshold temperature values.

We also made some runs to test this approach against a change in computer numerical precision and the order 
of integration; that is, we used float (32-bit) and long double (80-bit) precision, and found that although float and double 
differ from each other (float is wrong at all $\beta$), double and long double are almost the same at 
the parameters used to clarify this ($56\times14$, $\mu / T = 0.071, 0.143$, $\varepsilon = 0.0005$, 
$\sim 1800$ Langevin traj.). We also tested exponentialized Runge-Kutta integration at double 
precision, but results did not improve.

We have also checked the shape of the distributions, and these are shown in Fig. \ref{hist, cl, S}.
One can see that the distributions are narrower in the high temperature range.

\begin{figure}[H]
\begin{center}
\includegraphics[scale=0.6]{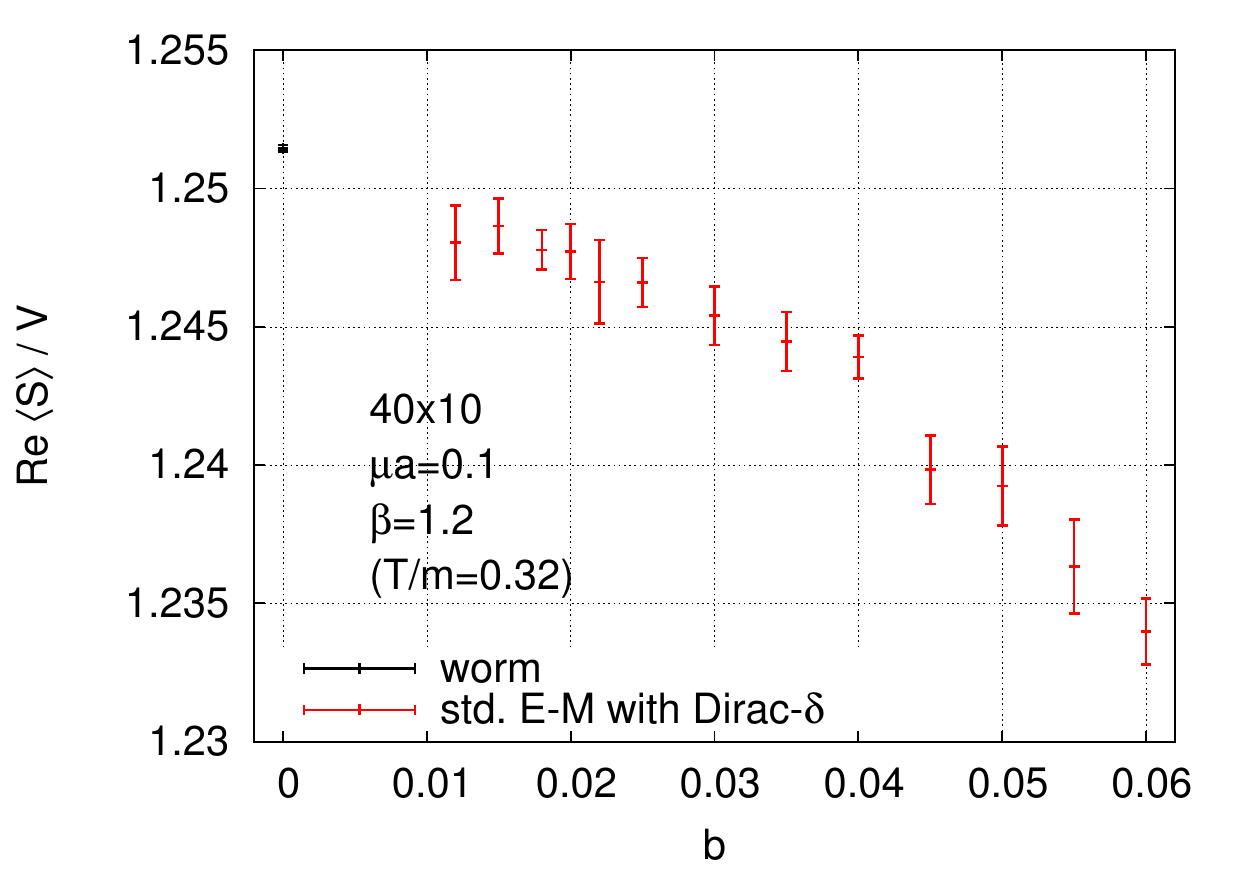}
\includegraphics[scale=0.6]{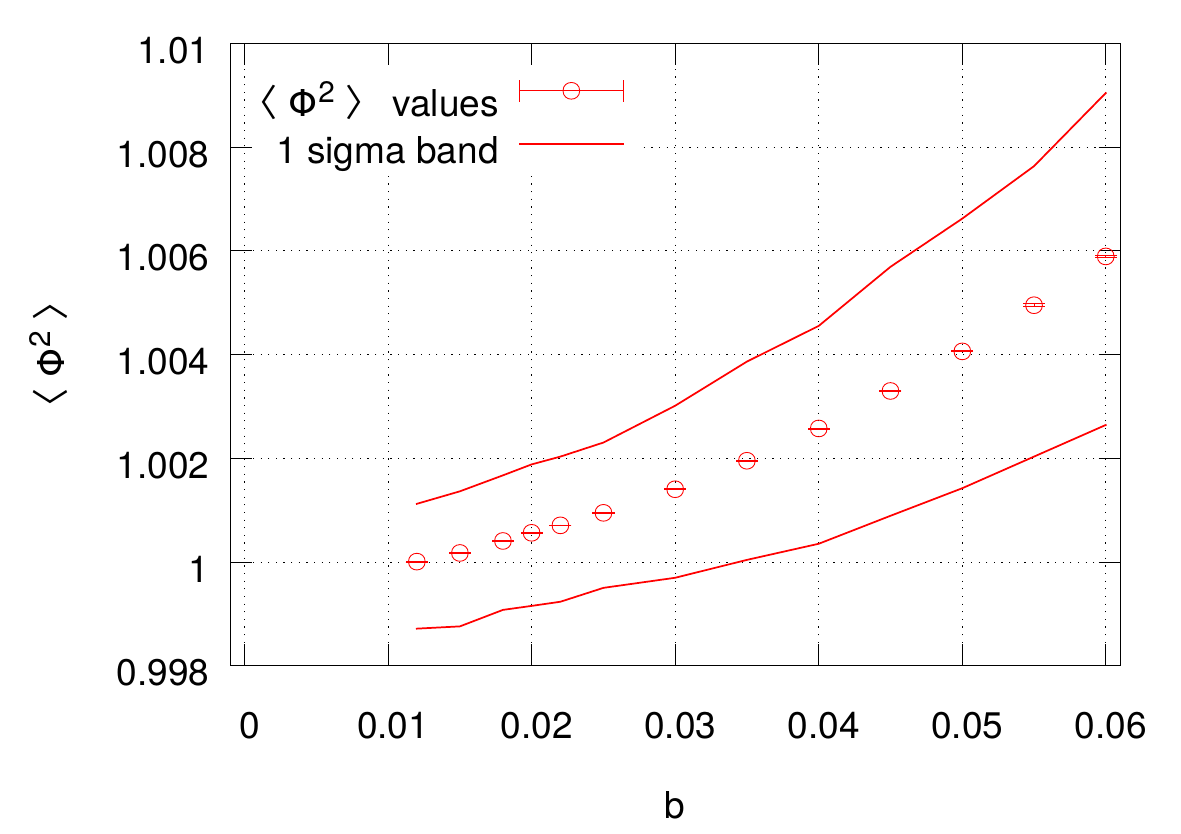}
\vspace{-0.1cm}
\caption{Top: The figure shows how taking the $b \rightarrow 0$ limit can help to obtain the 
correct results for the action with the standard E-M implementation (the starting $\varepsilon$ step size 
was $5 \cdot 10^{-5}$ during the simulations to obtain the data points of the figure).
Bottom: The effect of the reduction of the $b$ parameter on the average length of $\phi$ vectors.
The lines show the one sigma interval.}
\label{with Dirac delta, b_to_0}
\end{center}
\end{figure}

Finally, let us discuss the results obtained using the {\bf standard Euler-Maruyama discretization} 
with a Dirac $\delta$ approximated with a Gaussian.
We simulated with this algorithm at the parameters of Table \ref{Table: CL std E-M}.
We used several initial step sizes and at each step size several $b$ values ($0.01<b<0.06$).
Using double precision, we found that at a given step size,
below a low $b$ value, simulations became unstable, so there we used long double precision to set lower $b$ values. 
At each step size, we extrapolated to $b \rightarrow 0$, then used these results to extrapolate in $\varepsilon$.
As mentioned in Sec. \ref{subs:cl,std}, this algorithm did not keep the $\sum_i \phi_i^2 = 1$ 
constraint rigorously during the simulation. To characterize it quantitatively, 
we measured the length of the $\phi$ vectors over the lattice and found it is typically 
a bit larger than 1, but with smaller $b$ and $\varepsilon$ values it can be reduced (see the lower panel of 
Fig. \ref{with Dirac delta, b_to_0}).

\begin{figure}[H]
\begin{center}
\includegraphics[scale=0.6]{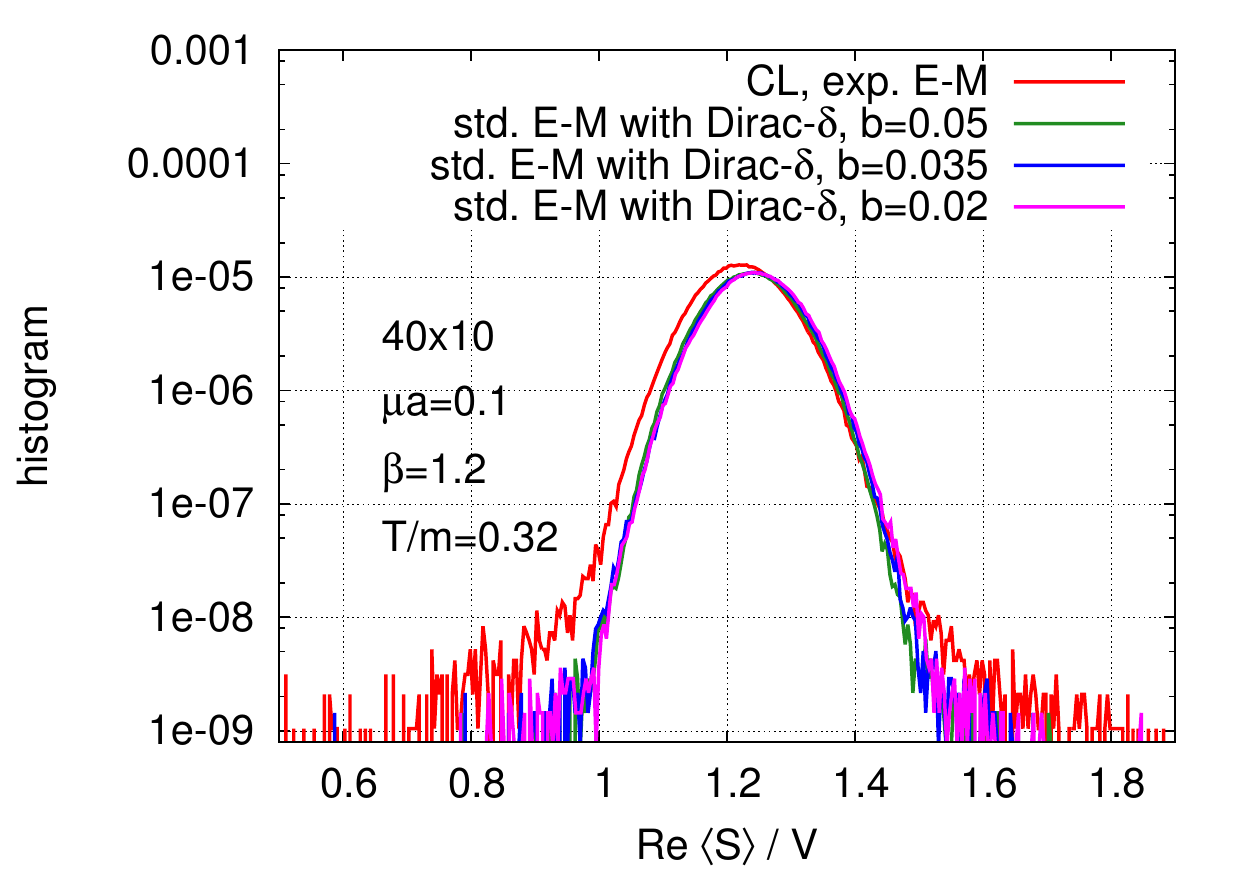}
\includegraphics[scale=0.6]{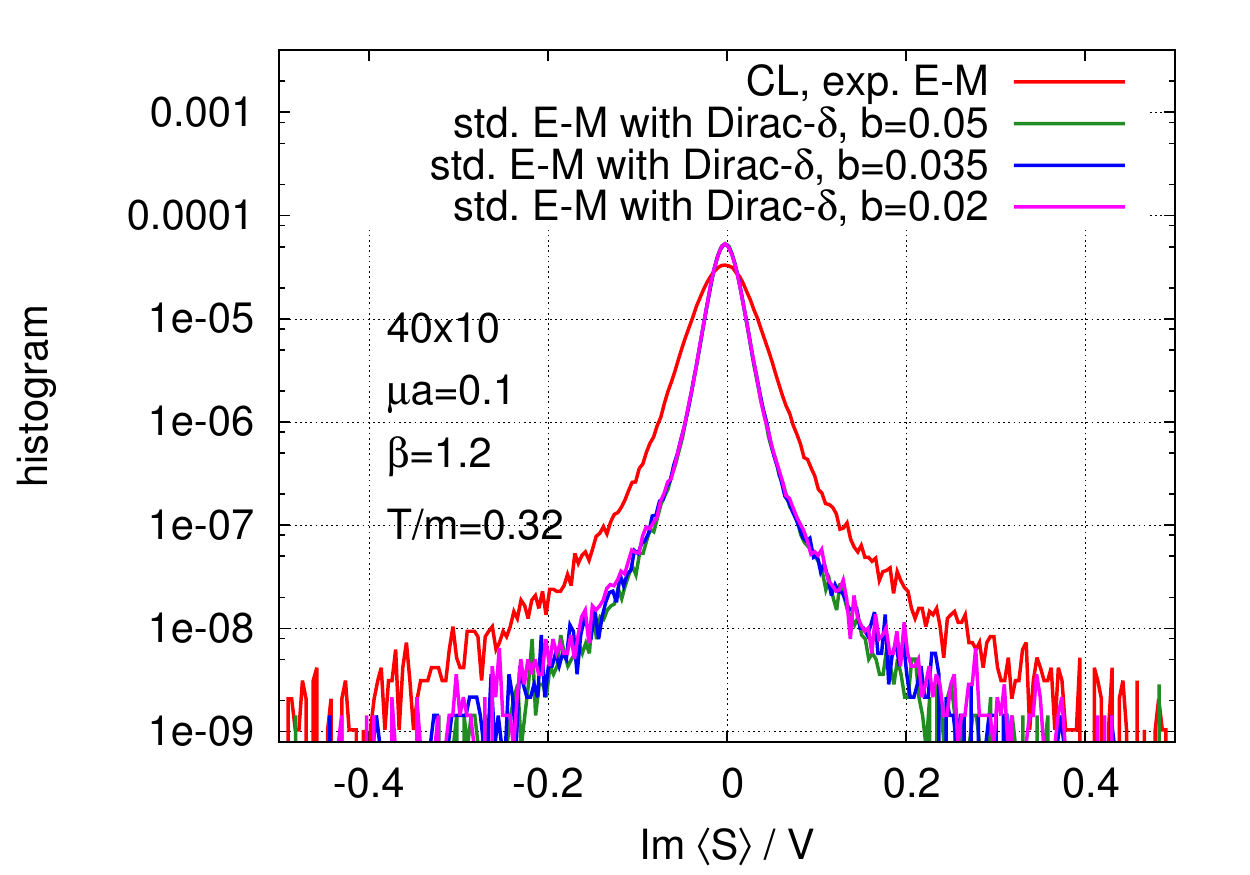}
\caption{Histograms of $\textrm{Re} \langle S \rangle/V$ (top) and $\textrm{Im}(S)/V$ (bottom) at $\beta=1.2$ obtained 
with the exp. E-M implementation and with the std. E-M implementation with Dirac $\delta$.
The std. E-M implementation develops quite similar distributions for the real part of the action, while 
narrower for the imaginary part. The shapes of the histograms do not change so much as decreasing $b$ 
(the width of the Gaussian that approximates the Dirac $\delta$), although results get closer to the 
correct ones.}
\label{hist, cl, S, std delta}
\end{center}
\end{figure}

We found that after taking both the $b$ to zero (Fig. \ref{with Dirac delta, b_to_0}, top), and 
$\varepsilon$ to zero limit, the results obtained by this method 
agree well with the correct \mbox{results (Figs. \ref{action, comp} and \ref{2sigma})}.
These results are interesting because when we accomplish the $b$ to zero and 
$\varepsilon$ to zero limits, the used data may have distributions also 
with some nonexponential decay, see Figure \ref{hist, cl, S, std delta}.
The obtained histograms were compared to those of the exp. E-M discretization 
and one can see that the decay of the standard E-M discr. with 
Dirac-$\delta$ seems to be sharper (Fig. \ref{hist, cl, S, std delta}).
For completeness, we mention here that the errors coming from the two extrapolations became significantly larger, 
especially at small $\beta$s as the chemical potential increased.

\vspace{-0.2cm}
\subsection{Density}

In the present subsection we review the results for the density (Eq. (\ref{dens_def})) obtained by the different 
algorithms. For the worm results and complex Langevin implementations, we used the same configurations 
as listed in the above subsections. 
For reweighting results we used the cluster algorithm to simulate at $\mu=0$ and used $3\times10^6$ updating steps.

We found that reweighting results agree well with the worm results below $\mu a \sim 0.16$ on $56 \times 14$ lattices, 
that is below $\mu / T \sim 2.2$ (see Fig. \ref{density_cl, 56x14}). 
At higher $\mu a$ values the results have large error bars and (apparent) deviations from correct results occur. 
This coincides with the fact that the sign problem became severe at these lattices 
around $\mu a \sim 0.15$ (Fig \ref{fig:rew_02}).

\begin{figure}[h!]
\begin{center}
\includegraphics[scale=0.65]{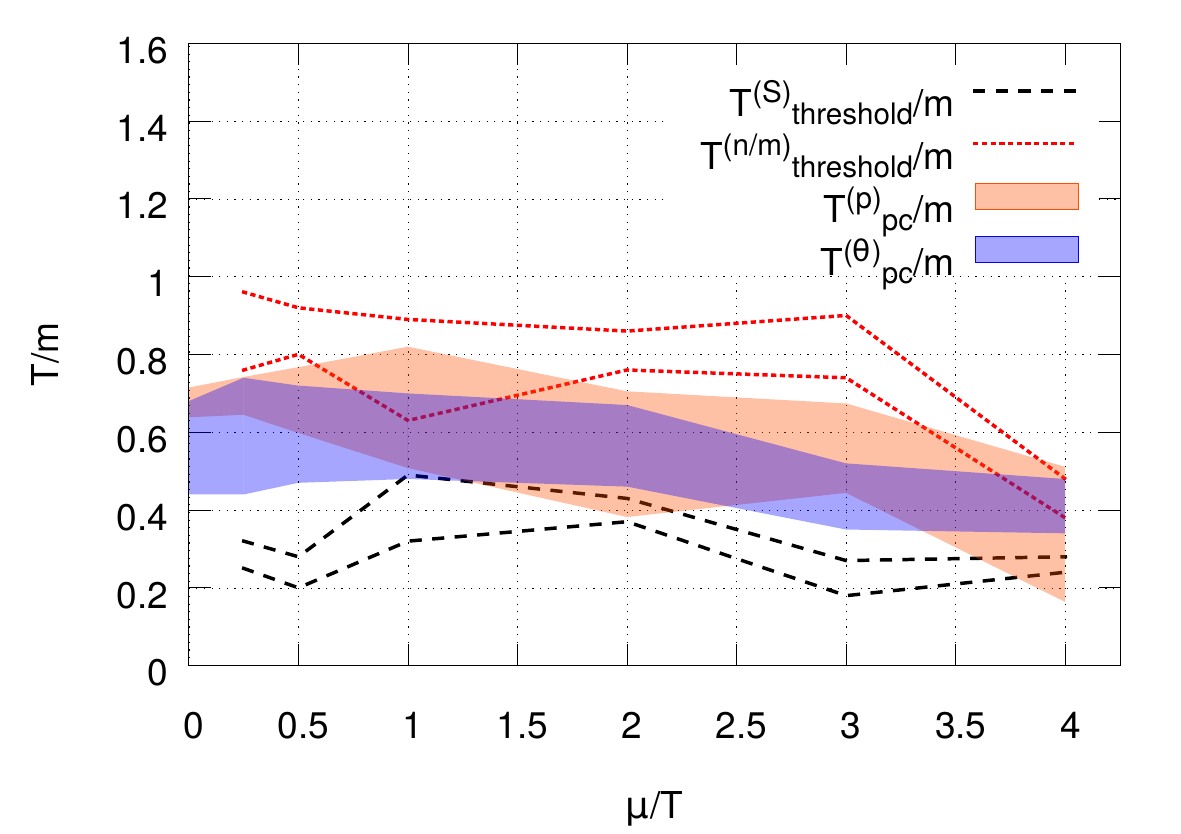}
\caption{The figure shows the temperatures $T_{\mathrm{threshold}}/m$ determined from the action differences and the 
density differences discussed in the text. Below $T^{(S)}_{\mathrm{threshold}}/m$, the exp. E-M discretization develops 
wrong results for the action. The std. E-M discretization with Dirac $\delta$ produces correct results 
for the action at all temperature. However, both implementations give wrong results for the density 
below $T^{(n/m)}_{\mathrm{threshold}}/m$. The blue and red solid bands show the pseudo-critical temperatures determined 
from inflection point of the pressure and from the maximum of the trace anomaly, $T_{\mathrm{pc}}^{(p)}/m$ and $T_{\mathrm{pc}}^{(\theta)}/m$, 
respectively.}
\label{T_threshold with dens}
\end{center}
\end{figure}

Regarding the different CL implementations we found that both the exp. E-M integration and 
the standard E-M with Dirac $\delta$ produced wrong results at low temperature (Fig. \ref{density_cl, 56x14}).
A threshold temperature ($T^{(n/m)}_{\mathrm{threshold}}/m$) can be defined similarly as we did in the case of the action:
this is the temperature below which continuum CL density results deviate from the continuum worm density results 
with 2 sigma significance. A comparison of the continuum results from the exp. E-M. CL and the worm algorithm can be seen 
in Fig. \ref{density_cl, contlim}. We note that in the low temperature region, the continuum extrapolation 
takes the CL results even further from the worm continuum results.
By increasing the chemical potential one can observe that $T^{(n/m)}_{\mathrm{threshold}}/m$ is approximately constant, 
then gets smaller (see Figs. \ref{dens_cl, mupT=1,4} and \ref{T_threshold with dens}). 
We note, that the $T^{(n/m)}_{\mathrm{threshold}}/m$ values are approximately the same for the exp. E-M. discretization 
and for the standard E-M. discretization with Dirac-$\delta$.

The spherical CL implementation also developed wrong results, but its threshold temperature seemed to be larger
than that of the others. However, we note that in the case of the spherical formulation we did not determine so 
carefully the threshold temperature.


The distributions of the real and imaginary parts of ${\partial S}/{\partial (\mu a)}$ are again 
narrower, when different complex Langevin algorithms produce correct results. 
Comparing the distributions for this quantity of the exp. E-M and std. E-M with Dirac-$\delta$ 
implementations, we see that, although the latter is narrower, the results do not converge to the correct results as 
in the case of the action they did.

\begin{figure}[h!]
\begin{center}
\includegraphics[scale=0.6]{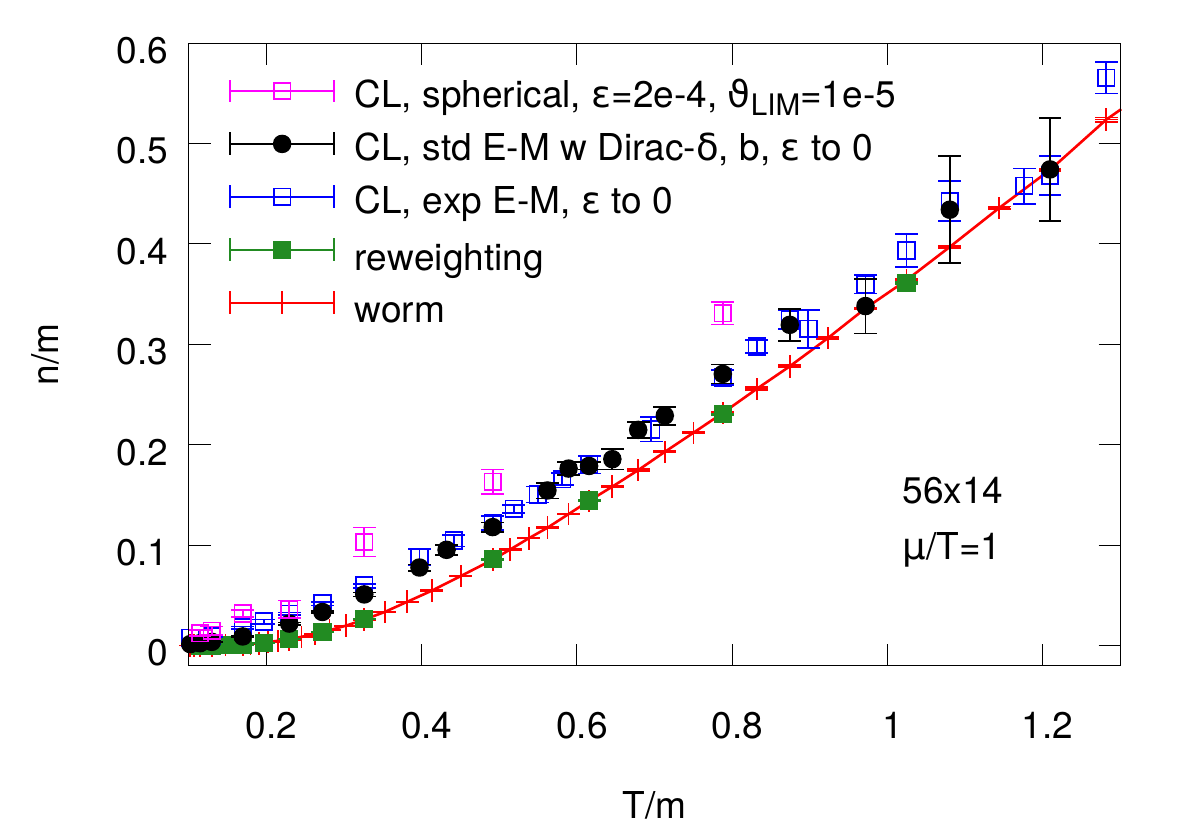}
\includegraphics[scale=0.6]{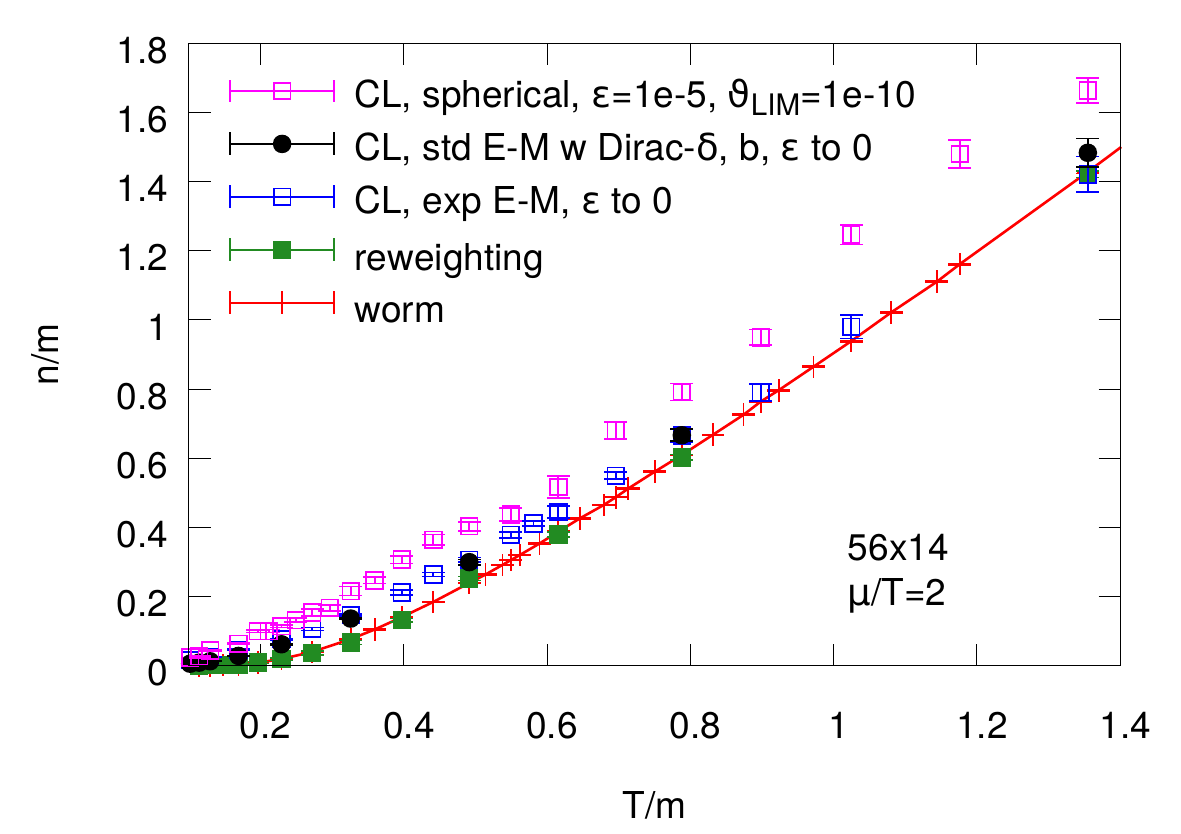}
\includegraphics[scale=0.6]{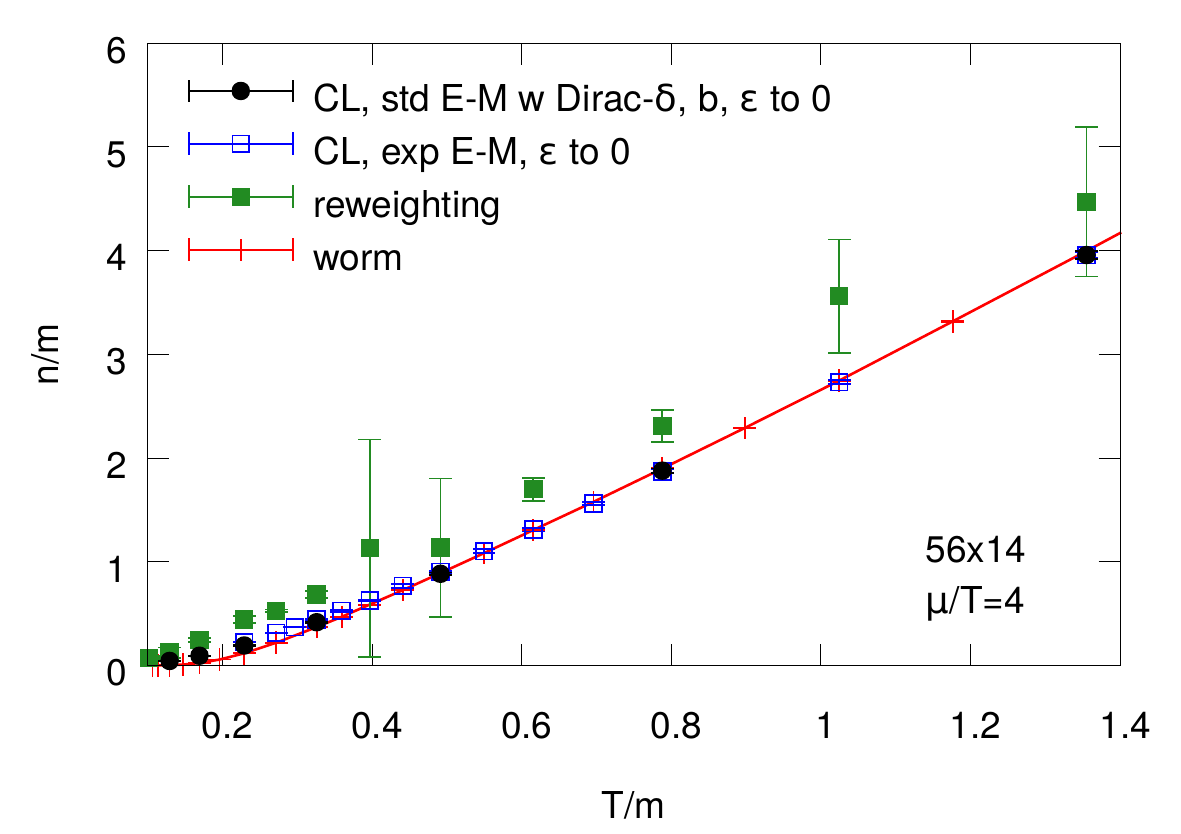}
\caption{The density at low temperature measured on $56\times14$ lattices, 
at $\mu/T=1$ (top), at $\mu/T=2$ (middle) and at $\mu/T=4$ (bottom).
At $\mu/T=4$ the reweighting results become unreliable, while CL remains correct
if the temperature is greater than 0.37$\ldots$0.5.}
\label{density_cl, 56x14}
\end{center}
\end{figure}

\begin{figure}[H]
\begin{center}
\includegraphics[scale=0.6]{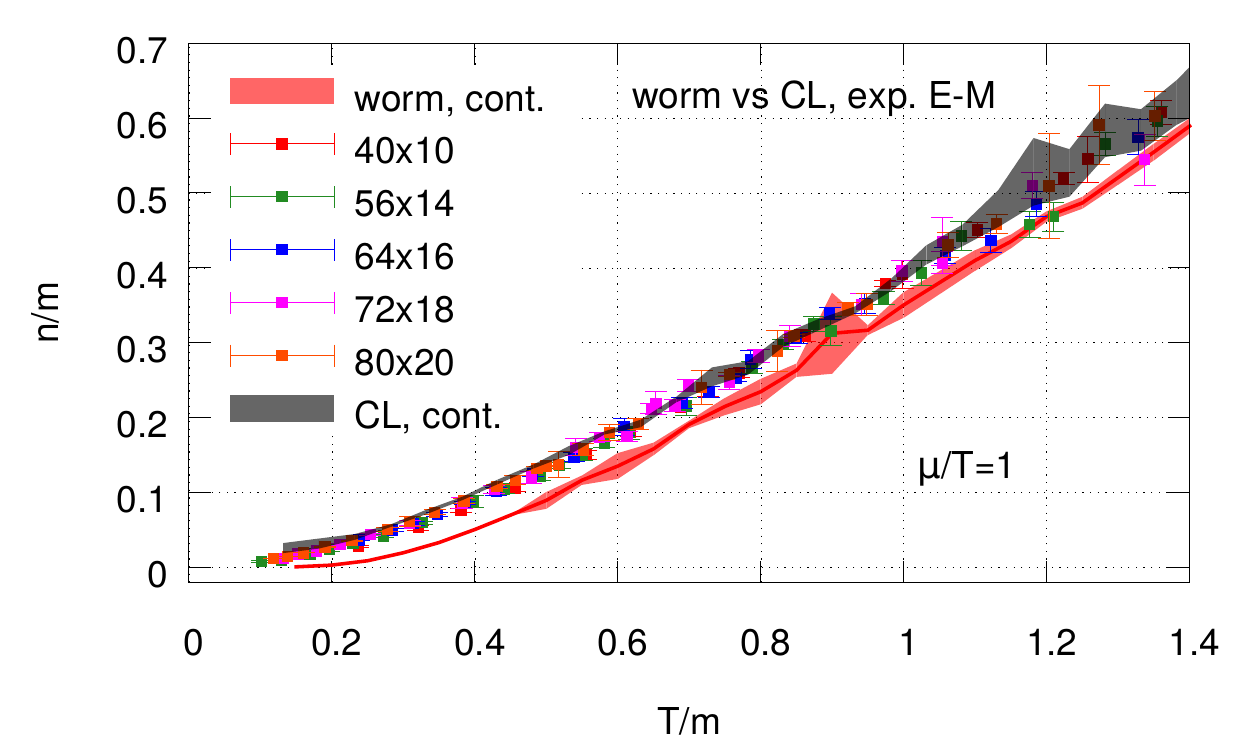}
\vspace{-0.3cm}
\caption{Comparison of the continuum results for the density obtained by the worm and the CL, exp. E-M 
discretization at $\mu / T=1$. The figure shows that the continuum extrapolation from the complex Langevin results differs 
from the worm continuum result.}
\label{density_cl, contlim}
\end{center}
\end{figure}
\begin{figure}[H]
\vspace{-0.7cm}
\begin{center}
\includegraphics[scale=0.6]{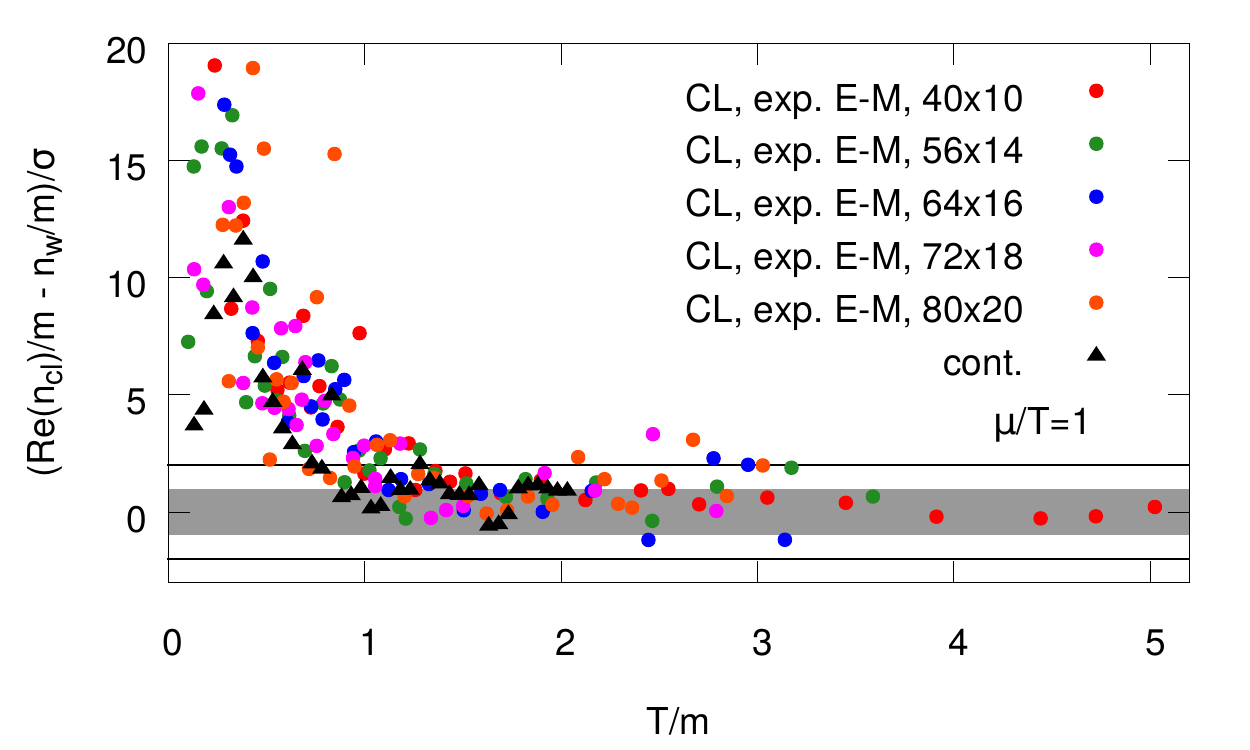}
\includegraphics[scale=0.6]{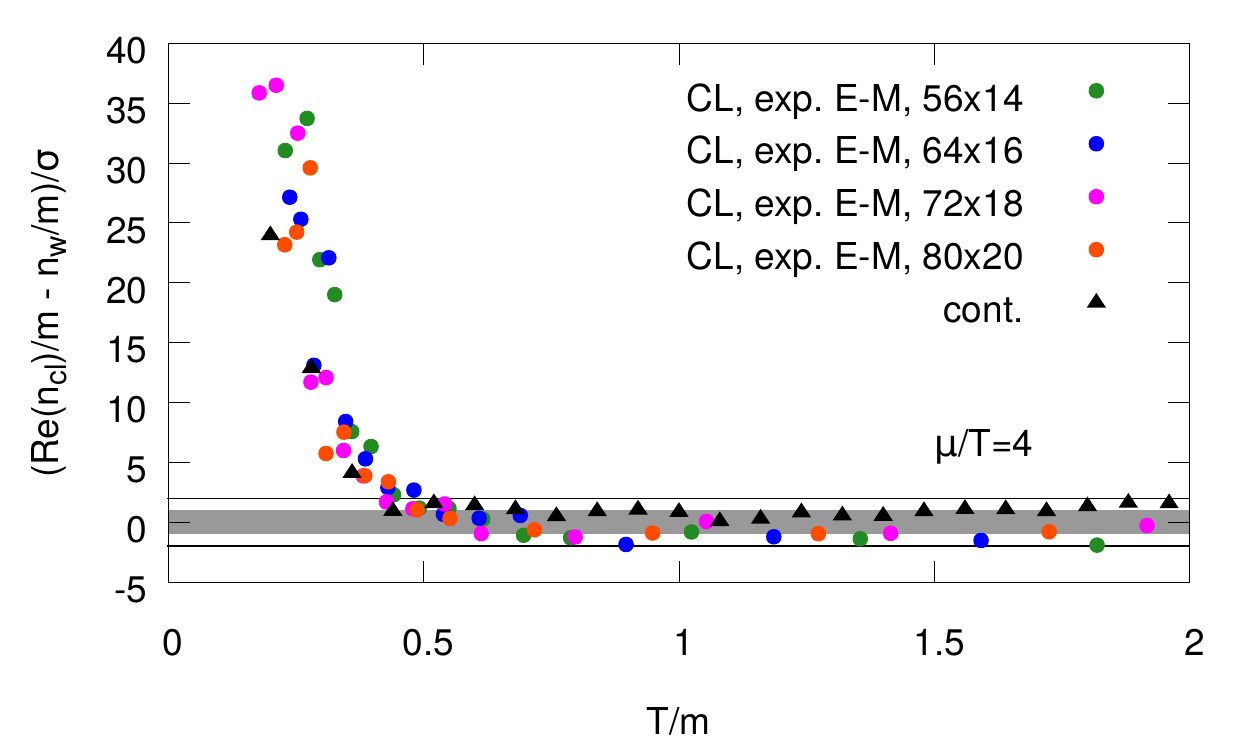}
\vspace{-0.3cm}
\caption{The density differences of different complex Langevin results ($n_{\mathrm{cl}}/m$) and the worm ($n_{\mathrm{w}}/m$)
divided by $\sigma=\sqrt{\Delta_{\mathrm{cl}}^2+\Delta_{\mathrm{w}}^2}$ at $\mu/T=1$ (top) and at $\mu/T=4$ (bottom).
The figure shows that the threshold temperature does not reduce toward the continuum.
The black triangles are calculated from the continuum CL density and the continuum worm density.}
\label{dens_cl, mupT=1,4}
\end{center}
\end{figure}

\begin{table*}
\begin{center}

\begin{tabular}{ |c|c|c|c|c|c|c| }
\hline
Method  & $N_x \times N_t$ & $\mu/T$ & $\beta$ & $\varepsilon$ & $\theta_{\rm{LIM}}$ & \# traj. \\
\hline
CL with spherical & 56$\times$14 & 0, 0.25, 0.5, 1, 2, 3, 4 & 0.9$\ldots$1.8 & 5, 2, 1$\times 10^{-4}, 10^{-5}$ & $10^{-5}$ & \pbox{3.0cm}{\vspace{0.3cm}1000$\ldots$1600} \\
\cline{2-6}
coordinates & 72$\times$18 & 1 & 0.9$\ldots$1.8 & 5, 1$\times 10^{-4}$ & $10^{-5}$ & \\
\hline
\end{tabular}
\caption{The set of parameters for the simulations with the complex Langevin algorithm using {\bf spherical coordinates}.
In the Table, $\varepsilon$ refers to the largest step size in the runs and $\theta_{\rm{LIM}}$ is the minimum distance between
any $\vartheta_x$ and $0$ ($\pi$). (See the text for the definition of $\theta_{\rm{LIM}}$.)
The $\theta_{\rm{LIM}}$ dependence of the results was analyzed on 56$\times$14 lattices at $\mu/T$=1 by using $\theta_{\rm{LIM}}= 10^{-3}, 10^{-5},$ $10^{-7}, 10^{-8}, 10^{-11}$. These are not listed in the table.}
\label{Table: CL spherical}
\vspace{0.5cm}

\begin{tabular}{ |c|c|c|c|c|c| }
\hline
Method & $N_x \times N_t$ & $\mu/T$ & $\beta$ & $\varepsilon$ & \# traj. \\
\hline
	& 32$\times$8 & 0.25 & 0.9$\ldots$1.8 & (5, 2, 1)$\times 10^{-4}$ & 4500$\ldots$5500 \\
\cline{2-6}
	& 40$\times$10 & 0.25, 0.5, 1 & 0.9$\ldots$1.8 & (10, 5, 2, 1, 0.5)$\times 10^{-4}$ & 3000$\ldots$5000 \\
\cline{2-6}
	& 56$\times$14 & 0, 0.25, 0.5, 1, 2, 3, 4 & 0.9$\ldots$1.8 & (10, 5, 2, 1)$\times 10^{-4}$ & 3000$\ldots$5500 \\
\cline{2-6}
CL with group integration & 64$\times$16 & 0.25, 0.5, 1, 2, 3, 4 & 1.1$\ldots$1.8 & (10, 5, 2, 1, 0.5)$\times 10^{-4}$ & 2000$\ldots$3500 \\
\cline{2-6}
(exp. E-M method) & 72$\times$18 & 0.25, 0.5, 1, 2, 3, 4 & 1.1$\ldots$1.8 & (10, 5, 2, 1)$\times 10^{-4}$ & 2500$\ldots$5500 \\
\cline{2-6}
	& 80$\times$20 & 0.25, 0.5, 1, 2, 3, 4 & 1.1$\ldots$1.8 & (10, 5, 2, 1)$\times 10^{-4}$ & 3000$\ldots$5500 \\
\cline{2-6}
	& 120$\times$30 & 0.5, 3 & 1.1$\ldots$1.8 & (1, 0.5, 0.2, 0.1)$\times 10^{-4}$ & 1000$\ldots$2000 \\
\cline{2-6}
	& 200$\times$50 & 0.5, 3 & 1.2$\ldots$1.8 & (1, 0.8, 0.5, 0.2, 0.1)$\times 10^{-4}$ & 800$\ldots$1200 \\
\hline
\end{tabular}
\caption{The set of parameters for the simulations with the complex Langevin algorithm 
using the {\bf group integration approach} (exponentialized Euler-Maruyama method). 
In the Table, $\varepsilon$ refers to the largest step size during the trajectories.}
\label{Table: CL group int}

\begin{tabular}{ |c|c|c|c|c|c|c| }
\hline
Method & $N_x \times N_t$ & $\mu/T$ & $\beta$ & $\varepsilon$ & $b$ & \# traj. \\
\hline
	& 40$\times$10 & 0.25, 0.5, 1 & 0.9$\ldots$1.8 & (5, 2, 1, 0.5, 0.2)$\times 10^{-4}$ & 0.01$\ldots$0.06 & 2000$\ldots$3000 \\
\cline{2-7}
CL with direct constraint & 56$\times$14 & 0, 0.25, 0.5, 1, 2, 3, 4 & 0.9$\ldots$1.8 & (5, 2, 1, 0.5)$\times 10^{-4}$ & 0.015$\ldots$0.05 & 1800$\ldots$3000 \\
\cline{2-7}
(standard E-M method with & 64$\times$16 & 0.25, 1, 2, 3 & 1.0$\ldots$1.8 & (1, 0.8, 0.5)$\times 10^{-4}$ & 0.02$\ldots$0.05 & 2000$\ldots$3500 \\
\cline{2-7}
Dirac $\delta$) & 72$\times$18 & 1, 2, 3 & 1.0$\ldots$1.8 & (1, 0.8, 0.5)$\times 10^{-4}$ & 0.018$\ldots$0.038 & 1600$\ldots$2500 \\
\cline{2-7}
	& 80$\times$20 & 1, 2 & 1.1$\ldots$1.8 & (2, 1, 0.8, 0.5)$\times 10^{-4}$ & 0.02$\ldots$0.04 & 1600$\ldots$2500 \\
\hline
\end{tabular}
\caption{The set of parameters for the simulations with the complex Langevin algorithm 
using the {\bf standard Euler--Maruyama discretization} with directly including the constraint by approximating 
the Dirac $\delta$. $\varepsilon$ refers to the largest step size during the trajectories and $b$ is the width of the
Gaussian approximating the Dirac $\delta$ (see Section \ref{subs:cl,std}).}
\label{Table: CL std E-M}

\end{center}
\end{table*}

\newpage
\section{Conclusion}
In the present paper we studied the sign problem in the O(3) model.
We used reweighting in order to investigate the severeness of the sign and overlap problems. 
We described a dual formalism and based on that a worm algorithm, which completely solves the sign problem.
Using this, we have calculated the pressure, the trace anomaly and the density at finite temperature.
At zero and low temperatures we reproduced the Silver Blaze phenomenon. 
We then analyzed the correctness of the complex Langevin as approaching the continuum limit.
The failure of the complex Langevin in certain parameter ranges is argued to be the 
consequence of the developing long tailed distributions. However, it is an interesting 
question whether the wrong convergence property happens at a specific $\beta$ value 
-- as was found in HDQCD \cite{Aarts:2013nja} -- or at a given temperature. In the former case 
continuum extrapolation could enable one to study the full phase diagram of the given model.
According to recent results \cite{Fodor:2015doa} the breakdown seems to prevent the exploration of 
the confined region in QCD. However, those simulations used only $N_t=4,6,8$ lattices. 
In the present paper our main goal was to investigate whether 
taking the continuum limit in the 1+1 dimensional O(3) model can help to overcome the 
wrong convergence property of complex Langevin.

For this purpose we have used three different approaches: 
describing the model with spherical coordinates; using the generators and integrating the 
CL equations in the group space (exp. Euler-Maruyama); and considering the $\sum_i \phi_i^2 = 1$ constraint
with a term containing the logarithm of a Dirac $\delta$ in the action (standard Euler-Maruyama discretization with Dirac $\delta$). 
We approximated the Dirac-$\delta$ by a Gaussian having a width $\sim 1/b$. 
We analyzed three observables: the action, the trace anomaly and the density. 
Regarding the action, we found that both the spherical coordinates and the exp. E-M discretization 
gave wrong results at the low $\beta$ (low temperature) range. For the exp. E-M discretization, 
we analyzed whether taking the continuum limit can help to improve the results and found that although the accessible 
temperature range becomes larger, one cannot explore the full low temperature region.
However, we established that the so-called standard Euler-Maruyama discr. with Dirac-$\delta$ method can give 
correct results for the action density after taking both the $b$ to zero and $\varepsilon$ to zero limits.
We found that for the density observable, the situation is different:
in this case the latter way of integrating the CL equations gave wrong results also at low $T/m$ and the results
did not seem to improve by taking larger and finer lattices.

\section{Acknowledgements}
The authors thank D\'enes Sexty, Erhard Seiler, and Falk Bruckmann for useful comments on the manuscript.
The work was supported by the Lend\"ulet program of the HAS (Program No. LP2012-44/2012) 
and by Grant No. OTKA-NF-104034 from OTKA.

\appendix
\section{Worm algorithm updating steps}

In this Appendix, we describe our updating steps of the worm algorithm in detail.
As in the beginning of Sec. \ref{sec:worm} (before Sec. \ref{subsec:num_res_worm}), we consider the general case $O(N)$ in 
$d+1$ dimensions.

\subsection{Updating steps of the worm algorithm} \label{subs:worm_upd_no_mu}

When $\mu$ is zero, we consider three simple updating steps.

{\bf 1.} Move the head of the worm from the position $u$ along a link $l$ to the new position $u^\prime$.
To keep the constraints, by this we increase or decrease the corresponding link variable $m=m_i^{(l)}$ by 1.
(Here, $i$ is the index of the worm.) The corresponding $2d$ possibilities are chosen with equal probabilities.

{\bf 1a.} Propose increasing the link variable, $m^\prime = m + 1$. The corresponding acceptance probability is
$p_{acc} = \mathrm{min}\{q,1\}$, where
\beq
  q = \frac{\beta}{m^\prime} \frac{k_i(u^\prime) + 1}{k(u^\prime) + N}.
\eeq

{\bf 1b.} Propose decreasing the link variable, $m^\prime = m - 1$. When $m = 0$, the move is rejected,
otherwise the corresponding acceptance probability is given by
\beq
  q = \frac{m}{\beta} \frac{(k^\prime(u) + N)}{(k_i^\prime(u) + 1)}.
\eeq

{\bf 2.} When the two heads of the worm coincide, the worm can jump to a new position, and change
its index $(u = v \rightarrow u^\prime = v^\prime, i \rightarrow j)$. In this case
\beq
  q = \frac{(k(u) + N - 2)( k_j(u^\prime) + 1 )}{( k_i(u) - 1 )( k(u^\prime) + N )} = \frac{ ( \hat{k}(u) + N )( \hat{k}_j(u^\prime) + 1 ) }{ (\hat{k}_i(u) + 1)(\hat{k}(u^\prime) + N) }
\eeq

{\bf 3.} Propose increasing or decreasing a link variable $m \equiv m_j^{(l)}$ by 2, without changing the
worm variables $u,v,i$.

{\bf 3a.} Propose increasing: $m \rightarrow m^\prime = m + 2$. The acceptance probability is given by
\beq
  q = \frac{\beta^2}{m^\prime(m^\prime-1)} \prod_{x \in \partial l} \frac{k_j(x)+1}{k(x)+N},
\eeq

{\bf 3b.} Propose decreasing: $m \rightarrow m^\prime = m - 2$. When $m < 2$ the proposal is rejected, otherwise
the acceptance probability is given by
\beq
  q = \frac{m(m-1)}{\beta^2} \prod_{x \in \partial l} \frac{k^\prime(x) + N}{k^\prime_j(x) + 1}.
\eeq

\subsection{Worm update with chemical potential}

According to the representation, Eq. (\ref{eq:scalar_prod}) of the scalar product, the index of the head and tail
of the worm could be $(i_u, i_v) = (-,+), (3,3), \ldots, (N,N)$; hence, we have to
distinguish two types of the worm. For the type $(i_u, i_v) = (j,j)$ the updating steps
described in A.1. remain unchanged. The same is true for updating a link variable
$m_j^{(l)}$ for $j=3, \ldots, N$. Below, we consider the case $(i_u, i_v) = (-,+)$.

\noindent {\bf 1.} Moving the $(-)$ end of the worm in direction $+\hat{\nu}$ form $u$ to $u^\prime = u+\hat{\nu}$

{\bf 1a.} Propose increasing the variable $m_+ \rightarrow m_+^\prime = m_+ + 1$ on the corresponding link:
\beq
  q = \eexp^{\mu_\nu} \frac{\beta}{m_+^\prime} \frac{k_{12}(u^\prime) + 1}{k(u^\prime) + N}
\eeq

{\bf 1b.} Propose decreasing the link variable $m_- \rightarrow m_-^\prime = m_- - 1$: if $m_- = 0$, then
the proposal is rejected; otherwise,
\beq
  q = \eexp^{\mu_\nu} \frac{m_-}{\beta} \frac{k^\prime(u) + N}{k^\prime_{12}(u) + 1}.
\eeq

\noindent {\bf 2.} Moving the $(-)$ end of the worm in direction $-\hat{\nu}$ from $u$ to $u^\prime = u - \hat{\nu}$:

{\bf 2a.} Propose increasing the link variable $m_- \rightarrow m_-^\prime = m_- + 1$:
\beq
  q = \eexp^{-\mu_\nu} \frac{\beta}{m_-^\prime} \frac{k_12(u^\prime) + 1}{k(u^\prime) + N}.
\eeq

{\bf 2b.} Propose decreasing the link variable $m_+ \rightarrow m_+^\prime = m_+ - 1$. If $m_+ = 0$, then
the proposal is rejected; otherwise,
\beq
  q = \eexp^{-\mu_\nu} \frac{m_+}{\beta} \frac{k^\prime(u) + N}{k_{12}^\prime(u) + 1}.
\eeq
The acceptance probabilities for moving the $(+)$ end of the worm are described by the same expressions,
the case $(+)$, $\pm \hat{\nu}$ is equivalent to $(-),\mp \hat{\nu}$ (both decrease/increase the $Q_{12} = +1$ line
by 1). However, because of the next updating step, one does not need to move the $(+)$ head to satisfy
ergodicity. Due to these facts, in our simulations we did not update the $(+)$ end, but only the $(-)$ end 
with $2/3$ probability.

{\bf 3.} When the two ends of the worm coincide ($u=v$) it can jump to a new position ($u^\prime = v^\prime$) and
change its index.

{\bf 3a.} $(i,i) \rightarrow (j,j)$:
\beq
  q = \frac{(\hat{k}(u) + N)(\hat{k}_j(u^\prime) + 1)}{(\hat{k}_i(u) + 1)(\hat{k}(u^\prime) + N)}
\eeq

{\bf 3b.} $(-,+) \rightarrow (-,+)$:
\beq
  q = \frac{(\hat{k}(u) + N)(\hat{k}_{12}(u^\prime) + 1)}{(\hat{k}_{12}(u) + 1)(\hat{k}(u^\prime) + N)}
\eeq

{\bf 3c.} $(-,+) \rightarrow (j,j)$:
\beq
  q = \frac{(\hat{k}(u) + N)(\hat{k}_j(u^\prime) + 1)}{(\hat{k}_{12}(u) + 1)(\hat{k}(u^\prime) + N)}
\eeq

{\bf 3d.} $(i,i) \rightarrow (-,+)$:
\beq
  q = \frac{(\hat{k}(u) + N)(\hat{k}_{12}(u^\prime) + 1)}{(\hat{k}_{i}(u) + 1)(\hat{k}(u^\prime) + N)}
\eeq

{\bf 4.} Changing the link variables $\pm$ on a given link simultaneously.

{\bf 4a.} $m_+ \rightarrow m_+^\prime = m_+ + 1, m_- \rightarrow m_-^\prime = m_- + 1$:
\beq
  q = \frac{\beta^2}{m_+^\prime m_-^\prime} \prod_{x\in\partial l} \frac{k_{12}(x) + 1}{k(x) + N}
\eeq

{\bf 4b.} $m_+ \rightarrow m_+^\prime = m_+ - 1, m_- \rightarrow m_-^\prime = m_- - 1$ (if $m_+ = 0$ or
$m_- = 0$, then the proposal is rejected):
\beq
  q = \frac{m_+ m_-}{\beta^2} \prod_{x \in \partial l} \frac{k^\prime(x) + N}{k^\prime_{12}(x) + 1}
\eeq

Note that these expressions with the chemical potential using the modified basis 
$\phi_+, \phi_-, \phi_3, \ldots, \phi_N$ look very similar to those discussed in the case without chemical potential.

\end{document}